\documentclass[a4paper,11pt]{article}
\pdfoutput=1 

\usepackage{jinstpub} 
\makeatletter
\gdef\@fpheader{Fermilab Technical Note: FERMILAB-PUB-25-0655-PPD\\
Accepted for publication in JINST}
\makeatother
\makeatletter
\let\orig@maketitle\maketitle
\renewcommand{\maketitle}{%
  \begingroup
    \renewcommand{\thefootnote}{\fnsymbol{footnote}}%
    \orig@maketitle
    \setcounter{footnote}{0}%
  \endgroup
}
\makeatother
\usepackage{lineno}

\usepackage{titlesec}

\usepackage{enumitem}
\usepackage{caption}
\usepackage{placeins}
\usepackage{afterpage}
\usepackage[utf8]{inputenc}
\usepackage{graphicx}
\usepackage{hyperref}
\usepackage{physics}
\usepackage{bm}
\usepackage{subcaption}
\usepackage{float}

\title{\boldmath Cryogenics and purification systems of the ICARUS T600 detector installation at Fermilab}

\author{
F.~Abd~Alrahman$^{8}$,
P.~Abratenko$^{27}$,
N.~Abrego-Martinez$^{5}$,
A.~Aduszkiewicz$^{8}$,
F.~Akbar$^{23}$,
L.~Aliaga~Soplin$^{26}$,
M.~Artero~Pons$^{18}$,
J.~Asaadi$^{26}$,
W.~F.~Badgett$^{7}$,
B.~Behera$^{6}$
% ...
}

\affiliation{
% ...
$^{5}$Centro de Investigaci\'on y de Estudios Avanzados del IPN (Cinvestav), Mexico City, Mexico\\
$^{6}$Colorado State University, Fort Collins, CO 80523, USA\\
$^{7}$Fermi National Accelerator Laboratory, Batavia, IL 60510, USA\\
$^{8}$University of Houston, Houston, TX 77204, USA\\
% ...
$^{18}$INFN Sezione di Padova and University of Padova, Padova, Italy\\
% ...
$^{23}$University of Rochester, Rochester, NY 14627, USA\\
% ...
$^{26}$University of Texas at Arlington, Arlington, TX 76019, USA\\
$^{27}$Tufts University, Medford, MA 02155, USA\\
% ...
}

\bigskip

\emailAdd{M. Geynisman (corresponding author): hope@fnal.gov}
\emailAdd{C. Montanari (corresponding author): cmontana@fnal.gov}

\abstract{
This paper describes the cryogenic and purification systems of the ICARUS T600 detector in its present implementation at the Fermi National Laboratory, Illinois, USA. The ICARUS T600 detector is made of four large Time Projection Chambers, installed in two separate containers of about 275 $m^3$ each. The detector uses liquid argon both as target and as active medium. For the correct operation of the detector, the liquid argon must be kept in very stable thermal conditions and the contamination of electronegative impurities must be consistently kept at the level of small fractions of parts per billion. The detector was previously operated in Italy, at the INFN Gran Sasso Underground Laboratory (LNGS), in a three-year run on the CERN to LNGS Long Baseline Neutrino Beam. For its operation on the Booster and NuMI neutrino beams at Fermilab, for the search of sterile neutrinos and measurements of neutrino--argon cross sections, the detector was moved from Gran Sasso to CERN for the upgrades required for operation at shallow depth with high intensity neutrino beams. The liquid argon containers, the thermal insulation and all the cryogenic equipment have been completely re-designed and rebuilt, following the schemes of the previous installation in Gran Sasso. The detector and all the equipment have been transported to Fermilab, where they have been installed, tested and recently put into operation. The work described in this paper has been conducted as a joint responsibility of CERN and Fermilab with the supervision provided by the ICARUS Collaboration. Design, installation, testing, commissioning and operation are the result of a common effort of CERN, Fermilab and INFN groups.
}

\arxivnumber{2509.18392}

\begin{document}
\maketitle

\section{Introduction}
\label{sec:Introduction}

\subsection{The ICARUS T600 detector and the Gran Sasso run.}

Argon, naturally present in the atmosphere at the 1\% level, has by long been recognized as an excellent
radiation detection medium. In the liquid form (Liquid Argon -> LAr) is characterized by high electron
mobility and dielectric strength and a relative density of 1.4, making it an ideal candidate as active
target for rare events searches such as proton decays, neutrino and dark matter interactions, etc.,
which require the deployment of detectors with large sensitive mass (hundreds or thousands of tons).
Electron-ion pairs are abundantly produced by the passage of charged particles in LAr: about 90,000
electron-ion pairs per centimeter are generated along the trajectory of a minimum ionizing single
charged particle (m.i.p.). By means of an appropriate electric field, electrons that survive the initial
recombination with their parent ions\footnote{Approximately 50\% of the electrons produced by a m.i.p. survive
recombination if a field of 500~V/cm is used.}
can be efficiently transported toward a set of electrodes (wires
or ``pixels'') placed at the boundary of the field region where their passage or arrival can be recorded in
the form of an electrical signal. If the electric field is reasonably uniform and the effects of LAr
turbulence or non-uniformity are negligible during the time required for the electrons to reach the
sensing electrodes, the geometry of the ionization tracks is preserved during transport in LAr. If wires
are used, multiple planes with different wires orientation can be employed to obtain different ``views''
of the same ionizing event, each view corresponding to a 2-dimensional projection of the event with
one coordinate orthogonal to the wires direction and the second one being the drift time (common to
all the views). Multiple views can be then combined to provide a 3-dimensional representation of the
event. From measurement of the amplitude of the signals, additional important information about the
event can be derived, such as energy deposition (for calorimetry), local ionization density (used e.g. for
particle identification), etc. Liquid argon is also an excellent inorganic scintillator. Scintillation light is
produced with about the same abundance as ionization charge, at a wavelength of 128~nm, in the
Vacuum UltraViolet (VUV) region. Detection of scintillation light provides, among others, the absolute
time of the ionizing event, allowing for the correct positioning of the ionization tracks along the drift
coordinate and a convenient prompt trigger signal.

\begin{figure}[htbp]
  \centering
  \includegraphics[width=0.8\textwidth]{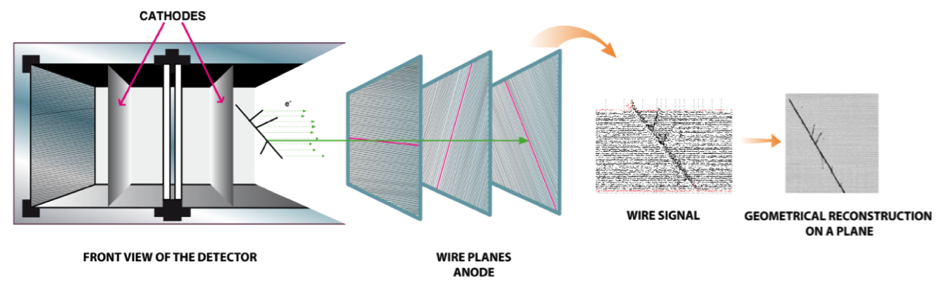}
  \caption{Sketch of the detector layout showing the LAr-TPC working mechanism.}
  \label{fig:intro1}
\end{figure}

In summary, the Liquid Argon Time Projection Chamber (LAr-TPC) is a continuously sensitive and self-triggering
detector that can provide an excellent 3D imaging and calorimetric reconstruction of ionizing
events occurring inside the sensitive volume. First proposed by C.~Rubbia in 1977~\cite{Ref1}, this detection
technique allows for a detailed study of the neutrino interactions, spanning a wide energy spectrum
(from few keV to several hundred GeV).

The ICARUS T600, with a total active mass of 476~ton, is the first large-scale operating LAr-TPC detector
\cite{Ref2}: it consists of a large cryostat split into two identical, adjacent modules. Each module houses two
LAr-TPCs separated by a common cathode and the maximum drift distance is 1.5~m, equivalent to $\sim 1$
ms drift time for the nominal 500~V/cm electric drift field. The anode is made of three parallel wire
planes (wire chambers). The other sides of the sensitive volume are covered by a set of parallel
electrodes (field cage), each one forming a rectangular ring, that are kept at a uniformly decreasing
voltage and ensure a uniform field between the cathode and the anodes. Behind the wire chambers
there are sets of photomultipliers (PMTs) to collect the scintillation light produced by the charged
particles in LAr and used for the trigger of the detector~\cite{Ref3}. ICARUS concluded in 2013 a very successful
3 years long run in the Gran Sasso underground laboratory (LNGS).

\begin{figure}[htbp]
  \centering
  \includegraphics[width=0.8\textwidth]{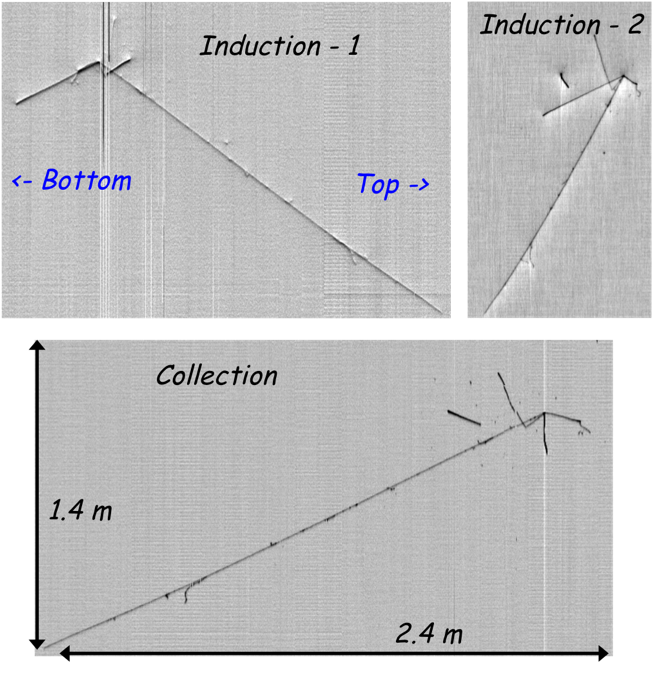}
  \caption{An atmospheric muon neutrino event ($E_{\rm DEP} \sim 1.7$~GeV) collected by ICARUS-T600 at LNGS as measured in the
three --- Induction-1, Induction-2 and Collection --- views is shown.}
  \label{fig:intro2}
\end{figure}

During the data taking the liquid argon was kept at an exceptionally high purity level ($< 50$~ppt of O$_2$
equivalent contaminants) reaching in 2013 a 16~ms lifetime corresponding to 20~ppt O$_2$ equivalent LAr
contamination~\cite{Ref4}, paving the way to the construction of larger LAr-TPC detectors with drift distances
up to 5~m. The detector has been exposed to the CERN to LNGS neutrino beam and to cosmic rays and
the recorded events demonstrate the high-level performances and the physical potentialities of this
detection technique: the detector showed a remarkable $e/\gamma$ separation and particle identification
exploiting the measurement of $dE/dx$ versus range~\cite{Ref5}. Furthermore, the momentum of escaping
muons has been measured studying the multiple Coulomb scattering with $\sim 15$\% average resolution in
the 0.4--4~GeV/c energy range, which is relevant to the next generation neutrino experiments~\cite{Ref6}. The
events related to cosmic rays have been studied to identify also the atmospheric $\nu$ interactions: 6 $\nu_\mu$CC
and 8 $\nu_e$CC events in a 0.43~kton~y exposure have been identified and reconstructed, demonstrating
that the automatic search for the $\nu_e$CC in the sub-GeV range of interest for the future short and long
baseline neutrino experiments is feasible~\cite{Ref7}.

\subsection{The sterile neutrino puzzle and the SBN program}

Some experiments, in particular LSND~\cite{Ref8}, and MiniBooNE~\cite{Ref9} have reported anomalous signals that
may imply the presence of an additional mass-squared difference $\Delta m^2_{\rm new} \sim 1.0$~eV$^2$ driving oscillations
at small distances and pointing toward the possible existence of non-standard heavier sterile
neutrino(s). A sensitive search for a possible $\nu_e$ excess related to the LSND anomaly in the CNGS $\nu_\mu$
beam has been performed using the neutrino events collected in the T600 detector during the Gran
Sasso run. 2650 CNGS neutrino interactions, identified in $7.9 \times 10^{19}$~pot exposure, have been studied in
details to identify the $\nu_e$ interactions. Globally, 7 electron-like events have been observed to be
compared with the $8.5 \pm 1.1$ expected from the intrinsic beam contamination and standard 3-flavor
oscillations: this study constrained the LSND signal to a narrow parameter region at $\sin^2 2\theta \sim 0.005$,
$\Delta m^2 < 1$~eV$^2$ that should be definitely investigated~\cite{Ref10}. The MicroBooNE experiment, which took data
between 2015 and 2021, did not observe any significant excess of events, further constraining the
allowed parameter space~\cite{Ref11}.

The SBN Short-Baseline Neutrino program at Fermilab can provide the required clarification of the LSND
anomaly. It is based on three LAr-TPC detectors (ICARUS-T600, MicroBooNE with 89~t active mass and
SBND with 82~t active mass) exposed at shallow depth to the $\sim 0.8$~GeV Booster neutrino beam at
different distances from the target (600~m, 470~m and 110~m respectively)~\cite{Ref12}. The used detection
technique will provide an unambiguous identification of the neutrino interactions, the measurement of
their energy and a strong mitigation of the possible sources of background. In addition, the study with
almost identical detectors at different distances from the source will allow to identify any variation on
the spectra, that will be a clear signature of neutrino oscillations. For these reasons, this experiment
will allow for a very sensitive search for $\nu_\mu \rightarrow \nu_e$ appearance signals, covering the LSND 99\% C.L. allowed
region at $\sim 5\sigma$ C.L. The high correlations between the event samples of the three LAr-TPC's and the
huge event statistics at the near detector will also allow for a simultaneous sensitive search in the $\nu_\mu$
disappearance and in the $\nu_e$ appearance channels.

During the data taking at Fermilab, the T600 detector will be also exposed to the off-axis neutrinos from
the NuMI beam, in the 0÷3~GeV energy range, with an enriched component of electron neutrinos (few
\%). The analysis of these events will provide useful information related to the detection efficiencies and
to the neutrino cross-sections at energies relevant to the future long baseline experiment with the
multi-kt DUNE LAr-TPC detector.

\subsection{The T600 in the SBN program}

The ICARUS-T600 detector at Fermilab takes data at shallow depth, protected by a 3~m concrete
overburden: O(10$^6$) $\nu$ interactions should be recognized amongst the $\sim 11$ cosmic muons that are
expected to cross the detector randomly in the 1~ms drift time corresponding to each triggered event.
In addition, the associated photons produced by cosmic rays can become a serious background source
for the $\nu_e$ search since the electrons produced via Compton scattering and pair production can mimic
$\nu_e$ CC events. To prepare the detector for this new SBN data taking, the T600 underwent an intensive
overhauling at CERN in the Neutrino Platform framework before being shipped to US in 2017,
introducing several technology developments while maintaining the already achieved performance.
The T600 is now equipped with an upgraded inner light detection system with 360 new 8'' photo-
multipliers, installed behind the TPC wire planes (90 PMTs in each TPC)~\cite{Ref13,Ref14}. The PMTs gain
equalization, and timing will be performed by 405~nm laser pulses flashing the PMTs via a fiber system.
This new light detection system will provide a sensitivity below 100~MeV of deposited energy, a $\sim$1~ns
time resolution with a high granularity: all these features are fundamental to effectively identify the
events associated to the neutrino beam and to measure the time of occurrence of each cosmic
interaction crossing the detector.

The overhauling of the T600 also gave the opportunity to design an upgraded ``warm'' TPC read-out
electronics, finalized to a better event reconstruction quality. This includes a front-end based on analog
low noise - charge sensitive pre-amplifier, serial 12-bit ADCs, one per channel, with 400~ns synchronous
sampling~\cite{Ref15}. A $\sim 1.5\,\mu$s faster shaping time allows to match the electron transit time in the wire plane
spacing, providing a better physical signals position separation, a drastic reduction of the undershoot
in the preamp response as well as the low frequency noise while maintaining a S/N $\geq 10$ ratio as
demonstrated with a 50-liter LAr-TPC prototype exposed to cosmic rays at CERN. Finally, a new compact
design allows to host both the analogue and digital electronics in a new A2795 CAEN board, directly
mounted on ad-hoc signal feed-through flanges and on each flange a custom mini-crate hosts nine
boards, corresponding to 576 channels.

To reject the cosmic background, during the new data taking the T600 will be surrounded also by an
external $\sim 4\pi$ segmented Cosmic Ray Tagging (CRT) system, composed by 3 subsystems each one with
two layers of plastic scintillators ($\sim 1100$~m$^2$). This additional system, in conjunction with the T600
internal PMT system, will allow to unambiguously identify cosmic rays entering the detector: the few
ns time resolution provided by this system, combined with the activity in the LAr volume will allow
measuring the direction of the particle propagation via time of flight and discriminate the
incoming/outcoming particles.

After the placement in the pit of the two ICARUS modules in August 2018, all the feedthrough flanges
for the TPCs and PMTs signals and for the injection of the laser flashes used to calibrate the PMTs have
been installed (December 2018). The gain and the dark rate for all the 360 PMTs have been measured
as a function of the applied voltage at room temperature. All the new TPC readout electronics in the 96
mini-crates and the low noise power supplies have been installed and verified: the full readout chain
has been tested injecting test pulses at far end of the chamber wires and reading out the signals by the
A2795 boards on the other end, to check the full system for noise monitoring purposes. In parallel all
the cryogenic equipment has been installed, welded and the complete system has been tested at 350
mbar overpressure. Then the cold vessels have been successfully evacuated at $10^{-5}$~mbar residual
pressure.

The detector installation has been completed at the beginning of 2020 and between February and April
the detector has been filled with Liquid Argon, and the liquid and gas recirculation systems that are
needed to purify the Argon have been activated. The LAr TPC and the LAr light detection system have
been fully activated in August 2020, starting to take cosmic rays data. Activation of the side CRT
followed immediately. First neutrino data have been recorded early 2021 and the first uninterrupted
neutrino run (RUN 0) has been conducted between May and June 2021. After verification that no
interventions requiring the use of the overhead cranes were needed, the experimental setup has been
completed with the installation of the top CRT (Dec 2021 -- Jan 2022) and of 3-meter-thick concrete
overburden (Jun 2022). Since its activation, in August 2020, the detector has been practically
continuously active, operating at the nominal field, and at nominal conditions of all sub-components.
No interruptions of data taking due to failures or instabilities have occurred. The target LAr purity
(concentration of electronegative impurities $< 0.1$~ppb of Oxygen equivalent molecules, corresponding
to a free electrons lifetime of 3~ms) was reached in the spring of 2021, and then stabilized at about two
times the nominal value in the following months. The stability of the cryogenics and argon purification
systems has been remarkable, with fluctuations in the absolute pressure within few mbar and
temperature uniformity across the entire sensitive volume within 0.1~K.

\begin{figure}[htbp]
  \centering
  \includegraphics[width=0.8\textwidth]{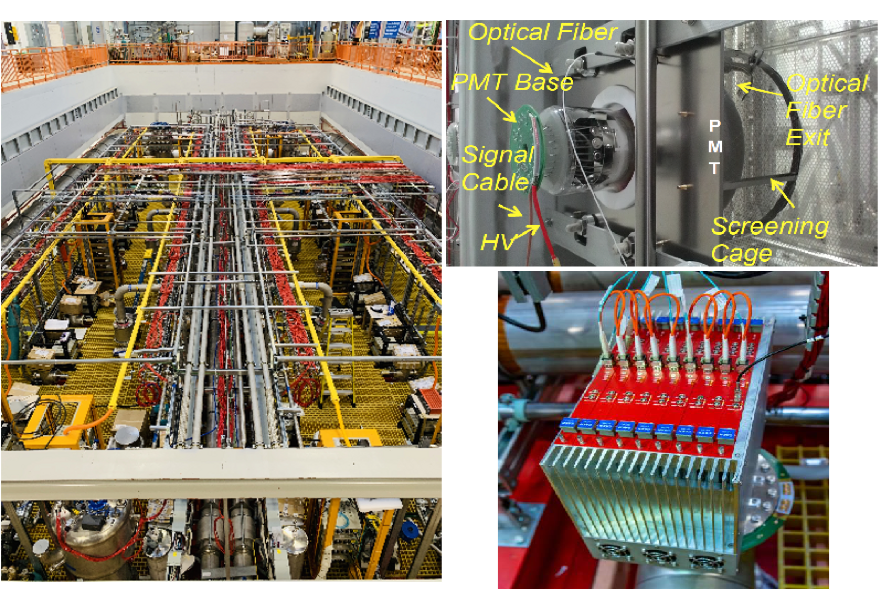}
  \caption{Left: ICARUS installation at SBN. Top right: one of the PMT installed inside the detector. Bottom right: one of the
new mini crates containing 9 CAEN A2795 boards used to collect the TPC wire signals.}
  \label{fig:intro3}
\end{figure}

\subsection{Specifications of the cryogenics and purification systems}

The goals of the cryogenic and purification systems are to maintain the LAr volume in stable thermal
conditions and to achieve and maintain purity levels, in terms of concentration of electronegative
impurities at the level of $\leq 0.1$~ppb (O$_2$ equiv.), to allow for the transport of the ionization electrons over
the distance between the cathode and the anode with the smallest possible attenuation. For reference,
at 0.1~ppb (O$_2$ equiv.) of e-negative impurities concentration, the corresponding free electrons lifetime
is $\approx 3$~ms and the ionization electrons attenuation, after 1~ms drift time (corresponding to 1.5~m drift,
at 500~V/cm) is about 28.5\%; for an impurities concentration of 0.05~ppb (O$_2$ equiv.) the attenuation is
$\approx 15$\%. The typical O$_2$ and H$_2$O concentration in standard commercial LAr is $\leq 1$~ppm.

Non-electronegative impurities present at small concentrations (ppm level) are, in general, not a concern for the operation of a LAr TPC. A notable case is nitrogen, which is typically present in commercial LAr at concentrations of the order of one to a few ppm. While nitrogen at these levels does not affect the transport of ionization electrons in liquid argon, it has been shown to suppress the slow component of the scintillation light. The suppression of the slow component is approximately 20\% for nitrogen contaminations around 3~ppm~\cite{Ref15_1}.

The implementation of a dedicated commercial system for nitrogen removal is, however, both costly and impractical for the large argon volumes involved in modern LAr TPC detectors, as it requires heating the carrier fluid (argon, in this case) to temperatures of the order of 400~$^\circ$C. Operational experience has demonstrated that the purity specifications guaranteed by the vendor for commercial LAr (see Table~\ref{tab:designparams}), combined with stringent leak-tightness requirements and established purification procedures for electronegative contaminants, are sufficient to ensure adequately low nitrogen concentrations over extended periods of operation (years).

Some polar molecules such as H$_2$O or CO$_2$, typically present at the ppm level in commercial LAr or desorbed from material surfaces, are effectively removed by molecular sieves. In contrast, the removal of O$_2$ and other electronegative contaminants requires chemical filters. Pure copper has been demonstrated to be an effective alternative to the original ICARUS choice (Oxysorb$^{\mathrm{TM}}$), whose use is not permitted in the United States.

During operation, the LAr purity is determined from the charge attenuation measured along cosmic muon tracks crossing the full drift volume and recorded by the TPCs. This method provides the highest sensitivity to electron attenuation over the maximum drift distance. For a drift length of 1.5~m (the maximum in ICARUS) and a drift field of 500~V/cm, distortions of the electric field due to space charge effects are small, and the corresponding corrections to the extracted electron lifetime are negligible for lifetime values up to several milliseconds.

During the filling phase, the argon purity is monitored at several locations, including the outlet of the storage dewar, points along the filling lines, and a few positions on the chimneys at the top of the detector, by means of gas analyzers (see Table~\ref{tab:gas-analyzers}). These analyzers are connected to the sampling points through stainless steel piping.

We note that radioactive contaminants naturally present in atmospheric argon, such as $^{39}$Ar and $^{85}$Kr, as well as radioactive decays from members of the $^{238}$U and $^{232}$Th chains present in detector and cryostat materials (primarily aluminum and stainless steel), are not a limiting factor for the present physics program. 

Another relevant requirement of the cryogenic and purification system is that it must not interfere with the operation of the TPC. The signals recorded by the TPC readout electronics are very small\footnote{The typical signal produced by a minimum ionizing particle on a single wire is of the order of 15,000 electrons}. Therefore we must make sure that any potential source of electromagnetic interference is appropriately shielded and grounded. The experimental setup is provided with two separate grounding systems:
\begin{itemize}
\item a detector ground, used as a reference for the detector components and the associated electronics. Also the main aluminum containers (Section~\ref{sec:LArContainers}) and the cold shields (Section~\ref{sec:ColdShields}) are connected to the detector ground, acting as a Faraday cage for the internal detector components;
\item a building ground, used as a reference for all other equipment. The cryogenic equipment (pumps, actuators, cryogenic sensors, transfer lines, auxiliary cryostats, cryogenic controls, etc.) must all be referenced to the building ground and remain electrically isolated from the detector ground under normal operating conditions. A safety bonding device automatically connects the two grounds if the potential difference exceeds a predefined threshold. 
\end{itemize}
To minimize electromagnetic interference the cryogenic controls must be on building ground hosted in shielding cabinets positioned outside the area occupied by the detector assembly (outside the warm vessel, see Section~\ref{sec:WarmVessel}). This includes the sources of auxiliary power (UPS).
Finally, all auxiliary equipment must be tested for potential electromagnetic interference before approval.

The ICARUS detector at the SBN facility is primarily devoted to the study of neutrino interactions in the $\gtrsim 100$~MeV energy range, including appearance and disappearance searches in accelerator neutrino beams, as well as detailed reconstruction of neutrino-induced final states. These analyses rely on the reconstruction of extended ionization tracks rather than on isolated low-energy deposits.

Two considerations make radioactive contaminants negligible for the present application:
\begin{enumerate}
\item The energies associated with the dominant decays of $^{39}$Ar and $^{85}$Kr ($\lesssim 600$~keV), as well as those from the $^{238}$U and $^{232}$Th chains ($\lesssim 3.5$~MeV), are well below the typical energy scale of neutrino-induced events of interest. Furthermore, these decays produce localized energy deposits that are readily distinguishable from the multi-track topologies characteristic of neutrino interactions.
\item The detector operates at shallow depth (under approximately 3~m of concrete overburden), and the overall background rate is dominated by cosmic radiation, primarily cosmic muons. The energy deposition rate in the liquid argon due to cosmic rays exceeds that from $^{39}$Ar decays by approximately a factor of 50, making the latter a subdominant contribution to the total background budget.
\end{enumerate}

For these reasons, while radioactive contaminants are of primary concern in low-background experiments searching for rare processes at the keV–MeV scale, they do not impose additional purity constraints for the ICARUS physics program. The required purity levels are therefore driven by the need to ensure adequate electron lifetime for long-drift charge collection rather than by radiological considerations.

As the drift velocity depends on the LAr temperature, a temperature uniformity to within 1~K is required
within the LAr volume. As a reference, the reduction in drift velocity corresponding to a temperature
variation from 88~K to 89~K, is about 2\%.

To prevent electrical discharges inside the detector (the voltage on the cathodes is -75~kV, in nominal
operating conditions), the risk of bubbles formation inside the LAr volume must be kept to as low as
possible. This translates into a requirement on the diffuse and localized heat loads on the LAr volume.
On the other side, to achieve an effective and uniform purification of the LAr volume, a reasonable
amount of convection is also required.

Another requirement arises from the mechanical stability of the wire chambers frames. The frames, entirely 
constructed from Stainless Steel AISI~304L,
were designed to achieve a precision in the relative positioning of the wires to $< 0.1$~mm and the overall
structure is aligned to $< 0.3$~mm. In addition, and most important, breaking of a single wire would
potentially compromise the operation of a large fraction of a wire chamber (see Reference~\cite{Ref16} 
for a detailed description of the detectors and its interfaces with the aluminum containers). Therefore, the mechanical
stress on the wire chambers during the cooldown and during operation must be minimized. The wire
chambers were specifically designed to minimize the stress with the introduction of elastically
compensating (spring loaded) rocking frames for the wire holders. The specification for the maximum
allowable thermal gradient on the wire chambers is 50~K. This has driven the design of the cooling
system and procedures.

\subsection{Functional description of the cryogenic and purification systems}

The reference layout of the T600 cryogenics and purification systems was developed after several years
of intense R\&D and first implemented and tested on prototypes during the nineties ([15] and references
therein). The first full scale implementation was performed in the T600 detector~\cite{Ref16,Ref17}. All large LAr-TPCs
use the same layout for the LAr purification and recirculation. In the T600, the process of initial
removal of air from the LAr containers is achieved by vacuum pumping. In other detectors, the same
goal is obtained by argon purging at room temperature; this requires a specific design of the internal
mechanical components, to achieve an effective removal of air at the level required for the argon purity
(typically at the level of $< 1$~ppm, before starting the cooling and filling process).

As the initial air removal is achieved by vacuum pumping, the main LAr volumes are required to be
vacuum tight (the specification is for localized Helium leak tightness $< 10^{-8}$~mbar~lt/s and for a global
leak rate $< 10^{-3}$~mbar~liter/s) and mechanically capable to withstand the atmospheric pressure. To
achieve an initial LAr purity $\approx 1$~ppb (O$_2$ equiv.), the residual pressure, before the initial cooling, has to
be $< 10^{-4}$~mbar with a corresponding global leak rate $< 10^{-1}$~mbar~lt/s (typically dominated by H$_2$O
outgassing of cables and other plastic detector components).

To minimize the contamination of the argon from outgassing, the cooldown must be as fast as possible,
compatibly with the requirement of maximum gradients on the wire chambers and on the LAr
containers (cold vessels)\footnote{The outgassing rate, for any given substance, is, in first approximation, an exponential
function of the temperature. In our case, outgassing is typically dominated by water and it is reduced by several
orders of magnitude when the temperature decreases from ambient to LAr temperature.}.
The cooling is performed by flowing LN$_2$ through heat exchangers (cold
shields) placed inside the thermal insulation and surrounding the two LAr containers. The function of
the cold shields is twofold:
\begin{enumerate}
\item Cool down the LAr containers and the internal materials to LAr temperature, before the start
of filling with LAr;
\item Prevent the heat, coming through the thermal insulation, from reaching the LAr volumes
(shielding function), thus preventing large thermal gradients in the liquid and the risk of
bubbles formation.
\end{enumerate}

The thermal insulation must take the heat load on the cold shields to a level which is low enough that
it can be handled by standard commercial equipment (cryogenic pumps, valves, piping, etc.). In our
case the average design thermal load was set at 10~W/m$^2$. Such a target value can be obtained using
standard commercial solutions based on pure passive, non-evacuated, insulating materials (foams,
powders, \dots). Passive thermal insulation is also attractive for practical, economical and safety reasons
and it has been the reference choice also for the previous installation at LNGS.

During operation, most impurities are coming to the argon gas phase from outgassing of materials
(mostly cables) and from micro-leaks possibly developed in some of the many feedthroughs located at
room temperature on top of the detector. To prevent these impurities to diffuse in the main LAr
volumes, the T600 is provided with units dedicated to the purification of the argon gas phase: these are
called gas recirculation units. These units are also used to stabilize the pressure of the LAr containers,
that must be kept at a safe level above the atmospheric pressure to prevent air back streaming from
leaks.

While the role of the gas recirculation system is to minimize the amount of impurities entering the
liquid from the gas phase, the purification of the bulk liquid is achieved in two steps:
\begin{enumerate}
\item during the filling, commercial LAr is flown through dedicated filters (a combination of O$_2$ and
H$_2$O filters) before being injected in the cold vessels; this removes the contaminants initially
present, typically at the ppm level, in commercial LAr;
\item during operation, LAr in the cold vessels is continuously recirculated, by means of cryogenic
pumps, and passed through filters; this removes the impurities continuously migrating in the
liquid from the gas phase or originating from other possible internal sources not removed
during the vacuum phase (this second source is purely hypothetical).
\end{enumerate}

In the following sections, a detailed description of the cryogenic and purification systems is provided,
including the thermal insulation and the cold vessels (Sections~\ref{sec:DescriptionCryo} to \ref{sec:ColdShields}). 
Section~\ref{sec:Controls} is dedicated to the architecture and implementation of the control process. 
The commissioning process and the results after one year of operation are reported in Section~\ref{sec:CryoComm}. 
Safety protocols and adopted standards are covered in Appendix~\ref{app:ProcessSafety}. 
\FloatBarrier

\section{Description of Icarus cryogenic plant}
\label{sec:DescriptionCryo}

The Icarus cryogenic plant is designed, built and installed at Fermilab by a scientific collaboration of
three international institutions, CERN, INFN and Fermilab, to support operations of the ICARUS liquid
argon time projection chamber (LAr-TPC). The original Icarus T600 LAr-TPC was installed at the INFN
Gran Sasso Underground Laboratory (LNGS), where it was operated from May 2010 to June 2013 taking
data on the CERN to Gran Sasso neutrino beam with extremely high argon purity, stability and detector
lifetime~\cite{Ref2,Ref16}. One of the most important key elements of the LAr TPC is the liquid argon purity:
residual compounds with high electron affinity have to be kept as low as 0.1 parts per billion (ppb) all
over the argon volume during the whole detector run, thus allowing the ionization tracks created by
interacting particles inside LAr to be transported practically undistorted in a uniform electric field up to
a multi-wire anodic structure placed at the end of the drift path. The cryogenic plant installed and
operated at LNGS supported all aspects of operations of the Icarus LAr-TPC, including cool down, fill
and purification of the approximately 760 tons of liquid argon divided over two identical vessels.
These vessels with internal dimensions of 3.6 (width) $\times$ 3.9 (height) $\times$ 19.6 (length)~m$^3$ are together
enveloped by a thermal shield cooled by a vaporizing liquid nitrogen flow~\cite{Ref17,Ref18}. The cryostat and
the cryogenic plant were dismantled in 2014. The detectors were moved to CERN for modifications to
match the design requirements for installation at Fermilab Short-Based Neutrino (SBN) experiment at
the SBN far detector building (SBN-FD). After these modifications, the detectors were installed in two
new cold vessels, and this installation has been delivered from CERN to Fermilab in 2017, installed in
2018, cooled down and filled with purified liquid argon in 2019--2020, and tuned to stability by May of
2021.

To achieve the physics goals of the ICARUS experiment at Fermilab, the Icarus cryogenic plant had to
fulfill the design and operational parameters as given in Table~\ref{tab:designparams}.

\begin{table}[htbp]
\centering
\caption{Design and Operating parameters}
\label{tab:designparams}
\begin{tabular}{p{0.42\textwidth} p{0.52\textwidth}}
\hline
\textbf{Parameter} & \textbf{Value} \\
\hline
Cryostat module type and size &
Two aluminum vessels with common LN$_2$ shields within a warm self-standing steel structure with
passive insulation; cold vessel internal dimensions: 3.6~m $\times$ 19.6~m $\times$ 3.9~m,
$\sim$380~tons each ($\sim$2\% ullage). Gas purge for insulation space. \\[4pt]

Contamination for LAr delivery &
O$_2 <$ 1~ppm, H$_2$O $<$ 1~ppm, N$_2 <$ 2~ppm \\[4pt]

LAr purity &
$<$100~ppt O$_2$ equivalent, N$_2 <$ 2~ppm (corresponding to 3~ms free electron lifetime) \\[4pt]

Cold vessels design pressure &
320~mbarg (also designed for high vacuum) \\[4pt]

Operating gas pressure &
1.070~bara with $\pm5\%$ \\[4pt]

Initial purification technique (warm) &
Pump cold vessels to full vacuum at warm conditions \\[4pt]

Cooldown technique &
Initial with LN$_2$ shield enveloping the cold vessels ($-2$~K/hr), final with LAr fill \\[4pt]

TPCs cool-down rate restriction &
$<70$~K/hr \\[4pt]

Liquid argon recirculation rate (normal operation) &
up to 2.5~m$^3$/hr per cold vessel \\[4pt]

Total heat load intercepted by nitrogen &
15~kW \\[4pt]

Number of side/bottom penetrations (in LAr) &
1 side and 1 bottom penetration per module (with use of external safety valves to prevent accidental
spillage of LAr) \\[4pt]

Included sub-systems &
LAr and LN$_2$ storage and transport, O$_2$--N$_2$--H$_2$O monitoring, LAr and GAr recirculation and filtration,
GAr condensing and recovery, safety and process controls \\[4pt]

Grounding for noise requirement &
Complete electrical isolation of the two cold vessels from external systems \\
\hline
\end{tabular}
\end{table}

\begin{figure}[htbp]
  \centering
  \includegraphics[width=0.9\textwidth]{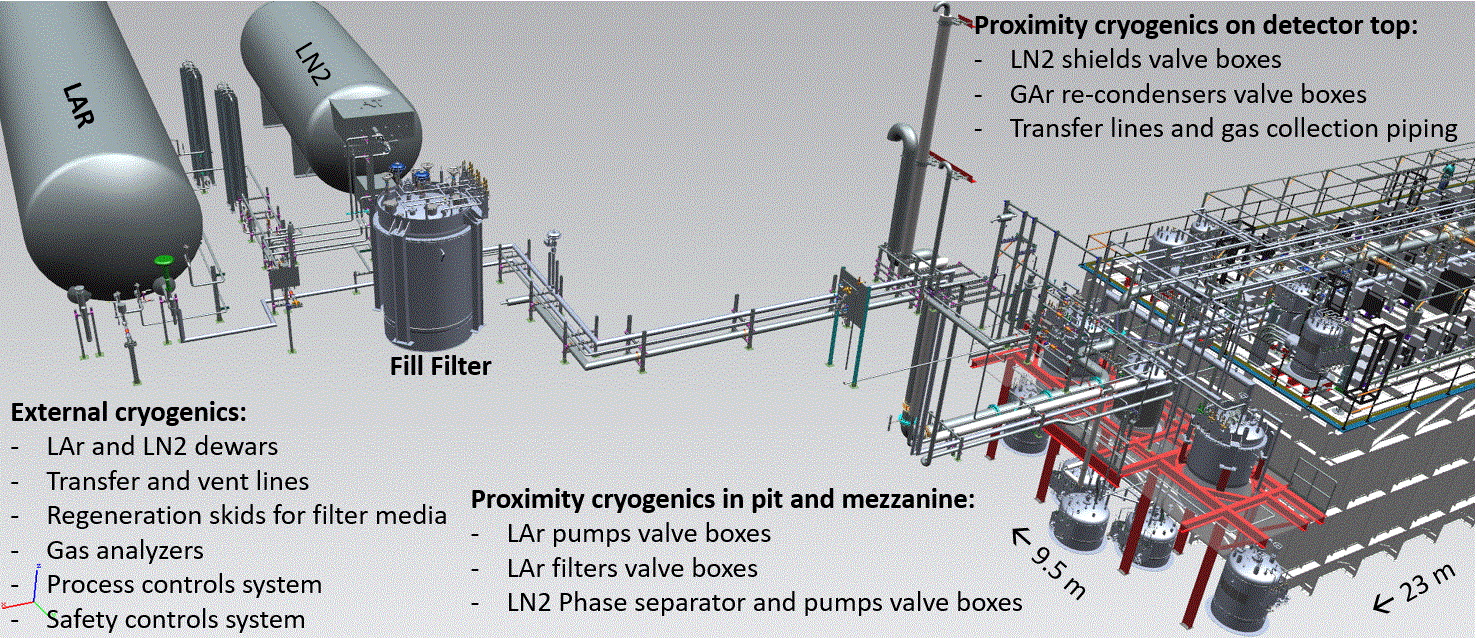}
  \caption{ICARUS Cryogenic Plant Physical Layout}
  \label{fig:sec2-04}
\end{figure}

\begin{figure}[htbp]
  \centering
  \includegraphics[width=0.9\textwidth]{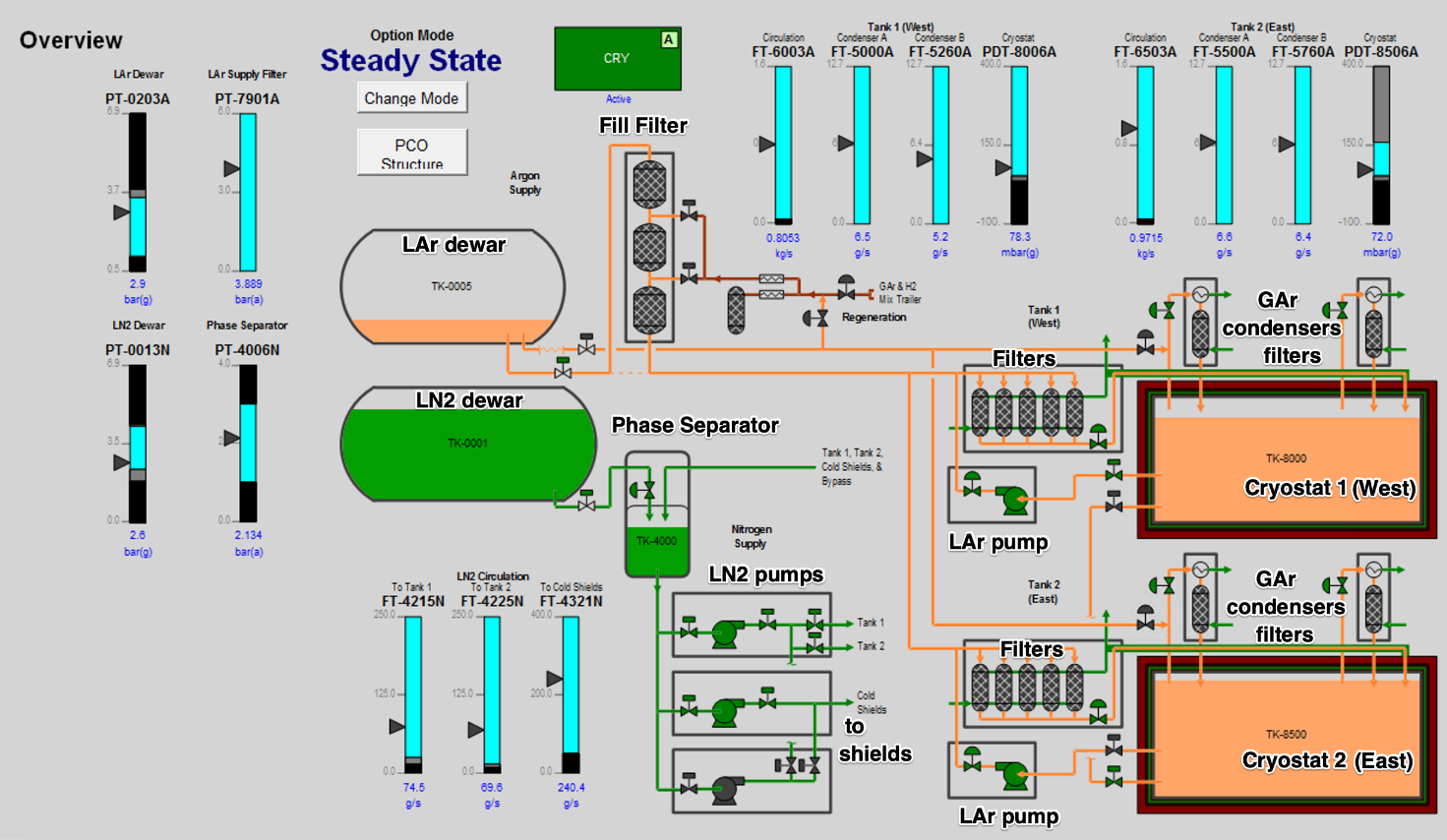}
  \caption{ICARUS Cryogenic Plant Schematical Layout}
  \label{fig:sec2-05}
\end{figure}

The components of the Icarus cryostat, warm support structure and cryogenic plant were delivered by
three institutions included in Icarus collaboration, Fermilab, CERN and INFN. By institutional delivery
responsibilities, the overall cryogenic plant was initially divided into three cryogenic systems, External,
Proximity and Internal as shown schematically in Figure~\ref{fig:sec2-04} and Figure~\ref{fig:sec2-05}.

\begin{itemize}[leftmargin=1.5em, labelsep=0.5em, itemsep=0pt, topsep=0pt, parsep=0pt]
    \item The \emph{External system} delivered by Fermilab is responsible for storage and supply of cryogens, venting of
gases, gas analysis, regeneration of filtration media, electrical power, process and safety controls systems. \\[2pt]
    \item The \emph{Proximity system} delivered by CERN is responsible for all transport and distribution of liquid argon
and nitrogen at the temperatures and pressures demanded by the internal cryogenic system, for removal
of impurities from argon, recirculation and filtration of boil-off argon gas. \\[2pt]
    \item The \emph{Internal cryogenic system} delivered by INFN and Fermilab is responsible for interfacing internal
volume of the modules with cryogenic plant, LN$_2$ shields, measurement of argon temperatures and
levels.
\end{itemize}

All three systems have been installed, connected at mechanical interfaces and integrated into Icarus
cryogenic plant with common electrical and controls systems in 2018--2019.

The ICARUS cryogenic plant at SBN-FD at Fermilab was fully designed, delivered, and installed by July
2019, with the commissioning phase started by January 2020. The equipment included into the ICARUS
cryogenic plant is schematically divided into the external components supplied by Fermilab, the
proximity components supplied by Demaco Holland B.V.\ under contract with CERN and components
internal to the cryostat supplied by INFN. Figure~\ref{fig:sec2-04} and Figure~\ref{fig:sec2-05} show the physical and schematic
layouts.

The main schematical description of the Icarus cryogenic plant is represented in its Process and
Instrumentation Diagram (P\&ID). While this 6-sheet drawing is not included in this paper, it is stored in
Fermilab Teamcenter F10041251 and in Fermilab DocDB SBN-doc-783.

\begin{figure}[htbp]
  \centering
  \includegraphics[width=0.9\textwidth]{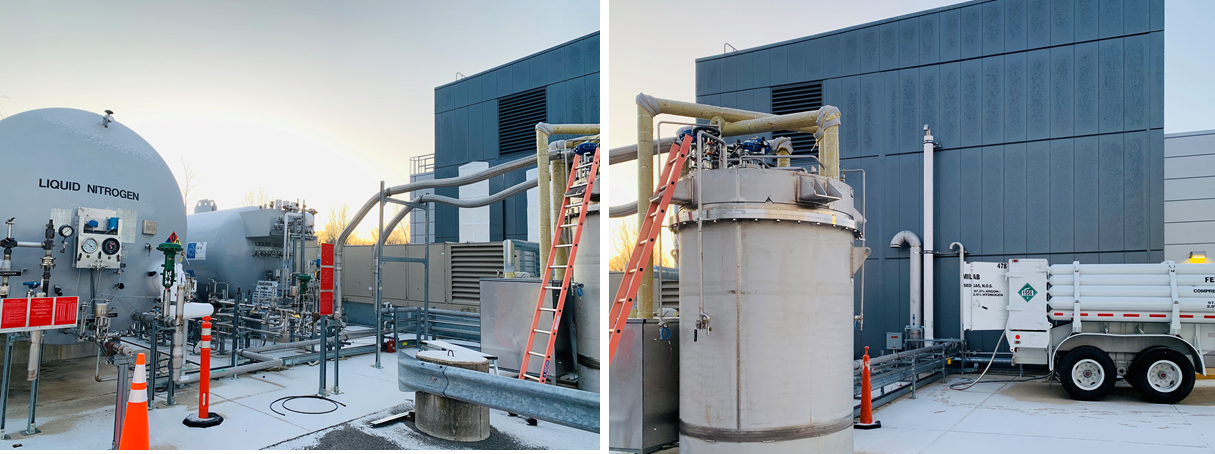}
  \caption{Cryogenic dewars and Fill Filter}
  \label{fig:sec2-06}
\end{figure}

\begin{figure}[htbp]
  \centering
  \includegraphics[width=0.9\textwidth]{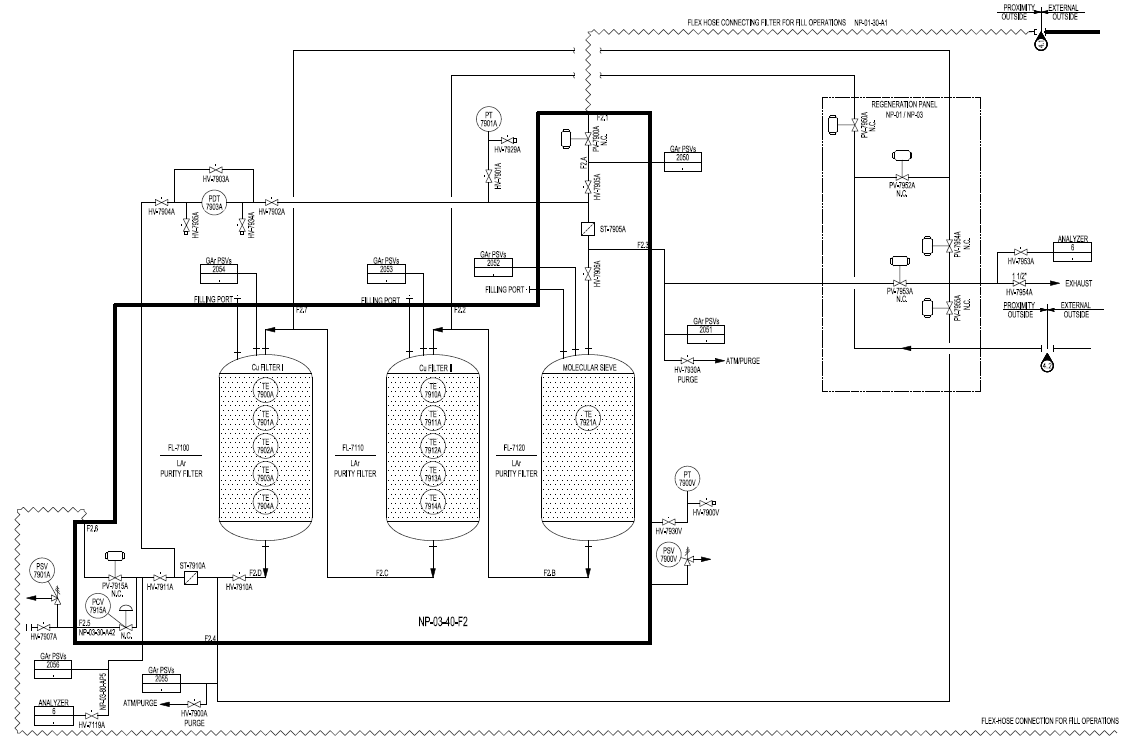}
  \caption{Fill Filter P\&ID}
  \label{fig:sec2-07}
\end{figure}

The main components of the Icarus cryogenic plant included:

\begin{itemize}[leftmargin=1.5em, labelsep=0.5em, itemsep=0pt, topsep=0pt, parsep=0pt]
     \item \emph{Liquid nitrogen storage dewar} with $\sim 75$~m$^3$ capacity (Figure~\ref{fig:sec2-06}). This vacuum jacketed and
perlite insulated tank rated and protected at 4.5~barg was built by Process Engineering in 1988,
operated at Fermilab and then moved to SBN-FD where it was completely refurbished with new piping,
reliefs and instrumentation in 2019. The dewar has been in operations since 2019. It is currently on
24/7 fill delivery by Matheson Gas to support LN$_2$ consumption by the Icarus cryogenic system of
approximately 360~liter/hr at 3.6~bara to compensate for a thermal load of approximately 14.6~kW.
The dewar is equipped with an external vaporizer that generates gaseous nitrogen used for purge of the
Icarus warm vessel.

    \item \emph{Liquid argon storage dewar} with $\sim 30$~m$^3$ capacity (Figure~\ref{fig:sec2-06}). This vacuum jacketed and perlite
insulated tank rated at 17~bara and protected at 4~barg was built by CRYOVAT in 2007, operated as part
of the Icarus cryogenic system at LNGS, and then moved to Fermilab via CERN. It was refurbished with
new piping, reliefs and instrumentation in 2019. The dewar has been in operations since October 2019.
The dewar is equipped with a vaporizer that generates gaseous argon used for regeneration of the
copper and molecular sieve media stored inside the liquid and gas filters. Additionally, the gaseous
argon could be used for any make-up gas needed by the Icarus system. While not being in continuous
active operations since completion of the fill of the modules, the dewar is maintained clean at positive
argon pressure. It is being re-cooled and partially refilled on demand when/if it is necessary to refill the
modules to compensate for vented argon.

    \item A \emph{vacuum insulated liquid argon purification vessel} (Figures~\ref{fig:sec2-06}, \ref{fig:sec2-07}) containing two filters filled
with Cu~0226 to remove oxygen and one vessel filled with molecular sieve to remove water from liquid
argon coming from the storage vessel. All the liquid argon transferred from the storage tank to the
ICARUS detector passes through this purifier system. The purity level of the liquid having passed this
purifier is in the high ppt level.

\begin{figure}[htbp]
  \centering
  \includegraphics[width=0.9\textwidth]{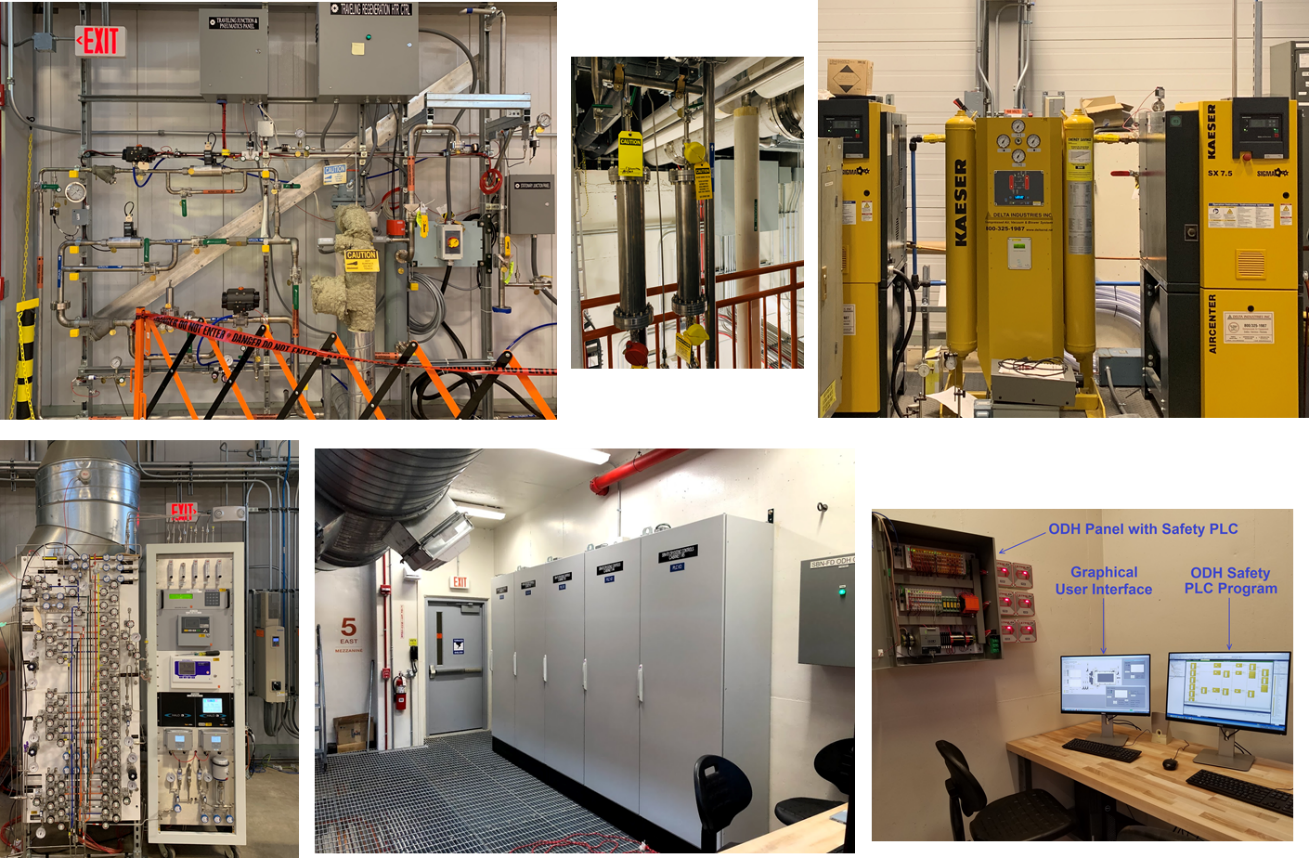}
  \caption{Top row – left to right: Regeneration panel, gas make-up filters, air skid; Bottom row – left to right:
analyzers panel, S7 process controls, Beckhoff ODH controls with access controls station.}
  \label{fig:sec2-08}
\end{figure}

    \item \emph{Regeneration system} (Figure~\ref{fig:sec2-08}), consisting of several piping skids incorporating valves and
instrumentation required for delivering, metering, mixing and heating pure argon gas from LAr dewar
and a 97.5\% Ar / 2.5\% H$_2$ mix from a high-pressure tube trailer parked outside. The gaseous mix is delivered
to external fill filter vessels or filter cartridges to remove water from molecular sieve and to activate or
regenerate copper filter media (Cu 0226) using the hydrogen gas mixture to reduce the copper oxide.
The system uses Brooks SLA5850 mass flow controllers and includes a 12~kW heater to preheat the
purifier medium to at least 160$^\circ$C. Once the correct temperature is reached, the hydrogen gas mixture
is introduced. The system was commissioned in late 2019.

    \item The \emph{gas analyzer system} (Figure~\ref{fig:sec2-08}) consists of six analyzers installed in a common switch panel,
plumbed to 14 different sample ports in the cryogenic system. The most prominent purpose is to sample
argon quality in the Air Products delivery tanks before accepting it to the liquid argon dewar.

\begin{table}[htbp]
\centering
\caption{Gas analyzers}
\label{tab:gas-analyzers}
\begin{tabular}{l l l l}
\hline
\textbf{Gas} & \textbf{Manufacturer} & \textbf{Model} & \textbf{Set range} \\
\hline
Oxygen   & Servomex & 310E-H05000 & 0--100~ppm O$_2$ \\
Oxygen   & Servomex & 560E        & 0--5~ppm O$_2$ \\
Nitrogen & Servomex & Plasma      & 0--100~ppm N$_2$ \\
Water    & Viasala  & DMT348      & -70--80$^\circ$C H$_2$O \\
Water    & Viasala  & DMT348      & -70--80$^\circ$C H$_2$O \\
Water    & Tiger Optics & HALO3-H$_2$O & 0--9~ppm H$_2$O \\
\hline
\end{tabular}
\end{table}

\begin{figure}[htbp]
  \centering
  \includegraphics[width=0.70\textwidth]{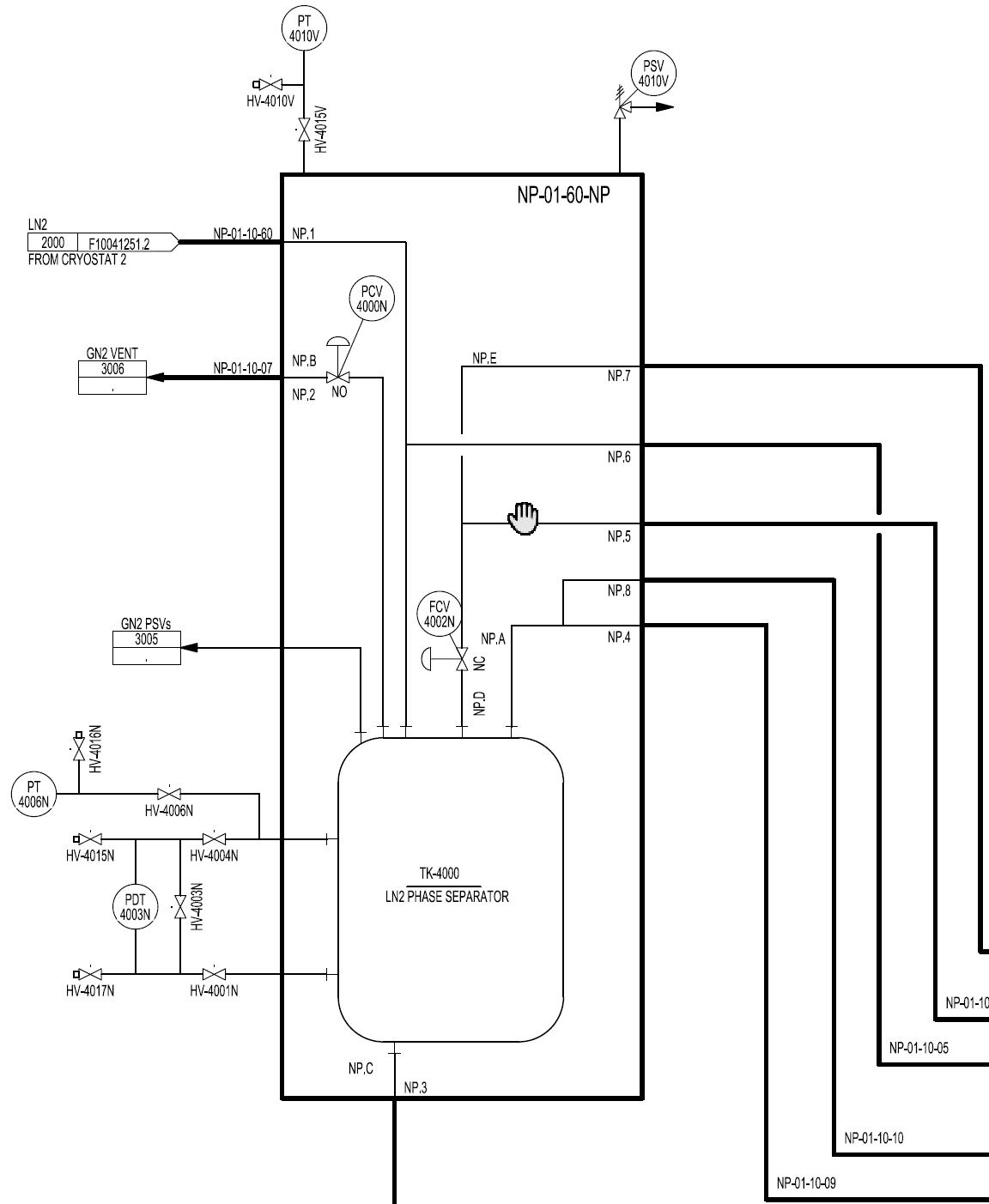}
  \caption{LN$_2$ phase separator box}
  \label{fig:sec2-09}
\end{figure}

\begin{figure}[htbp]
  \centering
  \includegraphics[width=0.9\textwidth]{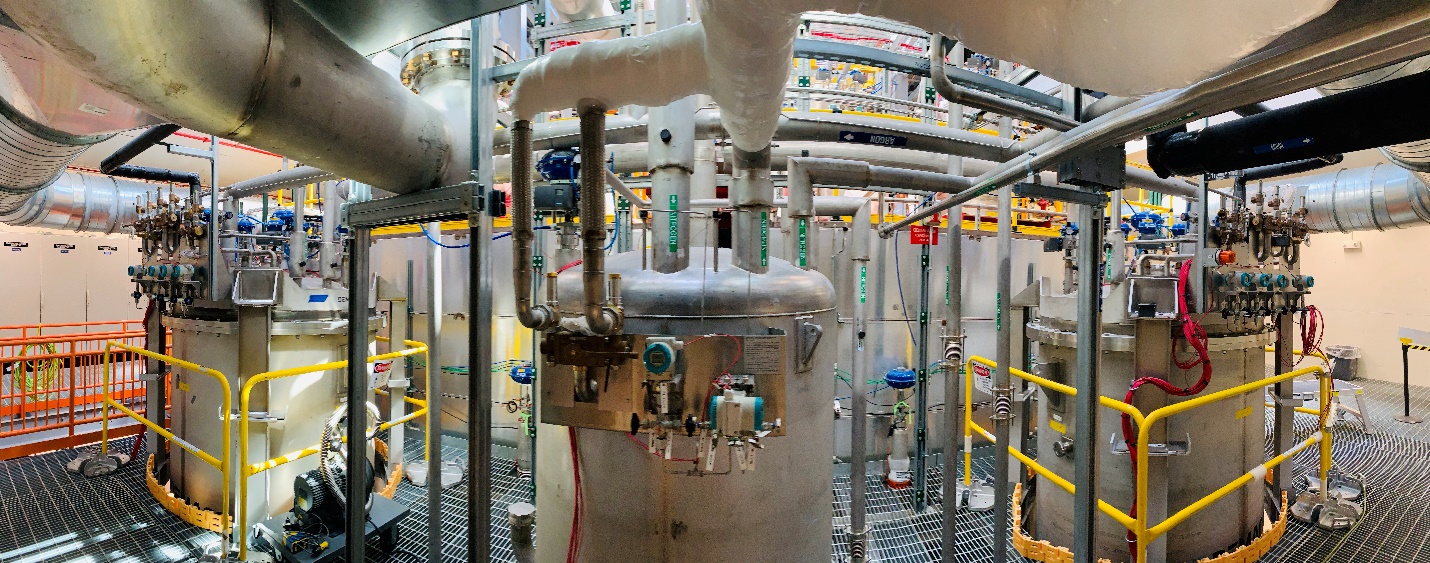}
  \caption{LN$_2$ phase separator and LAr filter boxes}
  \label{fig:sec2-10}
\end{figure}

    \item A vacuum insulated \emph{liquid nitrogen 1000-liter phase separator} (Figures~\ref{fig:sec2-09}, \ref{fig:sec2-10}) is placed
inside the building. It receives LN$_2$ from the storage tank, separates liquid from gas phase, and supplies
liquid nitrogen to the pumps located below. At the same time, two-phase LN$_2$ returning from the
cooling circuits (cryostat shields, gaseous argon condensers, liquid argon purifiers, and protected
transfer lines) is returned to this separator. The saturated nitrogen pressure in the separator is the
lowest in the nitrogen circuit and is regulated at 2.1~bar absolute, ensuring all temperatures in the
cooling circuits remain above the argon triple point.

    \item Three vacuum insulated \emph{LN$_2$ pump boxes} (Figures~\ref{fig:sec2-11}, \ref{fig:sec2-12}) are situated below the separator,
receiving gravity-driven LN$_2$. One pump supplies LN$_2$ to the cryostat thermal shields; a second pump supplies 
the liquid nitrogen to each of the four condenser boxes, to the shield of the LAr purification boxes and to the 
transfer line between these purification boxes and the cryostat inlet. The third pump box is backup pump, which can 
be used for any of the two systems discussed in case one of the two has problems or is in maintenance. All pumps are Barber
Nichols BNCP-68-M1. Mass flow and outlet pressure are regulated using bypass valves downstream of
each pump to ensure that load variations in one circuit do not affect the others.

\begin{figure}[htbp]
  \centering
  \includegraphics[width=0.9\textwidth]{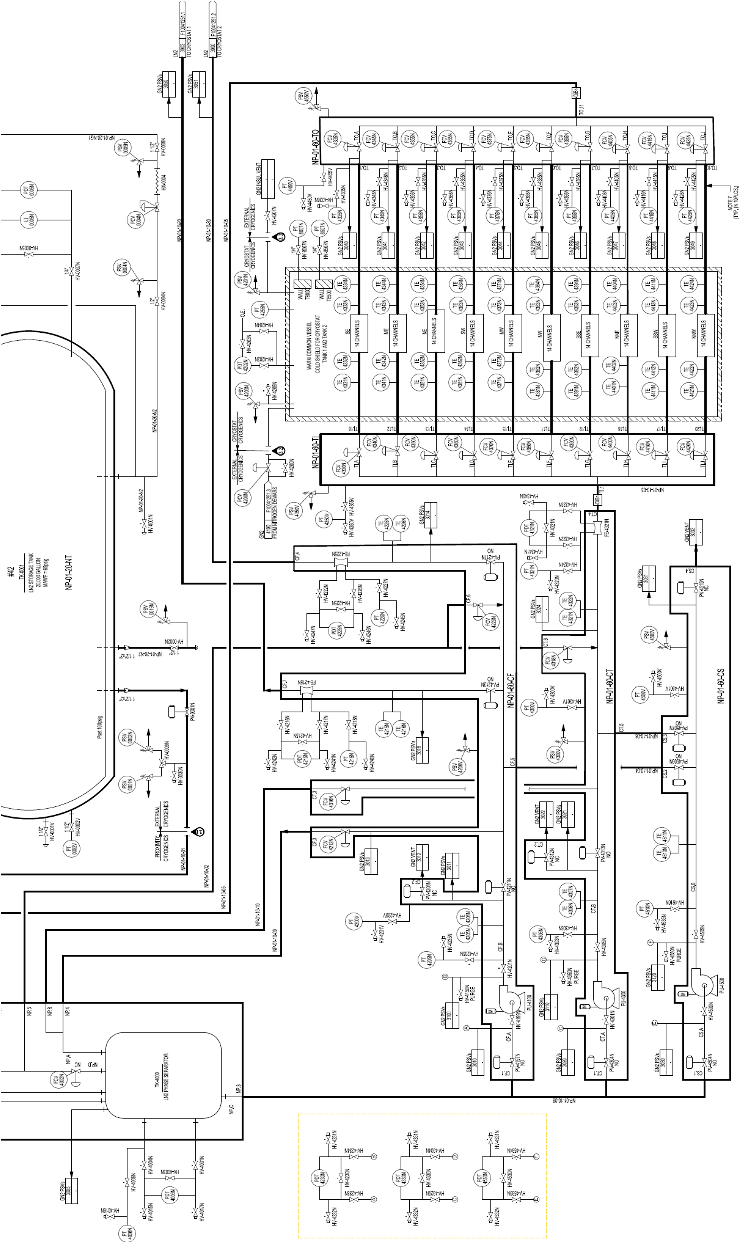}
  \caption{Phase separator, LN$_2$ pump boxes and cold shields}
  \label{fig:sec2-11}
\end{figure}

    \item The cold shields enveloping the two detector cold vessels (described in more detail in Section~\ref{sec:ColdShields})
have been equipped with a total of 10 nitrogen cooling circuits. The distribution of the cooling
circuits over the shields is given in Figure~\ref{fig:sec2-11}. The vacuum-insulated shield inlet box distributes
the liquid coming from the dedicated LN$_2$ pump over the 10 different heat exchangers. The amount
of flow going through each heat exchanger is regulated by a valve placed in each of the ten
individual circuits. The outlet of the ten different cold-shield circuits is collected in the vacuum
insulated heat exchanger outlet box. The pressure in, and thus the temperature of, each circuit can
be regulated by a valve placed in each of the ten circuits. The liquid/gaseous mixture coming from
the heat exchangers is returned to the nitrogen phase separator via the shield outlet box.

\begin{figure}[htbp]
  \centering
  \includegraphics[width=0.90\textwidth]{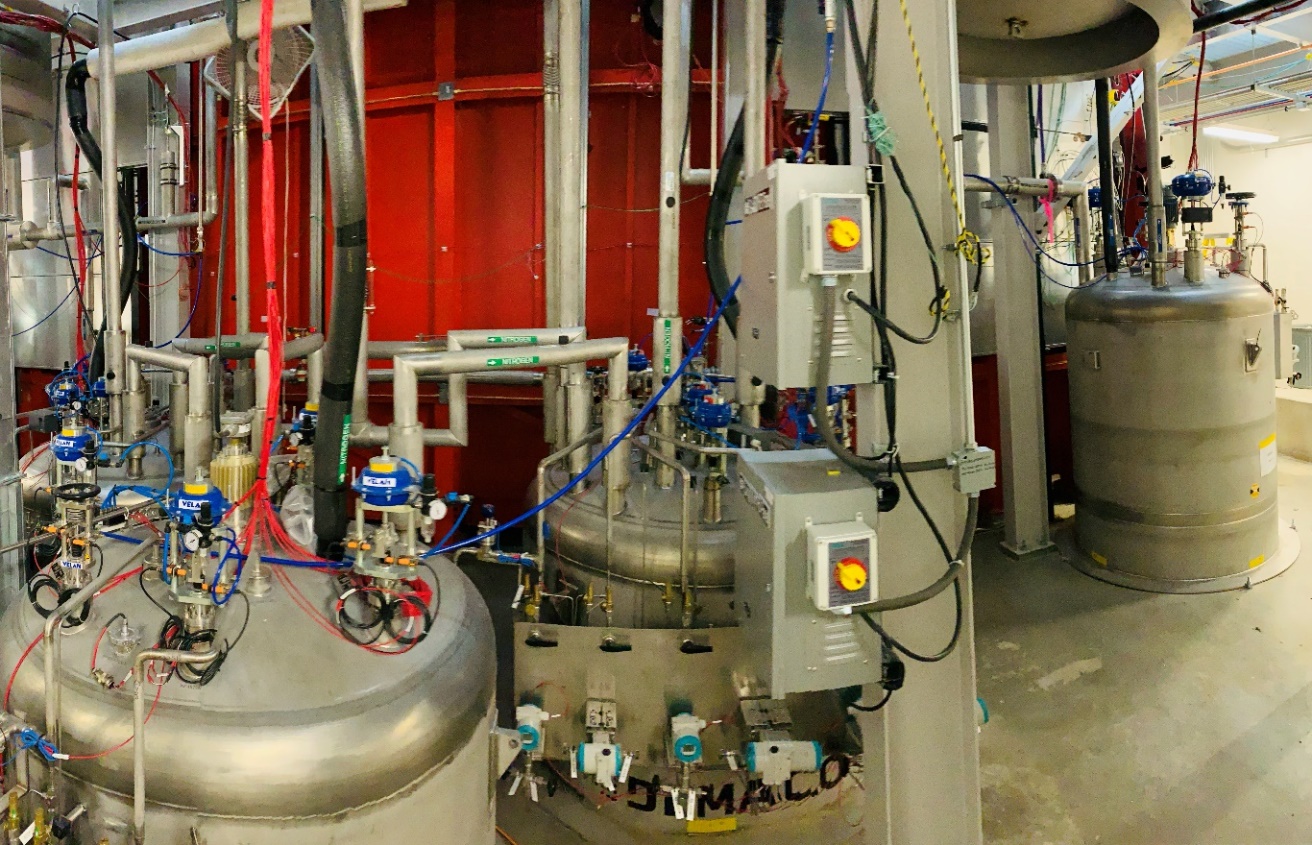}
  \caption{Three LN$_2$ pump boxes (left) and West LAr pump box (right)}
  \label{fig:sec2-12}
\end{figure}

    \item Each of the modules is connected to a vacuum insulated liquid argon pump box (see
Figure~\ref{fig:sec2-12}, Figure~\ref{fig:sec2-13} and Figure~\ref{fig:sec2-14}), placed at the bottom level of the module.
These pumps circulate a liquid argon flow from each module to a dedicated vacuum insulated
argon purification valve box. The West module is configured to take LAr from the top penetration,
while the East module takes LAr from the bottom penetration. This choice is due to the differences
in piping from the top or bottom penetrations inside the warm-vessel space. The piping
incorporates goosenecks to create a positive gas lock in case of interruption of flow, thus creating
various degrees of difficulty in starting the cold flow from the cold vessels to the pumps.

\begin{itemize}
\item \textbf{West 40-CP PU-6000 pump (BNCP-32C).}  
The pump is set to run at 41.8~Hz, 1.4~bara discharge while maintaining $\sim 0.81$~kg/s flow to
the West modules via the 40FC filter box. The operating point is set by adjusting the flow with
a discharge valve at $\sim 37\%$.

\item \textbf{East 50-CP PU-6500 pump (BNCP-32E).}  
The pump is set to run at 49.6~Hz, 2.0~bara discharge while maintaining $\sim 0.96$~kg/s flow to
the East modules via the 50FC filter box. The operating point is set by adjusting the flow with
a discharge valve at $\sim 44\%$.
\end{itemize}

\begin{figure}[htbp]
  \centering
  \includegraphics[width=0.90\textwidth]{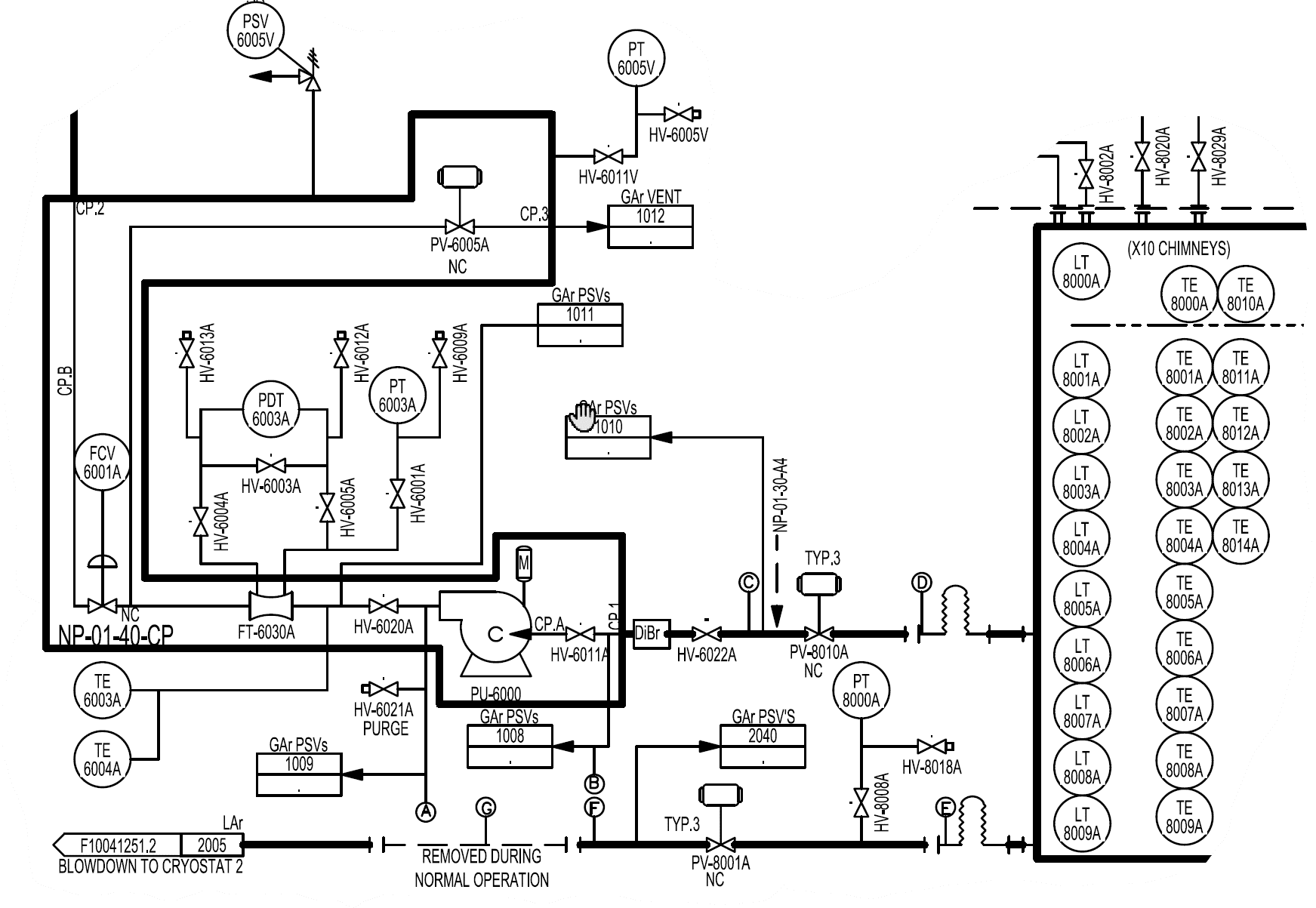}
  \caption{West module's LAr recirculation pump box (40-CP PU-6000) schematics. The inlet of the pump is connected to the top side penetration, about 1.5~m above the bottom of the module (right side of the image). The outlet of the pump goes to the inlet of the cold filter box shown in Figure~\ref{fig:sec2-15}. The diagram shows the pump, the various valves to control the flow and for the startup of the pump, and the associated instrumentation. The diagram also shows the bottom side penetration of the west module (right side of the image).}
  \label{fig:sec2-13}
\end{figure}

\begin{figure}[htbp]
  \centering
  \includegraphics[width=0.90\textwidth]{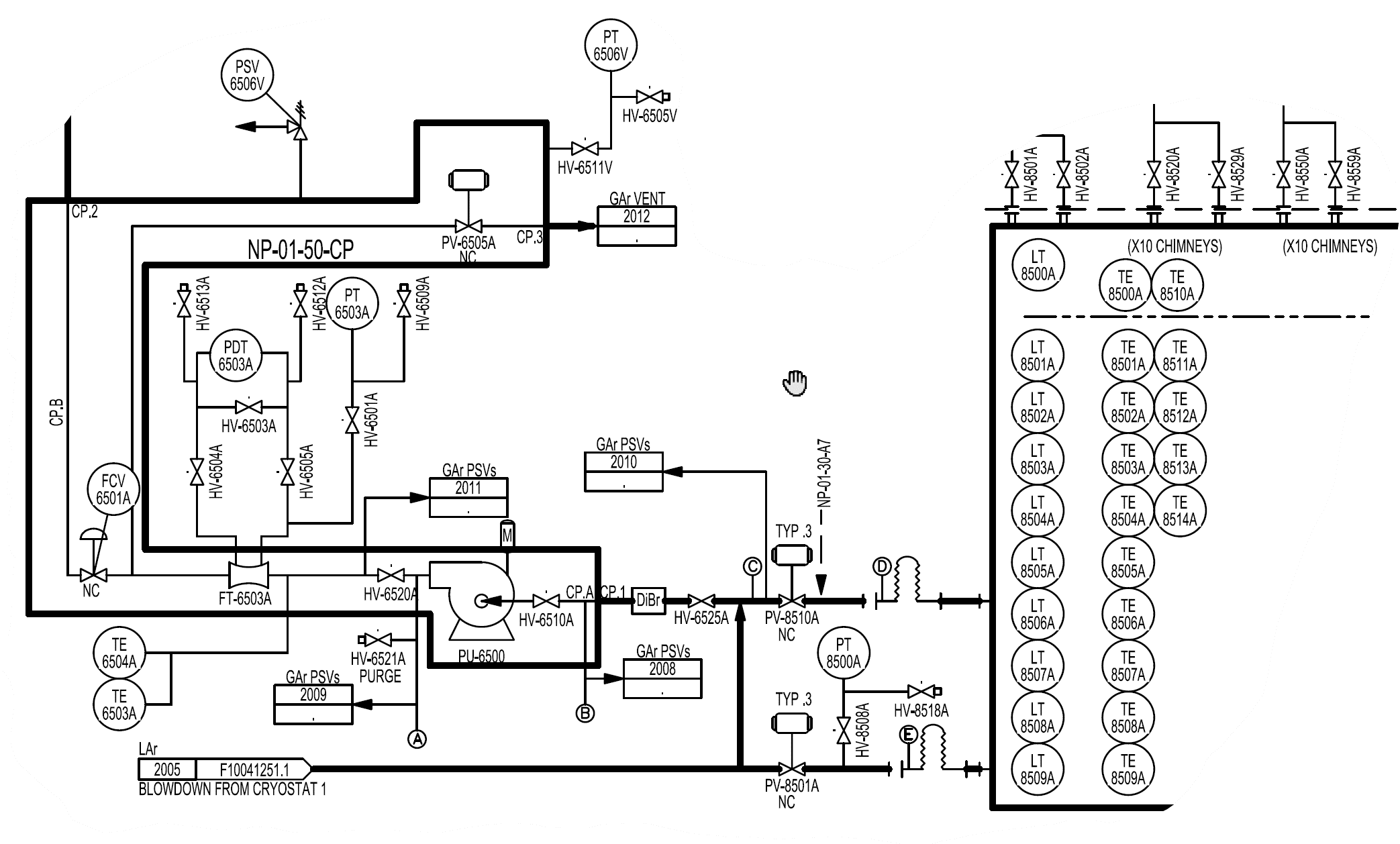}
  \caption{West module's LAr recirculation pump box (50-CP PU-6500) schematics. The layout is identical to the one of the west module (Figure~\ref{fig:sec2-13}) with one relevant exception: the bottom and the top side penetrations are interconnected after the intercept valves. This allows for recirculation from either level (or both, in principle). LAr recirculation in the east module is performed using the bottom side penetration as inlet (see Section~\ref{sec:CryoComm}); note that during operation the line marked as "BLOWDOWN FROM CRYOSTAT 1" is closed with a flange.}
  \label{fig:sec2-14}
\end{figure}

    \item Each of the LAr purification boxes (Figure~\ref{fig:sec2-15}) is equipped with five parallel active copper-filled
purifiers enclosed by nitrogen-cooled thermal shields. The inlets to the purification boxes are from
either the transfer line from the LAr dewar (during cooldown and fill) or from the LAr circulation
pump. The outlets of these argon purification valve boxes are connected to the modules via a
thermally shielded liquid argon transfer line. The shield in this line is also cooled using evaporating
liquid nitrogen.

\begin{figure}[htbp]
  \centering
  \includegraphics[width=0.75\textwidth]{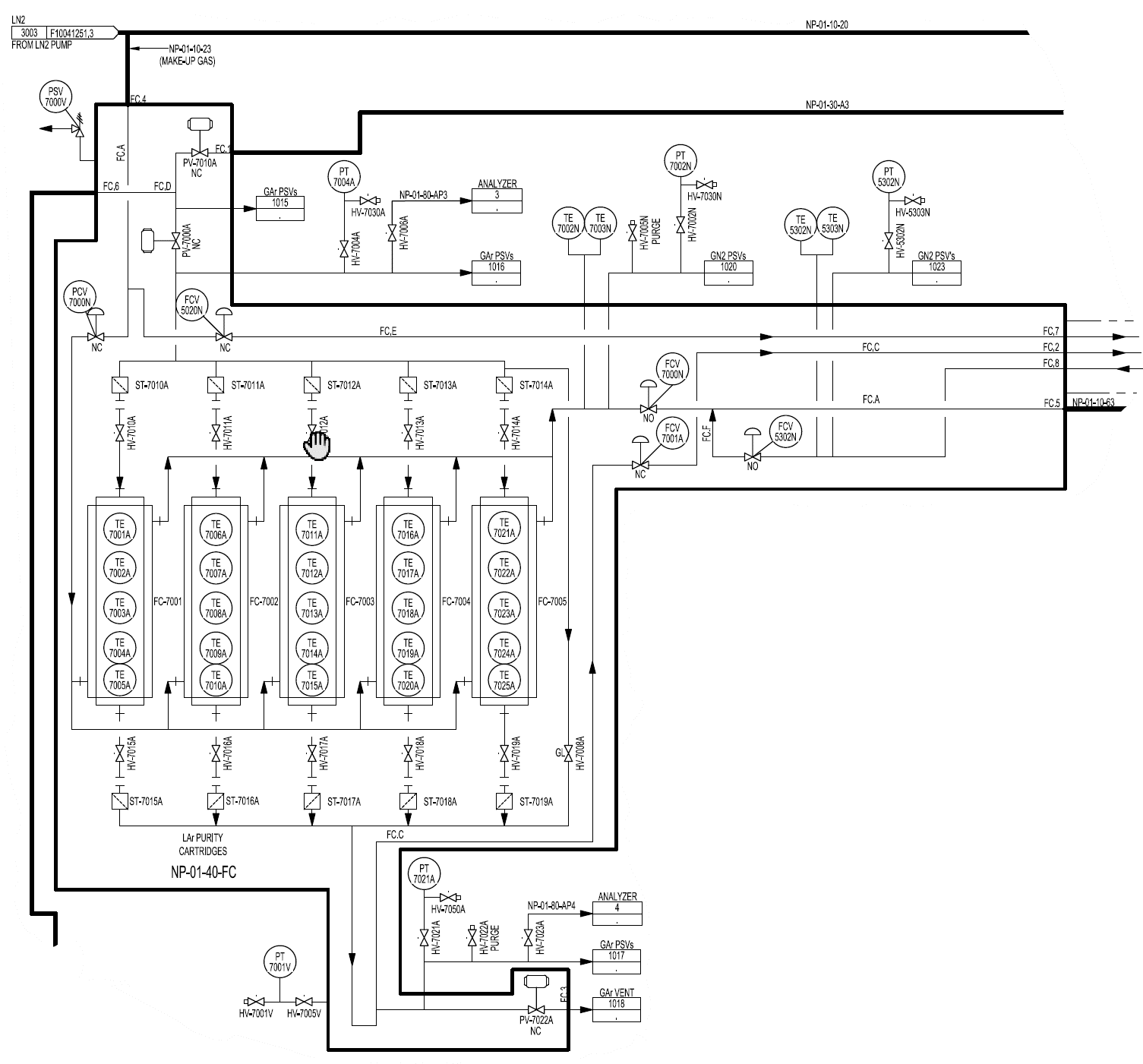}
  \caption{LAr purification box (West 40-FC, identical to East 50-FC). The diagram shows both the LAr and the LN2 circuits. LN2 is flown around the filter cartridges and then around the LAr output line (labels FC.7 and FC.8 for the LN2 line in and return, and FC.2 for LAr) to prevent gas formation in the LAr circuit. In the physical circuit, LAr is flown from the bottom of the cartridges to the top. Mechanical filters (labels ST-70XXA) are also shown at the inlet and outlet of each cartridge.}
  \label{fig:sec2-15}
\end{figure}

\begin{figure}[htbp]
  \begin{subfigure}[t]{0.40\textwidth}
    \centering
  	\includegraphics[width=\textwidth]{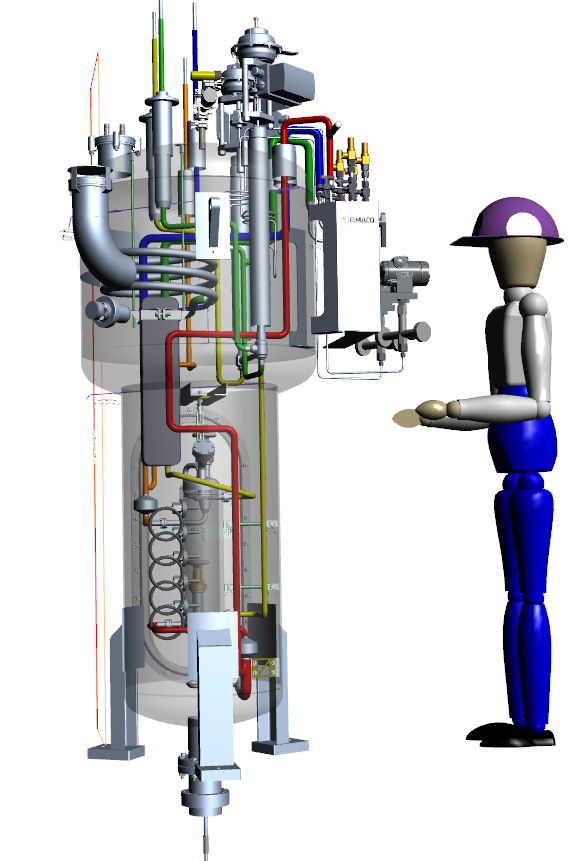}
 \end{subfigure}
\hfill
  \begin{subfigure}[t]{0.40\textwidth}
    \centering
	  \includegraphics[width=\textwidth]{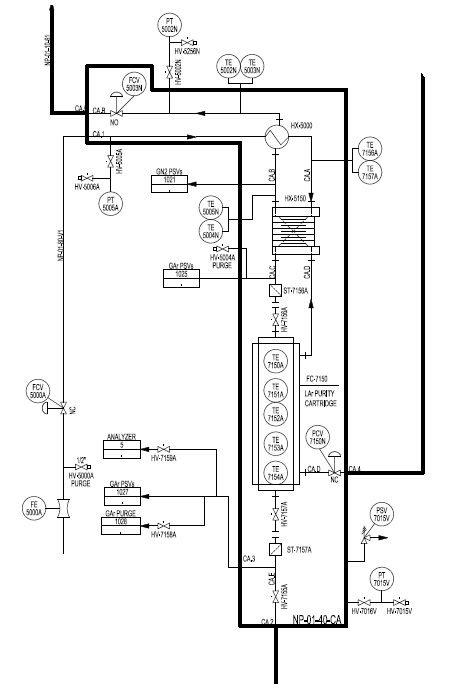}
 \end{subfigure}
  \caption{Gas recondensing unit. Two identical units are installed at the two ends of each module, for a total of four units. Each unit cools, recondenses, and purifies gaseous argon collected from half of the chimneys of a module. The rendered view shows the liquid nitrogen (LN$_2$) piping in red. The opening at the rear is used for installation and removal of the filter cartridge.}
  \label{fig:sec2-16}
\end{figure}

    \item A boiloff gas collection system forwards the argon boiloff gas from 40~$\times$~DN15 chimneys (per
module) to argon gas condensers via a Vortex flow meter and control valves. To ensure that the
gaseous argon in contact with the cables and other warm surfaces at the top of the cold vessels is
continuously purified, an argon flow is drawn via the signal feed-throughs mounted on the cold
vessels to the liquid argon condenser boxes (see Figure~\ref{fig:sec2-16}). Each cold vessel has been equipped
with two identical condenser boxes (40-CA and 40-CB for West; 50-CA and 50-CB for East). The room
temperature gas flow from the cold vessels is first condensed in these boxes using liquid nitrogen–
powered heat exchangers. After this re-liquefication, the liquid argon is passed through an active
copper purifier before being returned to the module volume. These condenser boxes serve as both
condensers and argon gas purifiers.

    \item A \emph{gas venting system} (Figure~\ref{fig:sec2-17}) consists of three separate piping networks: low- and
high-pressure argon, and high-pressure nitrogen, to vent argon from modules, as well as argon and
nitrogen from the proximity valve boxes, to outside the Icarus detector building. This is required to
mitigate oxygen deficiency hazards inside the building. The piping ranges from DN50 to DN350 and is
constructed using flexible joints and support systems to accommodate 3D movement of the vessels
during cooldown as well as thermal contraction during release of cold gases. To minimize electrical
noise in the electronics, all piping directly connecting to the modules incorporates dielectric joints to
separate the module electrical ground from the building electrical ground.
\end{itemize}

\begin{figure}[htbp]
  \centering
  \includegraphics[width=0.95\textwidth]{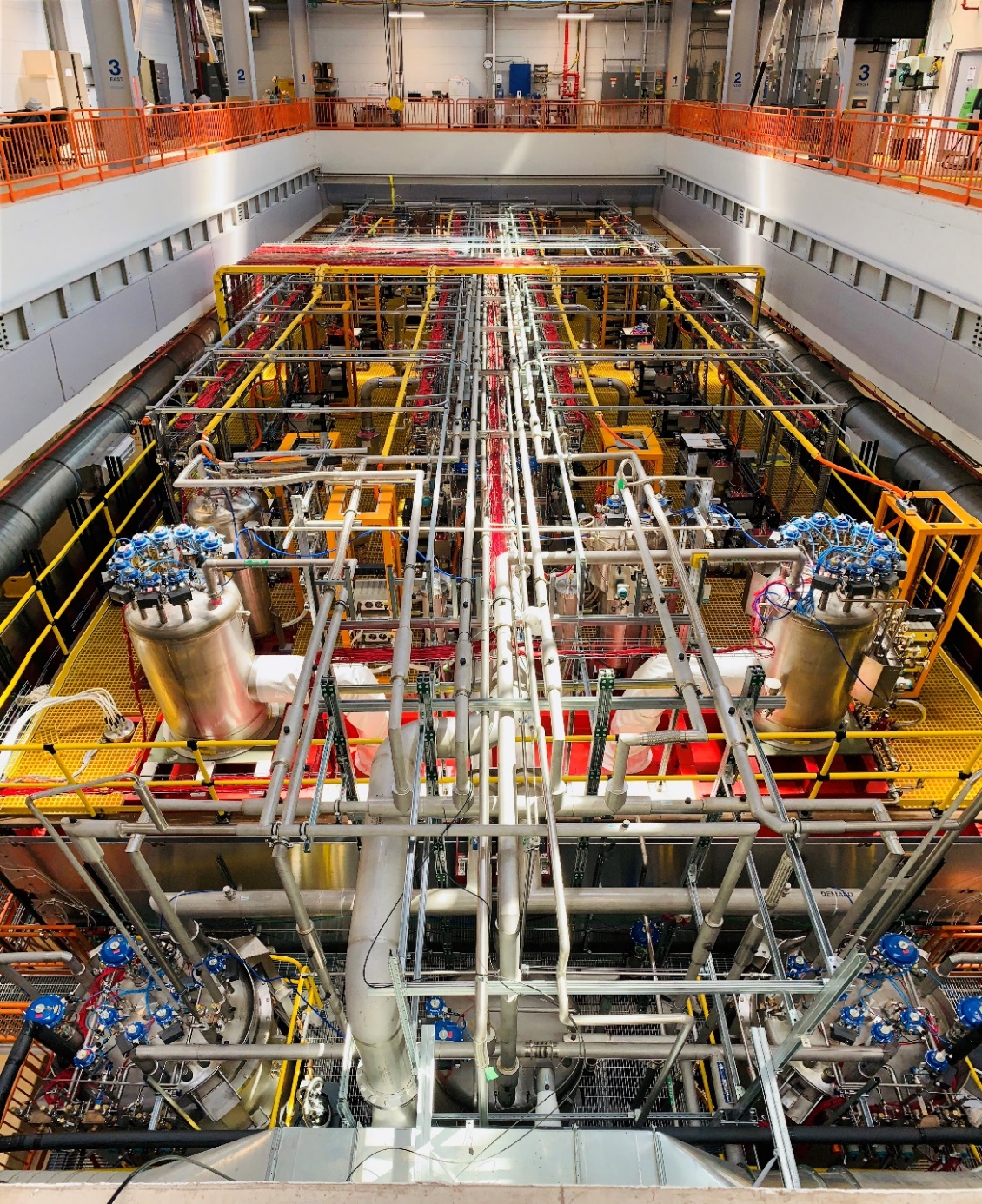}
  \caption{Top of Icarus detector with cryogenic components. Visible: LAr purification boxes 50-FC (left) and 40-FC (right);
LN$_2$ phase separator (center, North mezzanine); inlet 60-TI (left) and outlet 60-TO (right) LN$_2$ boxes for the shields
(North end detector); 50CA (left) and 40CA (right) condenser boxes; transfer lines and vent lines (center).}
  \label{fig:sec2-17}
\end{figure}

The \emph{instrument air system} is based on a dual Kaeser rotary screw air compressor skid (Figure~\ref{fig:sec2-08}),
with integrated refrigerated air dryer and adsorption swing desiccant dryer to supply all pneumatically
actuated valves with air dried below $-40^\circ$C dew point.

The \emph{control and safety instrumented systems} (Figure~\ref{fig:sec2-08}) are described in detail in Section~\ref{sec:Controls}.

\begin{figure}[htbp]
  \centering
  \includegraphics[angle=90, width=0.95\textwidth]{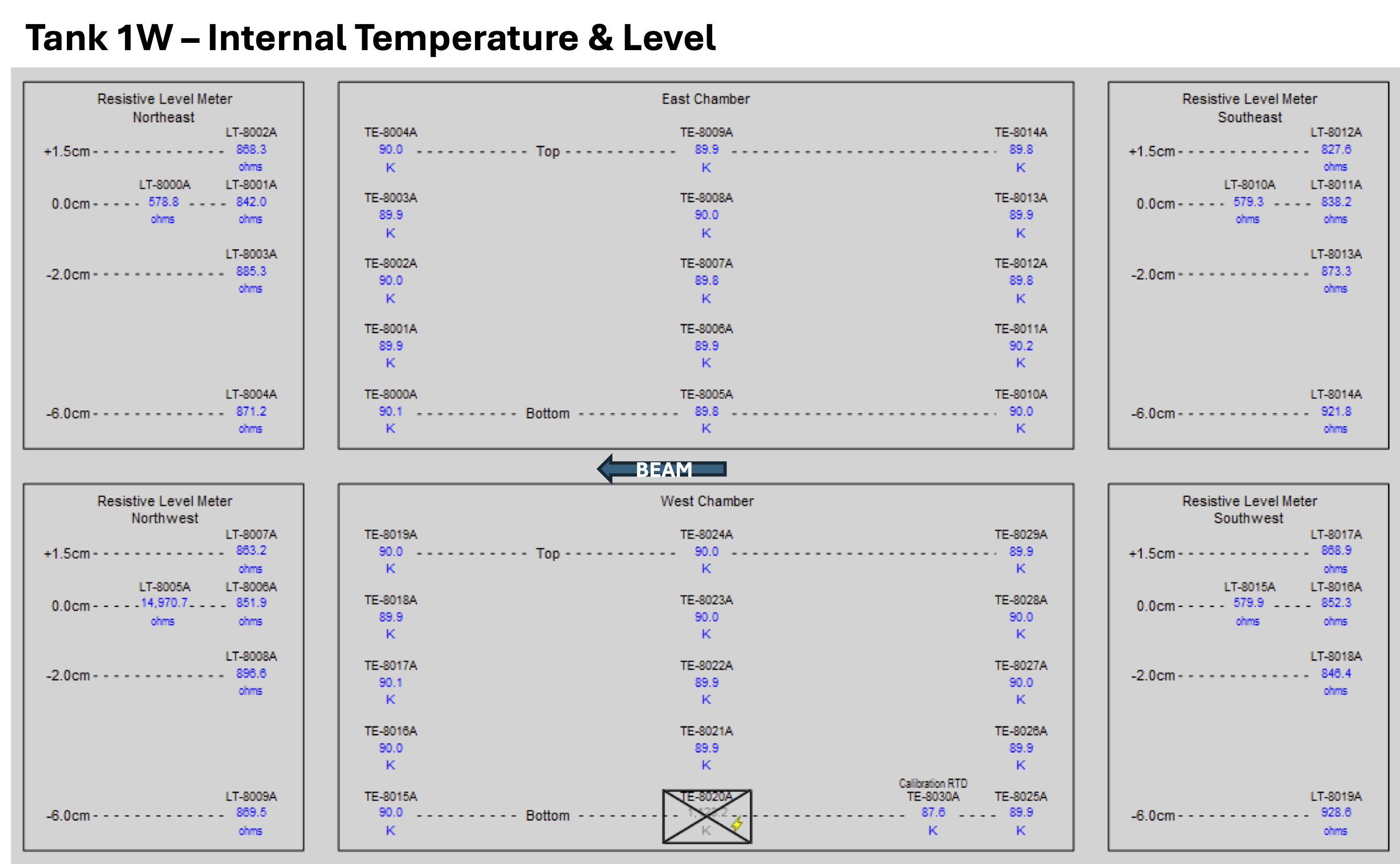}
  \caption{Internal cryogenic instrumentation for West modules (identical to East)}
  \label{fig:sec2-18}
\end{figure}

The \emph{internal cryogenic system} is comprised mostly of instrumentation located inside the Icarus cold
vessels (Figure~\ref{fig:sec2-18}). The instrumentation includes temperature resistors and diodes to measure
accurately the temperature on the TPC frames and level of the modules within $(-6)$~cm to $(+1.5)$~cm of
the design level.

Additional information related to the Icarus cryogenic system at Fermilab is given in references~\cite{Ref19} through~\cite{Ref25}.

\FloatBarrier

\section{LAr main containers (cold vessels)}
\label{sec:LArContainers}

As part of SBN Program at Fermilab and Neutrino Platform at CERN, two identical Icarus Cold Vessels
T300, each holding TPC, are installed in one common warm vessel T600 at the SBN-FD building
(references [26] to [30]). In this paper we refer to them as West and East modules.

\begin{figure}[htbp]
  \centering
  \includegraphics[width=0.85\textwidth]{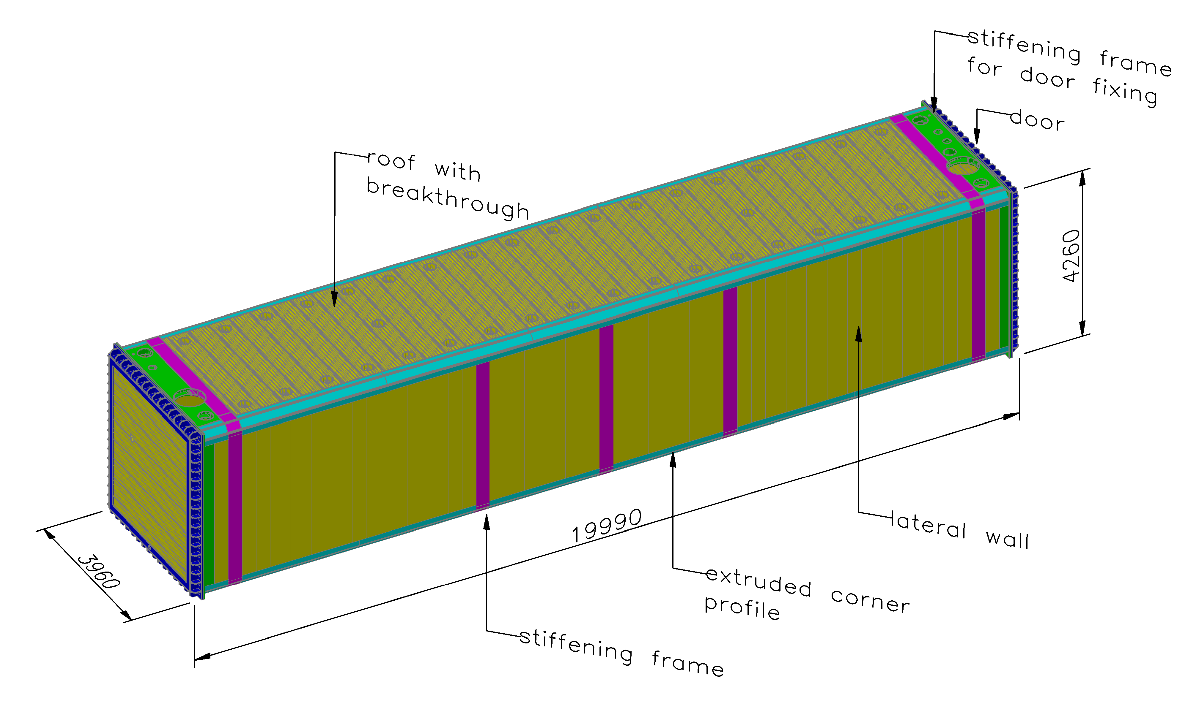}
  \caption{Modules external dimensions}
  \label{fig:sec3-20}
\end{figure}

The T300 internal dimensions of each vessel are 3.62~m $\times$ 3.92~m $\times$ 19.65~m (see Figure~\ref{fig:sec3-20})
with volume 279~m$^3$ filled with $\sim$380~tons of liquid argon at maximum 350~mbarg gas pressure in the
ullage (based on relief set pressure). The vessels are pressurized during ``normal'' operations, but they
were also pumped down to full vacuum for leak checking and cleanup before commissioning into
``normal'' operations.

The vessels were originally designed by Finzi Associati, Milano, Italy for design pressure of 350~mbarg
at the ullage. The design used EN-AW~6082~T6 aluminum as material for all structural elements. The
choice of AW~6082~T6 was justified based on the maximum capabilities of the best available extrusion
presses so to maximize size of the extruded panels and minimize number of welds. The original design
by Finzi was based on Eurocode EN~1999 ``Design of aluminium structures''.

\begin{figure}[htbp]
  \centering
  \includegraphics[width=0.85\textwidth]{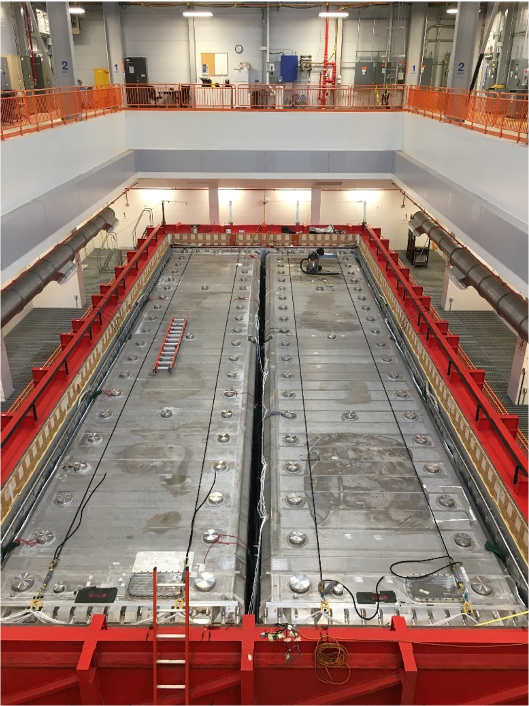}
  \caption{T300 cold vessels installed (Aug. 2018)}
  \label{fig:sec3-19}
\end{figure}

The rectangular shell of the vessels is closed on each side with the doors. The doors are attached with
studs and nuts. This bolted connection is responsible for the structural integrity under internal pressure.
The roof of the vessels has multiple flanged penetrations, including 40 chimneys (per vessel), plus 2
flanged penetrations for re-condensed argon, 1 flanged connection for liquid argon return from pumps,
and 3 flanged connections for magnetic safety devices. Each vessel has 2 $\times$ 800~mm $\times$ 420~mm manholes
blanked with flanges and secured with clamps. Each vessel also has 2 flanged side penetrations to
extract liquid argon to the argon circulation pumps. The safety impact from having penetration below
the liquid argon level is considered in design of the cryogenic piping system and in the ODH analysis;
double safety-interlocked isolation valves leading to the pumps are installed to mitigate probability of
liquid spill into the building space.

\begin{figure}[htbp]
  \centering
  \includegraphics[width=0.85\textwidth]{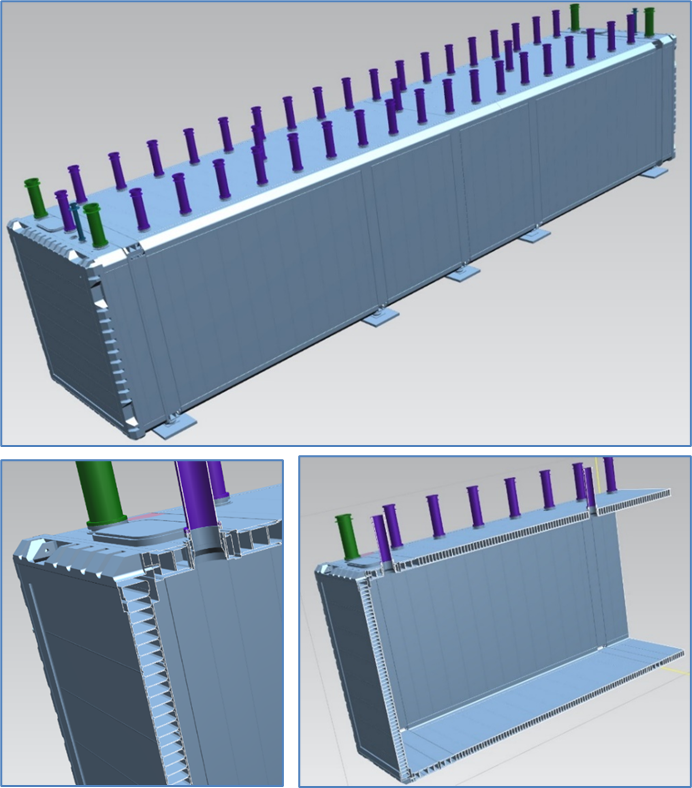}
  \caption{Details of construction of T300 cold vessels}
  \label{fig:sec3-21}
\end{figure}

The fabrication and welding of the extruded panels comprising the vessels were performed by ST
Extruded Products Group --- STEP-G, Germany GmbH. ST Extruded Products Group --- STEP-G supplied
CERN with material certificates listing destructive test data for extruded profiles. That data confirms
material properties for EN-AW~6082~T6 with strength and ductility to be better than listed properties.
The final welding and assembly have been performed at CERN by CERN personnel. The final closure
weld between the vessel and the doors on both sides, as well as closure welds for the manhole covers,
are performed by Fermilab weld shop.

CERN did the final assessment of the design and construction. CERN presented a methodology~\cite{Ref26}
to demonstrate compliance with harmonized European standards. Though the design pressure of
350~mbarg places the cold vessels outside the scope of US and EU pressure vessels standards, CERN chose
EN~13445 ``Unfired Pressure Vessels'' as the design code for both, internal pressure case and vacuum
case. Specifically, CERN chose Finite Element Analysis (FEA) as the calculation method used for the
design of the vessel per methodology of EN~13445-3 Annex~B -- Design-by-Analysis ``direct route''. CERN
issued documents to validate methodology~\cite{Ref26}, FEA calculation results~\cite{Ref27}, buckling analysis results~\cite{Ref28}
and door bolts calculations~\cite{Ref29}. The safety assessment of the cold vessels was done per Fermilab
ESH Manual and described in EN02601 and EN02615~\cite{Ref30}. That included verification of design
calculations, manufacturing data and installation QA/QC.

It is important to mention that both, Finzi and CERN listed two loading cases for the T300 vessels:
operational case at internal pressure of 350~mbarg and vacuum test case at external pressure of
1~bar. CERN further tested the T300 vessels at full vacuum (EDMS Doc. No: 1761013 and 1769691) and
verified that results of FEA calculations agreed with measured stresses and strains.

The vessels were further tested and accepted for vacuum operations at Fermilab in summer of 2018
per Engineering Notes EN02613 and EN02614. The vessels were then equipped with sets of strain
gauges (like those used at CERN for vacuum tests) and then installed in the T600 warm vessel (Figure~\ref{fig:sec3-19}). The
vessels were pneumatically pressure tested at 350~mbarg in April 2019, and then successfully filled with
liquid argon during commissioning in February--April 2020. The pressure control for the cold vessels is
implemented with gas re-condensers at $\sim$1.050~bara. Additionally, two vent valves per cryostat are set
to control the pressure at 1.150~bara. Additionally, a single control valve with Cv=70 is set to control
the pressure at 175~mbarg with full opening at 250~mbarg based on analogue pressure readout or
pressure limit switch status. Finally, the pressure safety for each of the T300 cold vessels is implemented
with three (3) magnetic safety devices by VELAN (France), model 2N2573 with orifice 168.3~mm
(22156~mm$^2$ flow area) and discharge coefficient 0.68. They are calculated to protect the vessels from
at least these four main over-pressurization scenarios:
\begin{enumerate}
\item Loss of LN$_2$,
\item Overfill from LAr dewar,
\item Runaway makeup gas,
\item External fire engulfing warm vessel.
\end{enumerate}

\FloatBarrier

\section{Passive insulation and supporting structure}
\label{sec:WarmVessel}

\begin{figure}[htbp]
  \centering
  \includegraphics[width=0.90\textwidth]{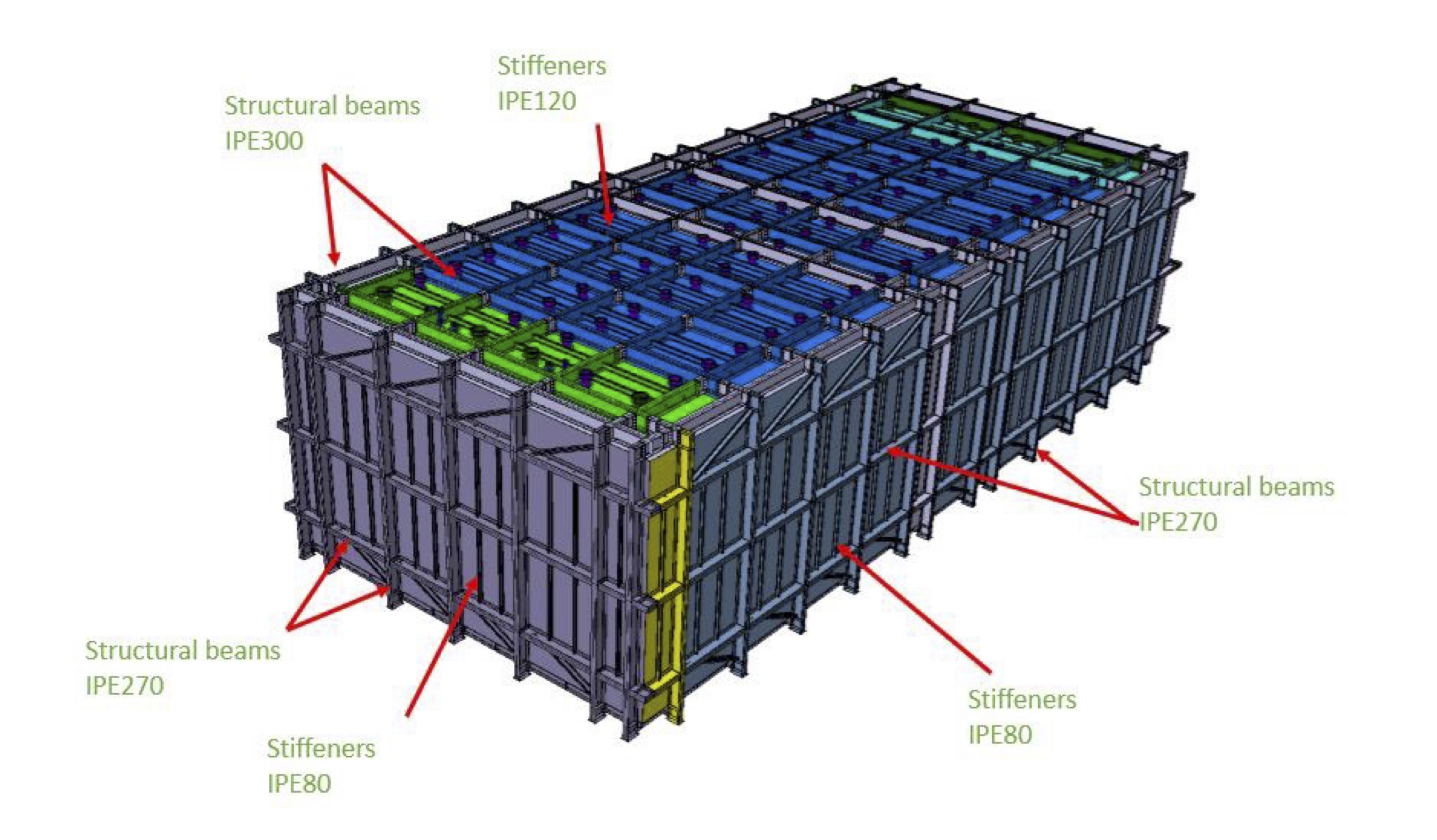}
  \caption{Support structure layout (top view)}
  \label{fig:sec4-22}
\end{figure}

\begin{figure}[htbp]
  \centering
  \includegraphics[width=0.90\textwidth]{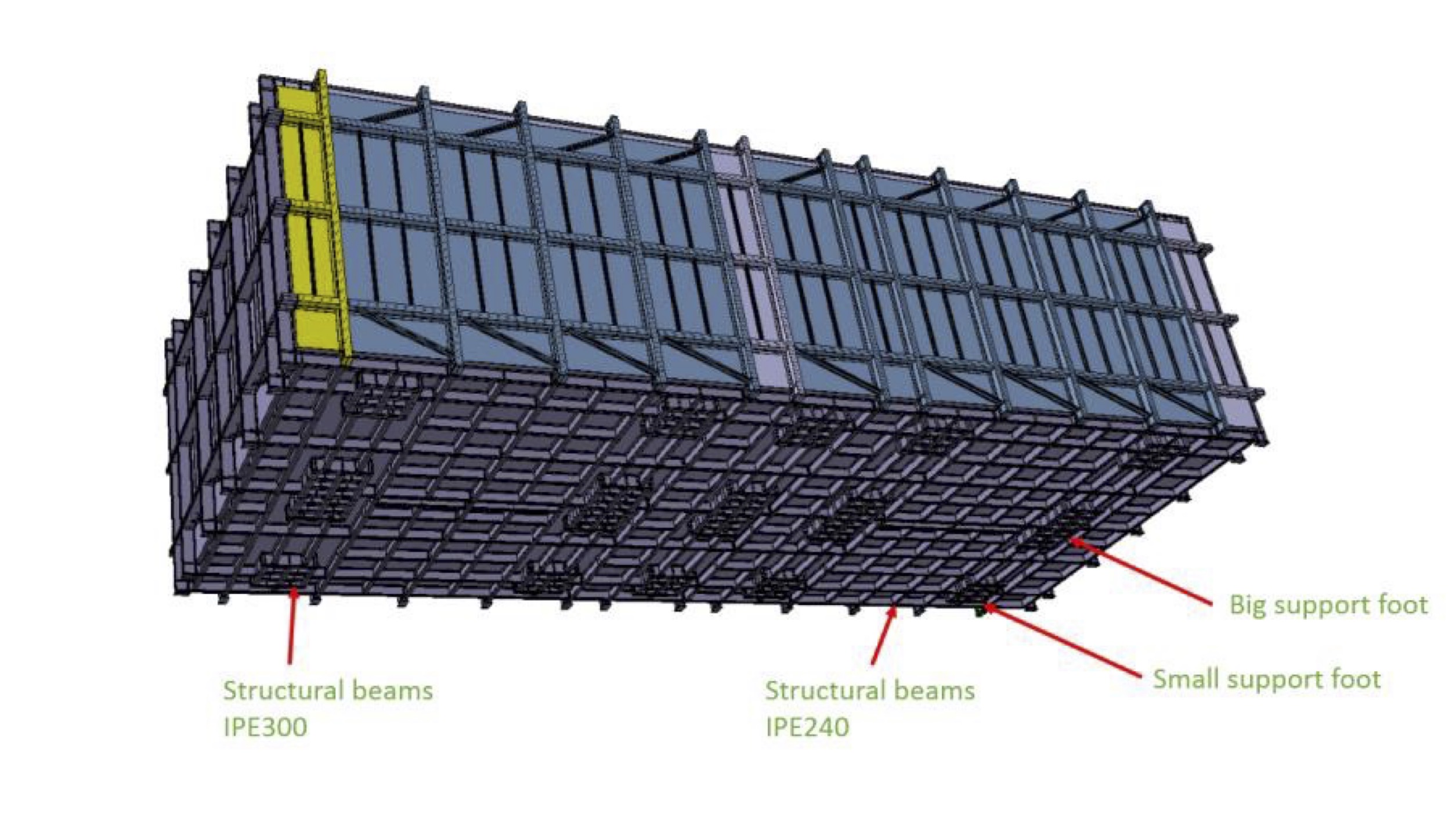}
  \caption{Support structure layout (bottom view)}
  \label{fig:sec4-23}
\end{figure}

CERN designed the WA104 warm vessel~\cite{Ref31,Ref32,Ref33}. The structure (Figure~\ref{fig:sec4-22},
Figure~\ref{fig:sec4-23}) is classified as execution class~2 (EXC2) in accordance with EN~1090. It provides the
mechanical support of the inner insulation and containment of low-pressure nitrogen purge gas.

The structure consists of vertical beams alternated with horizontal beams and extra stiffeners to
support and stiffen the structural steel plate. Inside the steel structure, a skin of structural steel plate
is welded. On top of the plates a layer of insulation is installed. The insulation is glued on the side
walls, which means that the weight of it concerning the side walls as well as the weight of the plate
are directly taken by them, and no extra load is transferred to the floor of the steel structure.

The dimensions of the warm vessel structure are shown in Table~\ref{tab:warmstructure}.

\begin{table}[htbp]
\centering
\caption{Warm Vessel Structure geometrical parameters}
\label{tab:warmstructure}
\begin{tabular}{lccc}
\hline
\textbf{Dimensions} & \textbf{Width (mm)} & \textbf{Length (mm)} & \textbf{Height (mm)} \\
\hline
External & 10268 & 22638 & 6218 \\
Inner (structural steel plate to structural steel plate) & 9720 & 22090 & 5664 \\
Inner (from insulation to insulation) & 8520 & 20890 & 4664 \\
Middle axis of beam to middle axis of beam & 9998 & 22368 & 5948 \\
\hline
\end{tabular}
\end{table}

CERN subcontracted the manufacturing (including welding) and delivery of 84 modules + 93 connecting
elements, following the specification drawings provided by CERN. The total weight of the structure is
around 87~T. Each subassembly is constructed of IPE beams welded together. The subassemblies are
delivered equipped with pre-welded steel plates.

Acceptance tests (witnessed by CERN) per EXC2 requirements as per EN~1090, plus multiple additional
tests per EU regulations were conducted prior to delivery.

The structure was erected in the SBN-FD building in 2017 (with the exception of the roof) by a joint 
CERN/INFN/Fermilab team. The prefabricated components were assembled by bolting the structural 
elements together, followed by TIG welding of the external skin to the beams and of the individual skin 
panels to form a continuous, leak-tight surface.

\begin{figure}[htbp]
  \centering
  \includegraphics[width=0.90\textwidth]{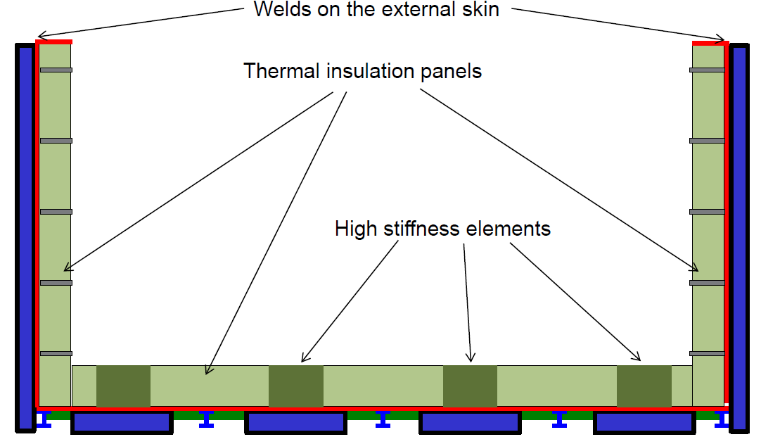}
  \caption{Warm vessel assembly schematics}
  \label{fig:sec4-24}
\end{figure}

\begin{figure}[htbp]
  \centering
  \includegraphics[width=0.90\textwidth]{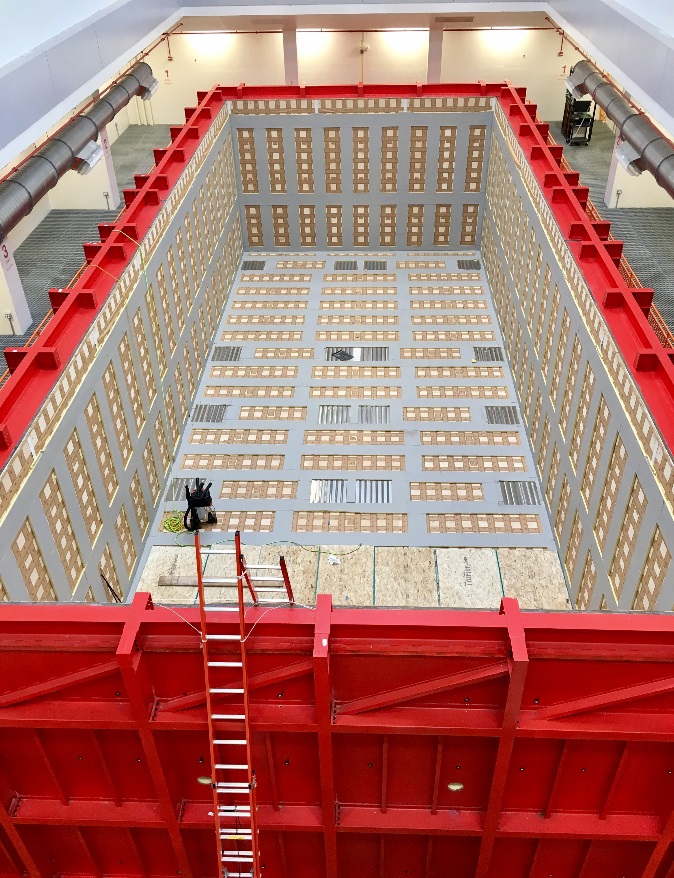}
  \caption{Warm vessel assembly (2018)}
  \label{fig:sec4-25}
\end{figure}

In addition to standard weld quality assurance procedures—including welder certification, visual inspection, 
and liquid penetrant testing—the leak tightness of the welds between skin panels was verified using 
dedicated vacuum test fixtures. These tools enabled the application of vacuum to sections of weld 
approximately 40 cm in length, allowing localized leak detection. All weld qualification and leak tests 
were completed prior to installation of the insulation panels.
The insulation panels were then installed on the floor and walls (see Figure~\ref{fig:sec4-24},
Figure~\ref{fig:sec4-25}).

Two cold vessel tanks (see Section~\ref{sec:LArContainers}) were installed inside the warm vessel in July--August~2018, after
the installation of the lower and vertical side elements of the cold shields (Section~\ref{sec:ColdShields}). Each tank sits on
ten feet. The insulation below those feet is a denser one so that it is suitable to withstand the total
load. For that reason, on the floor of the support steel structure special metallic feet were designed to
support the total weight coming from the tanks. The rest of the area of the floor only supports the
insulation.

\begin{figure}[htbp]
  \centering
  \begin{subfigure}[t]{0.9\textwidth}
    \centering
  \includegraphics[width=\textwidth]{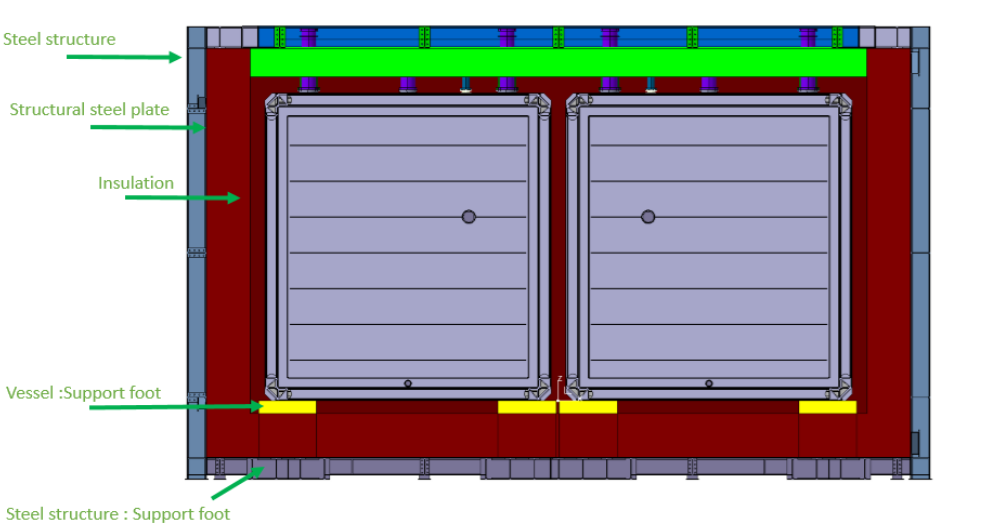}
  \end{subfigure}
  \begin{subfigure}[t]{0.9\textwidth}
    \centering
  \includegraphics[width=\textwidth]{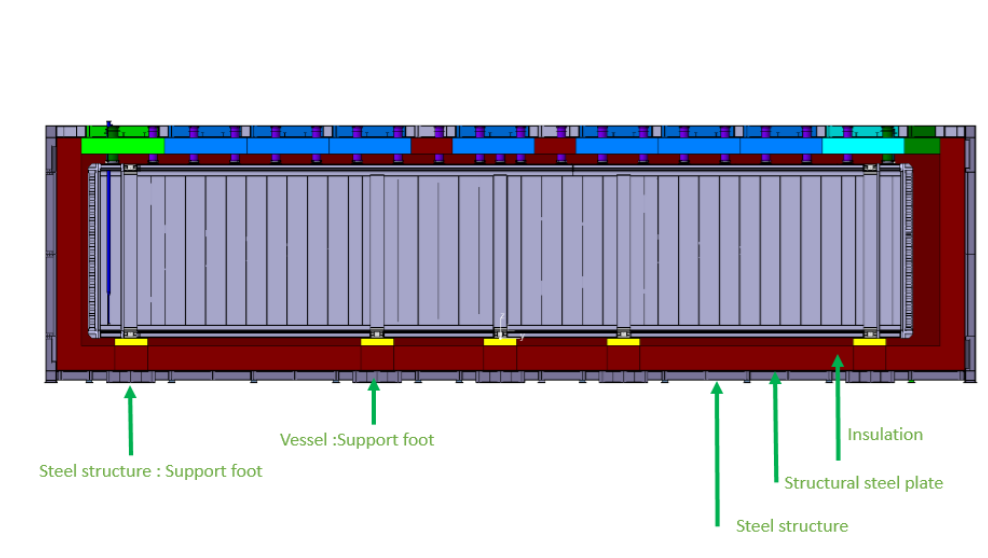}
  \end{subfigure}
  \caption{Inner tanks \& Support feet}
  \label{fig:sec4-26}
\end{figure}

The top of the steel structure was accessible for installation of nitrogen shields and cold vessels and
was closed later, after the T300 vessels were installed. The feedthroughs are only supported on the roof
of the inner tanks and not on the roof beams or the roof steel plates of the steel structure. The
insulation thickness is 600~mm on the bottom and side walls, and 400~mm on the roof. The structural
steel skin thickness is 4~mm on the side walls and roof, and 10~mm on the bottom wall. See
Figure~\ref{fig:sec4-26} for an illustration of cold vessels installation inside the structure.

\begin{figure}[htbp]
  \centering
  \includegraphics[width=0.90\textwidth]{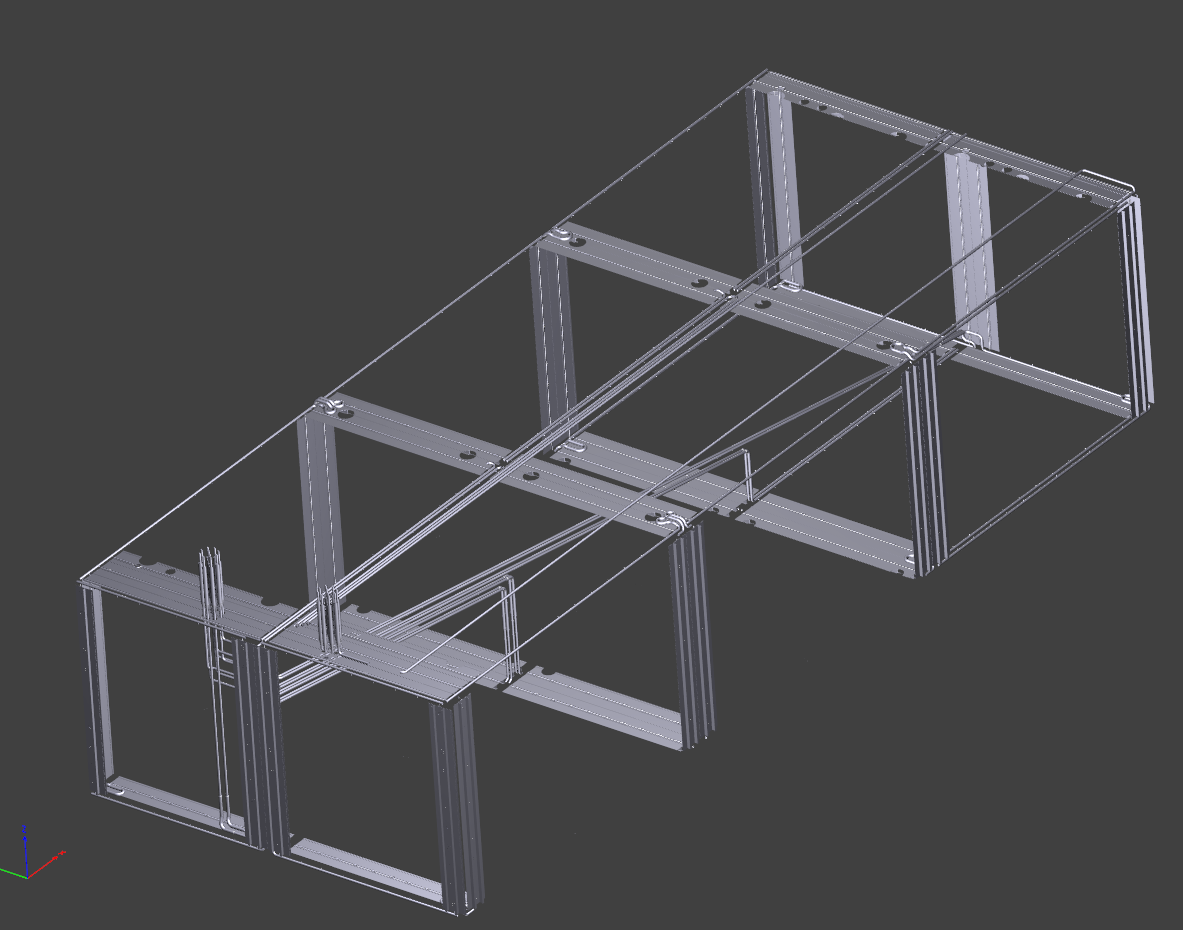}
  \caption{Cold nitrogen shields assembly}
  \label{fig:sec4-27}
\end{figure}

The cold nitrogen shields (see Section~\ref{sec:ColdShields} assembled inside the warm structure (see Figure~\ref{fig:sec4-27})
and positioned between the cold vessels and the foam insulation, thus effectively intercepting the heat
leak from atmosphere through the passive insulation. The weight load from the nitrogen shields is
accounted for in the design of the structure, while a possible rupture and subsequent leak of nitrogen
into the warm-structure space is accounted for in the venting calculations.

The support structure therefore serves a dual purpose: it houses the modules inside the common
insulated space and maintains the inert nitrogen environment inside the structure to prevent any
moisture content plating on the cold surfaces. The penetrations on the roof (chimneys, transfer lines,
etc.) are sealed with plastic covers to minimize leakage of purging gas to atmosphere.

The safety assessment of the warm structure was done per Fermilab ESH Manual and described in
EN02602. That included verification of design calculations, manufacturing data and installation QA/QC.

The pressure safety included the following over-pressurization scenarios:
\begin{enumerate}
\item Internal failure of GN$_2$ purge supply and release of max available 10~m$^3$/hr,
\item Failure of LN$_2$ shield supply pipe and release of full LN$_2$ pump discharge flow into the warm vessel,
\item Failure of a module or its component and release of LAr into the volume of the warm vessel.
\end{enumerate}

\begin{figure}[htbp]
  \centering
  \includegraphics[width=0.90\textwidth]{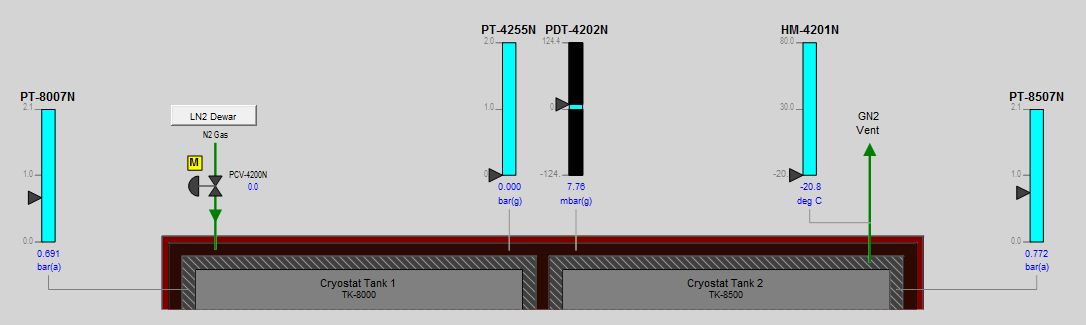}
  \caption{Warm structure volume process diagram}
  \label{fig:sec4-28}
\end{figure}

The warm structure (with nitrogen shields inside it) is schematically shown in Figure~\ref{fig:sec4-28}. It
shows that the volume is purged with gaseous nitrogen at $\sim 7$~mbarg and is protected from
overpressure with two pressure reliefs set at 10~mbarg. The reliefs are Construzioni Meccaniche Lupi
S.r.l., Model~493 -- 10'' (DN250). It is set at 100~mm H$_2$O (10~mbar) and has a rated flow of
$\sim 3813$~Nm$^3$/hr at 11~mbar.

The warm structure was pneumatically pressure tested at $\sim 33$~mbarg in November~2019.

\FloatBarrier

\section{Cold shields}
\label{sec:ColdShields}

\subsection{Functional description of the cold shields}

The cold shields are a set of heat exchangers placed in the space between the internal surface of the
thermal insulation and the external surface of the two cold vessels. They surround the ensemble of the
two cold vessels: there is no cold shield in between the two cold vessels. The heat exchangers are made
of St.\ Steel pipes (AISI~316L), flown with liquid and gas nitrogen, mechanically attached to aluminum
(AL~6101B) extruded profiles.

The cold shields serve three main purposes:
\begin{enumerate}
\item During the initial phases of the cryogenic commissioning: cool down the cold vessels and the
internal detector components in less than 10 days, while maintaining the thermal gradients on the
St.\ Steel detector structures to within the specifications ($\Delta T_{\rm Max} = 50$~K). During this phase, the
temperature profiles across the thermal insulation are also established.
\item During the filling and operation: define the bulk temperature, and therefore the saturation
pressure, of the liquid argon mass.
\item During operation: avoid that localized heat inleaks, coming through the thermal insulation, induce
boiling or large convective motions in the liquid argon volume and establish a temperature
uniformity in the liquid argon volume to within the specifications ($\Delta T_{\rm LAr,Max} = 1$~K).
\end{enumerate}

Thermal exchange between the cold shields and the thermal insulation, on one side, and the cold
vessels, on the opposite side, is provided mainly by thermal convection of the nitrogen gas at
atmospheric pressure, that fills the volume between the thermal insulation and the cold vessels. On the
top, there is also thermal contact between the cold shields' aluminum profiles and the cold vessels.

To intercept heat inleaks, the cold shield panels must cover the largest possible fraction of the internal
surface of the thermal insulation, in particular on the bottom and vertical sides. The requirement on
the temperature uniformity ($\Delta T_{\rm LAr,Max} = 1$~K) sets a constraint on the minimal conductance of the
nitrogen circuit that must be large enough to prevent heating of the liquid nitrogen beyond the above
specification.

\begin{figure}[htbp]
  \centering
  \includegraphics[width=0.70\textwidth]{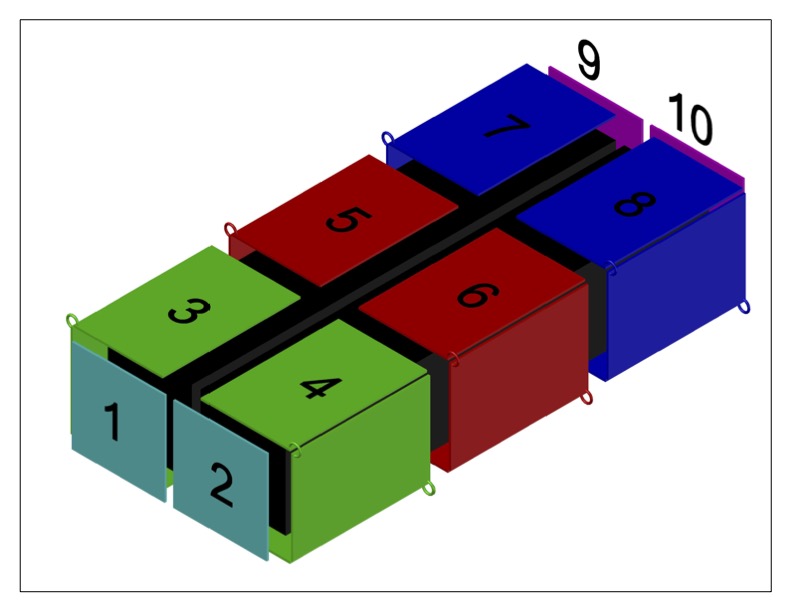}
  \caption{Layout of LN$_2$ shields}
  \label{fig:sec5-29}
\end{figure}

Following the original scheme that was implemented in Gran Sasso, the cold shields are divided in ten
sub-circuits according to the layout shown in Figure~\ref{fig:sec5-29}. Six sub-circuits, three for each cold vessel, cover
the bottom, the East, the West and the top sides. Each one of these six sub-circuits is obtained by joining
in series a horizontal heat exchanger panel on the bottom with a vertical heat exchanger panel on the
side (East or West) and then with a horizontal heat exchanger panel on the top to form a C-shaped
circuit. The remaining four sub-circuits (two per cold vessel) cover the North and South vertical sides,
each one made of a single vertical heat exchanger panel.

Liquid nitrogen is provided to the cold shields either by pressure transfer from the external storage
dewar or from the phase separator, through a Barber Nichols cryogenic liquid circulation pump. The
common input is separated through the ten sub-circuits in a valve box where a set of 10 proportional
valves is used to regulate the LN$_2$ flow in each sub-circuit. Another valve box collects the output of cold
shields and another set of 10 proportional valves is used to regulate the LN$_2$ pressure in each sub-circuit
before merging the flows and returning them to the phase separator.

At the beginning of the cool down, the input to the cold shields is coming, by pressure transfer, from
the external LN$_2$ storage dewar. In this initial phase, all the LN$_2$ is gasified before reaching the output
valves and it is exhausted to the atmosphere, outside the building. When the temperatures are
sufficiently low, a fraction of the cooling nitrogen will return in the liquid state and will start to fill the
phase separator. At this point, the LN$_2$ circulation pump can be started, and the cooling will proceed
with forced circulation through the pump.

\subsection{Cold Shields: design and construction specifications}

\begin{figure}[htbp]
  \centering
  \includegraphics[width=0.70\textwidth]{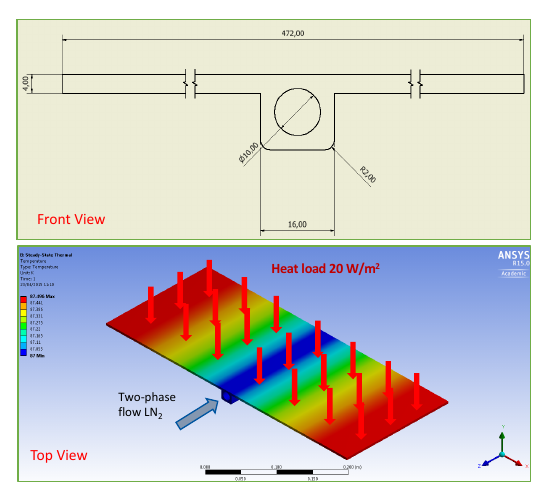}
  \caption{View of the panel geometry and thermal loads used in the thermal simulation. Dimensions are in mm.}
  \label{fig:sec5-30}
\end{figure}

The design of the cold shields was initially developed assuming a structure completely made of
aluminum, with plates with a circular hole in the middle where LN$_2$ at the desired temperature (87~K)
was flown (Figure~\ref{fig:sec5-30}). The minimal design specifications are: $< 1$~K temperature gradient across the
panel, with a heat load of 10~W/m$^2$ uniformly distributed on the panel surface (the specifications for
the thermal insulation require an average heat inleak of $< 10$~W/m$^2$).

\begin{figure}[htbp]
  \centering
  \includegraphics[width=0.70\textwidth]{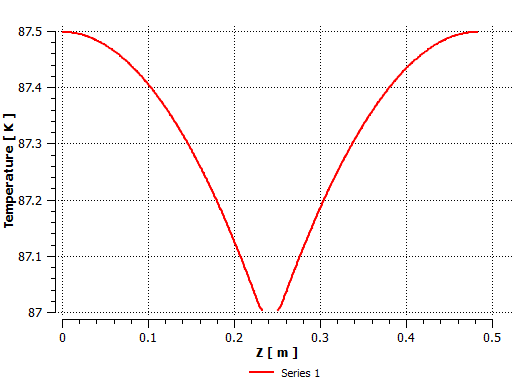}
  \caption{Calculated temperature profile across the panel for a heat load of 20~W/m$^2$ uniformly distributed on the
panel's surface.}
  \label{fig:sec5-31}
\end{figure}

The geometry illustrated in Figure~\ref{fig:sec5-30} was simulated with the ANSYS code, taking as reference material
the alloy AL~1050A (thermal conductivity = 270~W/m/K at 87~K). The results (Figure~\ref{fig:sec5-31}) show that a
temperature difference of $< 0.5$~K is achieved with a thermal load of 20~W/m$^2$: the chosen geometry
provides a comfortable safety factor of 4 with respect to the specifications. Similarly, the pressure drop
across the nitrogen circuit was calculated for two different flow conditions: single phase (liquid) and
two phases with gas to liquid mass ratio 1/5.

\begin{figure}[htbp]
  \centering
  \includegraphics[width=0.85\textwidth]{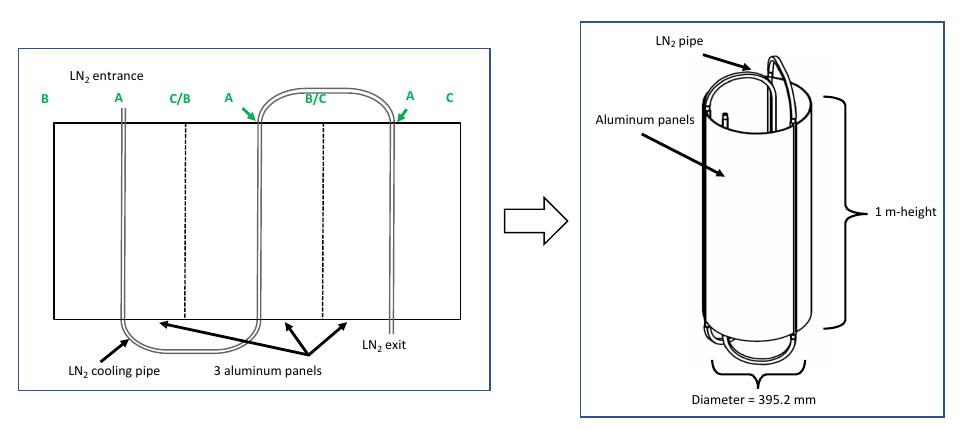}
  \caption{Schematic drawing of the panel tested at CERN. Each one of the three sections shown on the drawing on the left has the same width of the panel in Figure~\ref{fig:sec5-30}.
  The arrangement in a cylindrical shape (right drawing) ensures that the thermal load is on the external side only. This configuration provides a thermally equivalent boundary condition to a single panel with length equal to three times the length of one section laying flat on a surface providing a heat load of 20 W/m$^2$.}
  \label{fig:sec5-32}
\end{figure}

\begin{figure}[htbp]
  \centering
  \includegraphics[width=0.75\textwidth]{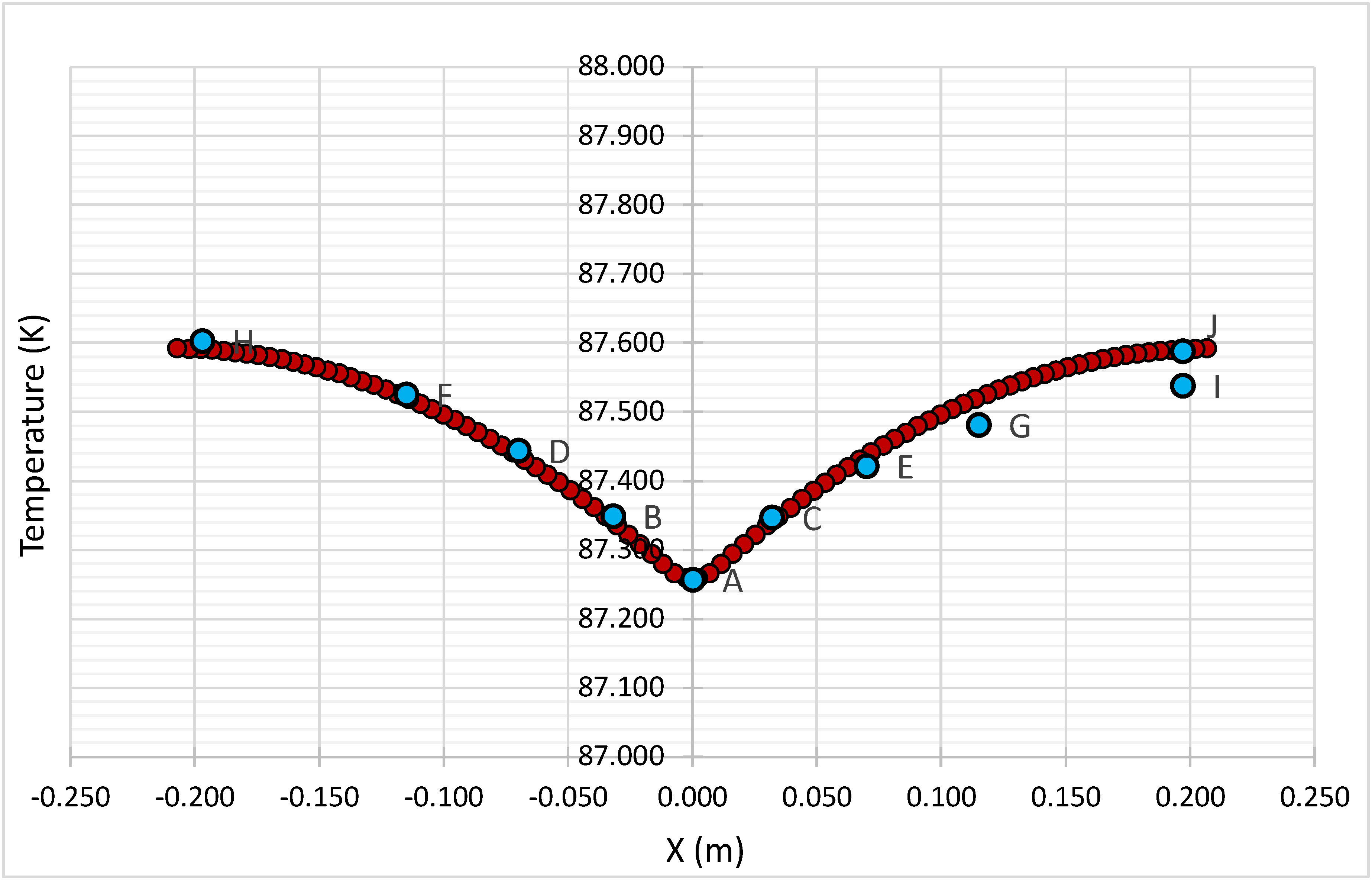}
  \caption{Results of the measurements performed on the test panel (blue circles) compared with the theoretical predictions (red circles), demonstrating the validity of the theoretical calculations. The theoretical curve differs from that reported in Figure~\ref{fig:sec5-31}, as the calculation was adapted to the geometry and thermal configuration of the mock-up test setup.}
  \label{fig:sec5-33}
\end{figure}

The theoretical calculations were then validated with a test conducted at CERN on a mock-up sample
of three panels arranged in a cylindrical shape (Figure~\ref{fig:sec5-32}) so that they could be hosted inside a
cylindrical dewar. The measurements confirm the prediction from the simulations (Figure~\ref{fig:sec5-33}).

\begin{figure}[htbp]
  \centering
  \includegraphics[width=0.75\textwidth]{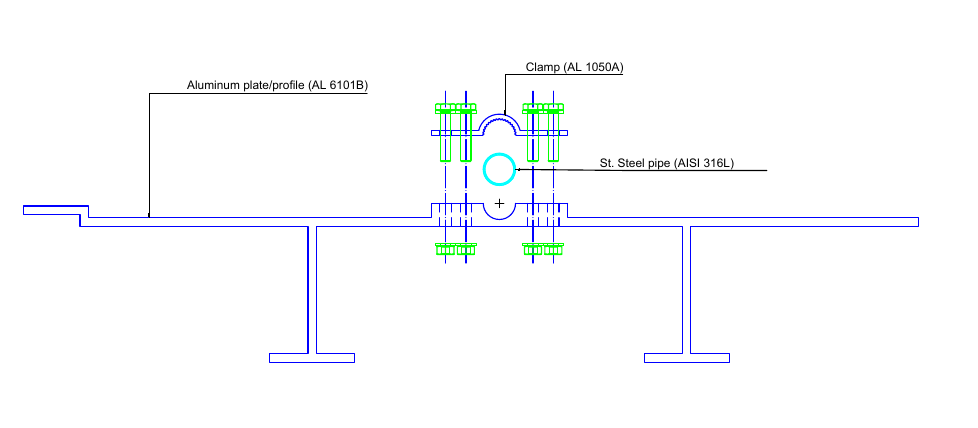}
  \caption{Coupling between an aluminum panel and the St.\ Steel LN$_2$ distribution pipe.}
  \label{fig:sec5-34}
\end{figure}

Due to the technical difficulty and cost in realizing the entire cold shields in aluminum, a hybrid design
was developed, with stainless steel pipes, for LN$_2$ distribution, mechanically coupled (clamped) to
aluminum plates as illustrated in Figure~\ref{fig:sec5-34}.

\begin{figure}[htbp]
  \centering
  \includegraphics[width=0.95\textwidth]{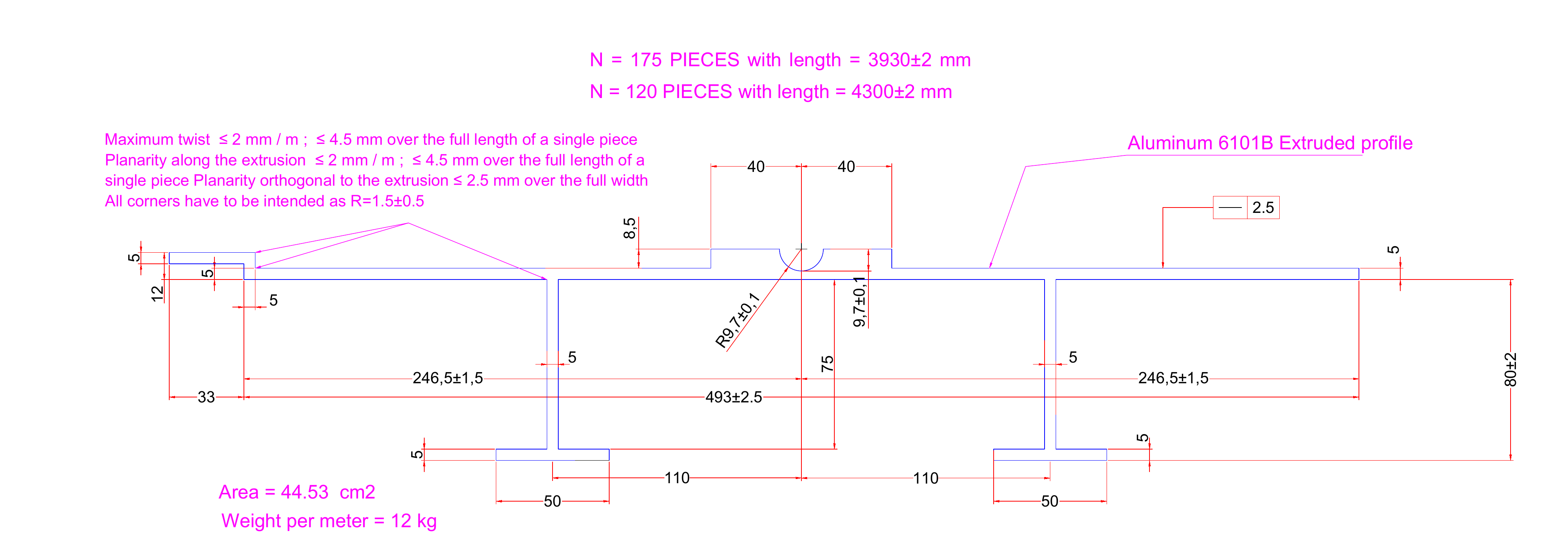}
  \caption{Technical drawing of the aluminum panels cross-section used for the extrusion.}
  \label{fig:sec5-35}
\end{figure}

All the extruded aluminum panels have the profile shown in Figure~\ref{fig:sec5-35}. The groove at the top center of
the profile hosts the stainless-steel pipe where LN$_2$ is flown. The ribs in the lower part of the panels
ensure mechanical stiffness and create some distance from the adjacent surface allowing for a more
uniform thermal exchange. Also, the aluminum thickness was increased from 4~mm to 5~mm, to partially
compensate for the thermal resistance added by the stainless-steel pipes and to increase mechanical
stiffness. During installation, operators need to walk on the cold shields to perform various tasks,
including welding and positioning relatively heavy equipment.

\begin{figure}[htbp]
  \centering
  \includegraphics[width=0.75\textwidth]{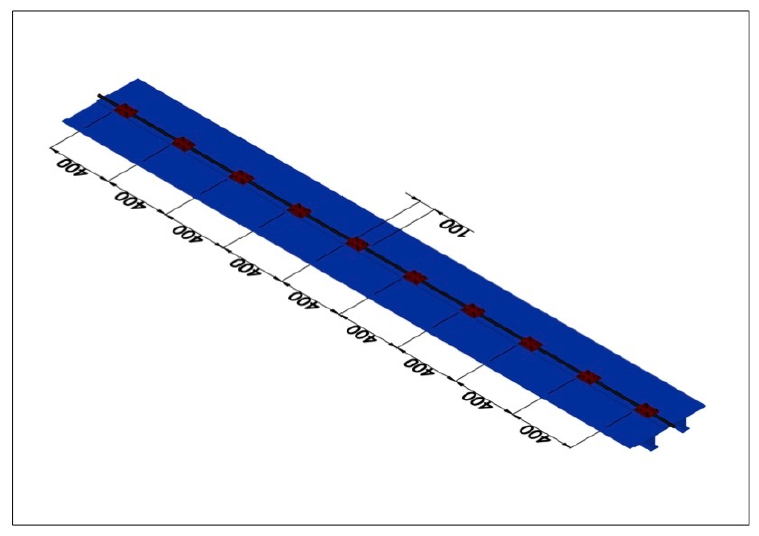}
  \caption{Illustration of the assembly of the piping on the aluminum plates.}
  \label{fig:sec5-36}
\end{figure}

\begin{figure}[htbp]
  \centering
  \includegraphics[width=0.75\textwidth]{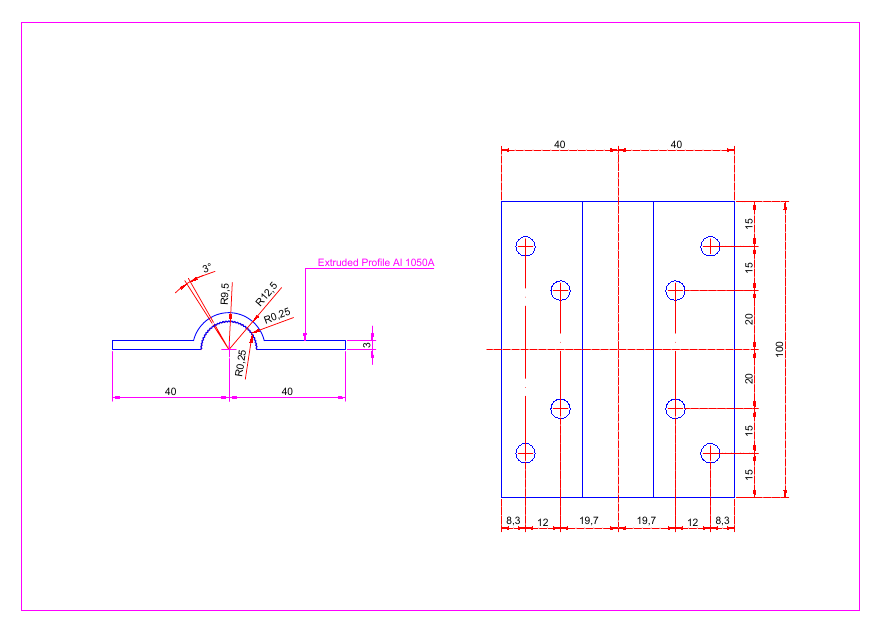}
  \caption{Technical drawings of the clamps for the cold shield panels.}
  \label{fig:sec5-37}
\end{figure}

To ensure a good thermal contact between the pipe and the aluminum panel, 100~mm wide clamps are
placed every 400~mm (Figure~\ref{fig:sec5-36}). The cylindrical inner surface of the clamps is grooved, to ease the
relative displacement of the pipes with respect to the panels due to thermal contractions
(Figure~\ref{fig:sec5-37}).

\begin{figure}[htbp]
  \centering
  \includegraphics[width=0.75\textwidth]{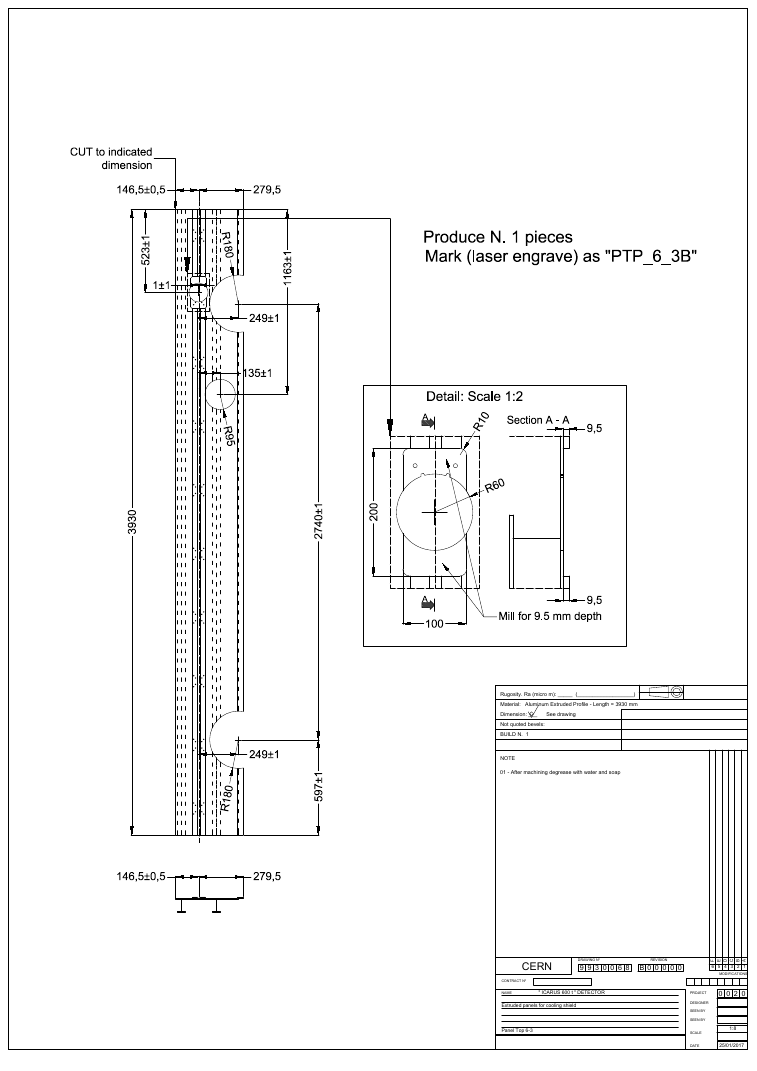}
  \caption{Machining drawing of one of the extruded panels for the top side.}
  \label{fig:sec5-38}
\end{figure}

\begin{figure}[htbp]
  \centering
  \includegraphics[width=0.90\textwidth]{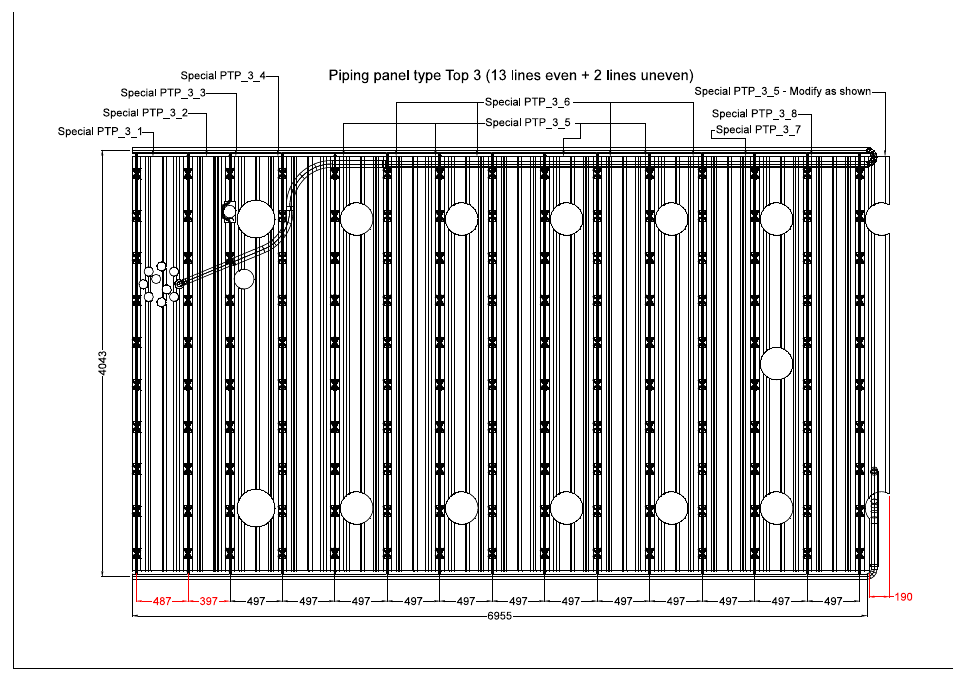}
  \caption{Assembly drawing of one of the cold shield top panels.}
  \label{fig:sec5-39}
\end{figure}

The extruded plates are cut to two lengths, one of 3930~mm for the horizontal (top and bottom) sides,
and one of 4300~mm for the vertical (lateral) sides. All the holes for installation of the clamps, passage
of LAr pipes, cold vessel supports, chimneys for the passage of the wire chambers cables, etc., as well
as the reduction in width of some panels to adapt the assembly to the required lengths, are realized by
machining of the extruded plates (see Figure~\ref{fig:sec5-38} for an example of machining specification and
Figure~\ref{fig:sec5-39} as an example of assembled panels of one of the circuits on the top side).

The nitrogen distribution circuit is made by sets of parallel AISI~316L stainless steel pipes of 16~mm~ID,
19~mm~OD, connected at both ends to manifolds, also AISI~316L stainless steel, with 50~mm~ID,
54~mm~OD. Each set of pipes, mechanically coupled (clamped) to the appropriate set of extruded
aluminum plates, realizes one of the cold shields' planar sub-elements (22 in total) illustrated in
Figure~\ref{fig:sec5-29}.

\subsection{Cold Shields: procurement, assembly, installation and test}

The aluminum plates and the clamps were ordered by CERN to two aluminum extrusion companies:
one operating in Turkey for the plates, and one operating in Israel for the clamps. The two companies
also provided all the required machining (cutting, drilling of holes, etc.) and QA/QC documentation. To
ensure a good thermal contact between the plates and the LN$_2$ distribution pipes, the central groove
was machined to the required dimension and tolerance starting from a smaller groove realized by
extrusion. Standard CERN specifications for material conformity were applied (Certificates type~3.1,
according to EN~10240). For both components, the material is aluminum type~6101b.

The 22, pre-assembled planar LN$_2$ circuits, made of stainless steel type AISI~316L, were also ordered by
CERN to an external company. Specifications referred to applicable European codes for metallic
industrial piping (EN~13480-2,3,4,5), stainless steel (EN~10088-1), stainless steel piping for pressurized
circuits (EN~10216-5; EN~10217-7), welding (ISO~3834; EN~ISO~15614-1; EN~ISO~15609-1; EN~ISO~14732;
EN~ISO~9606-1), cryogenic vessels (EN~13458-2), non-destructive testing (ISO~17637; ISO~17636-1;
EN~ISO~5817; EN~ISO~9712; ISO~13585), certificates (EN~10204). In particular, the pre-assembled circuits
were tested for helium leak tightness at $1 \times 10^{-9}$~mbar~liters/s at room temperature. All pre-assemblies
were also pressure tested at 10~bar for at least 1~hour.

Additional components, such as pipes, joints, braided flex hoses, to realize input and output lines and for
panels interconnections, were procured by Fermilab.

Funding for all cold shields parts was provided by INFN.

All components were delivered at Fermilab for the final assembly and installation.

\begin{figure}[htbp]
  \centering
  \includegraphics[width=0.75\textwidth]{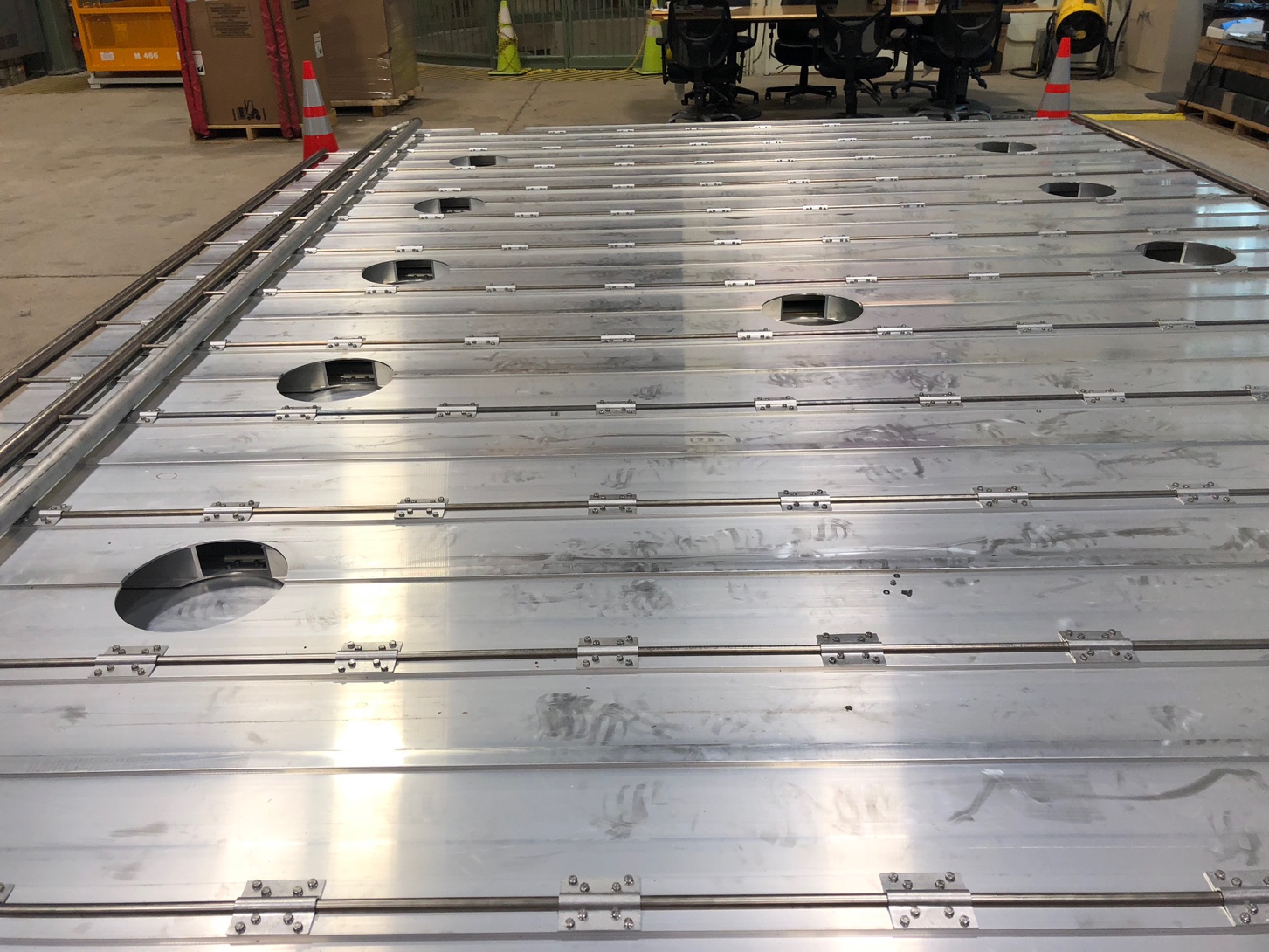}
  \caption{Photograph of one of the top cold shield panels after assembly.}
  \label{fig:sec5-40}
\end{figure}

The first step has been the clamping of the aluminum plates to the pipes of the LN$_2$ circuit to realize
the basic panels of the cold shields (Figure~\ref{fig:sec5-40}). The first panel was thermal shocked in liquid nitrogen,
to verify that the clamping was allowing for the reciprocal displacement between the pipes and the
aluminum plates due to thermal contractions.

\begin{figure}[htbp]
  \centering
  \includegraphics[width=0.75\textwidth]{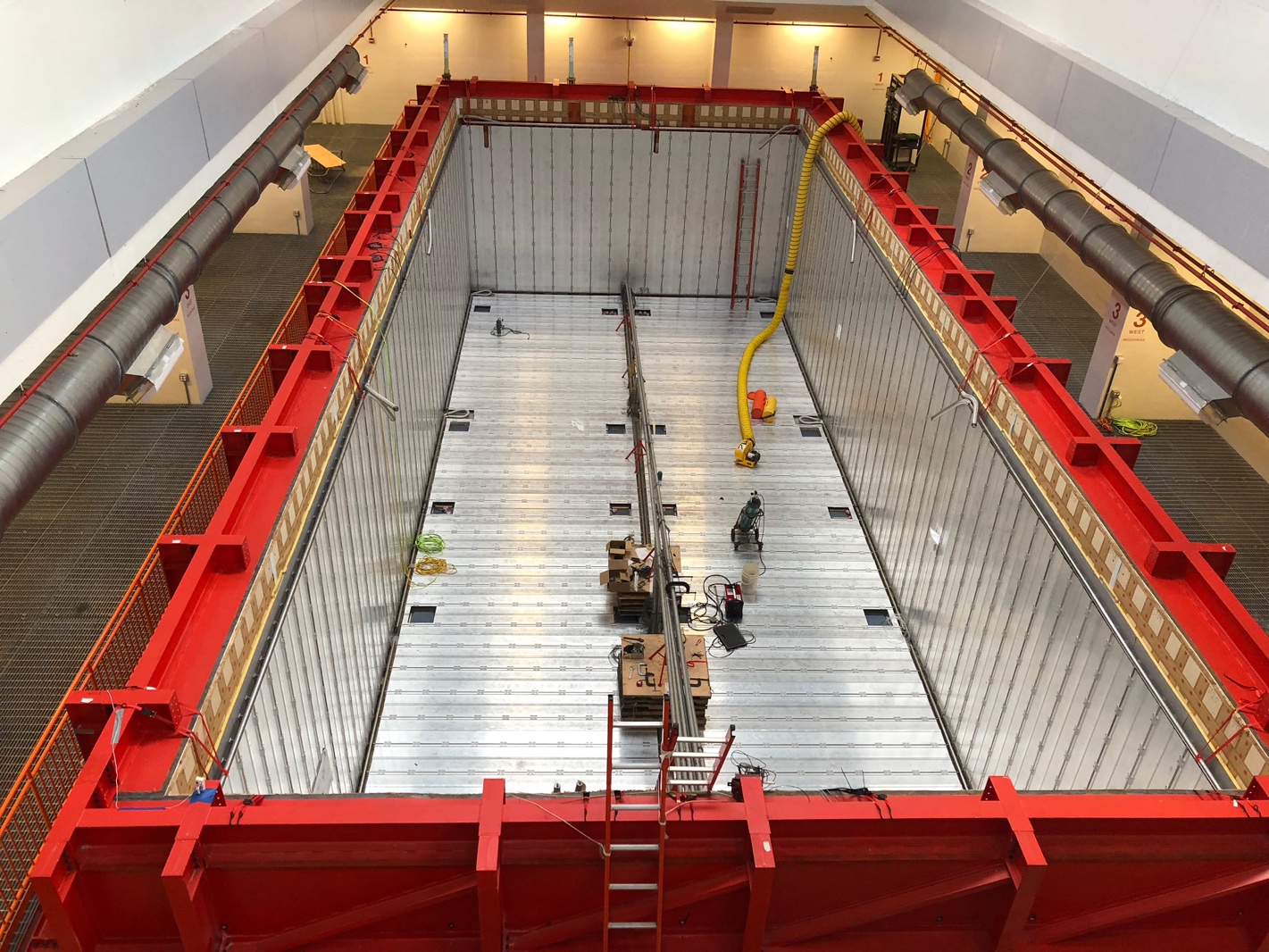}
  \caption{Photograph of ICARUS cryostat after installation of the bottom and vertical cold shields panels.}
  \label{fig:sec5-41}
\end{figure}

The bottom and the vertical panels were then moved inside the thermal insulation. The bottom panels
were connected with braided flex hoses to the lateral panels (side panels). The inlet pipes, running at
the center, in the region between the two cold vessels and welded to the lower part of the cooling
circuit, were installed (Figure~\ref{fig:sec5-41}). Temperature gauges were glued to the aluminum panels in
correspondence of the inlets and cabled. To ensure accessibility of the parts for inspection and repair,
the half-completed circuits were sealed and tested for helium leak tightness before positioning of the
cold vessels. They were also pressure tested at 10~bar for 1~day.

\begin{figure}[htbp]
  \centering
  \includegraphics[width=0.75\textwidth]{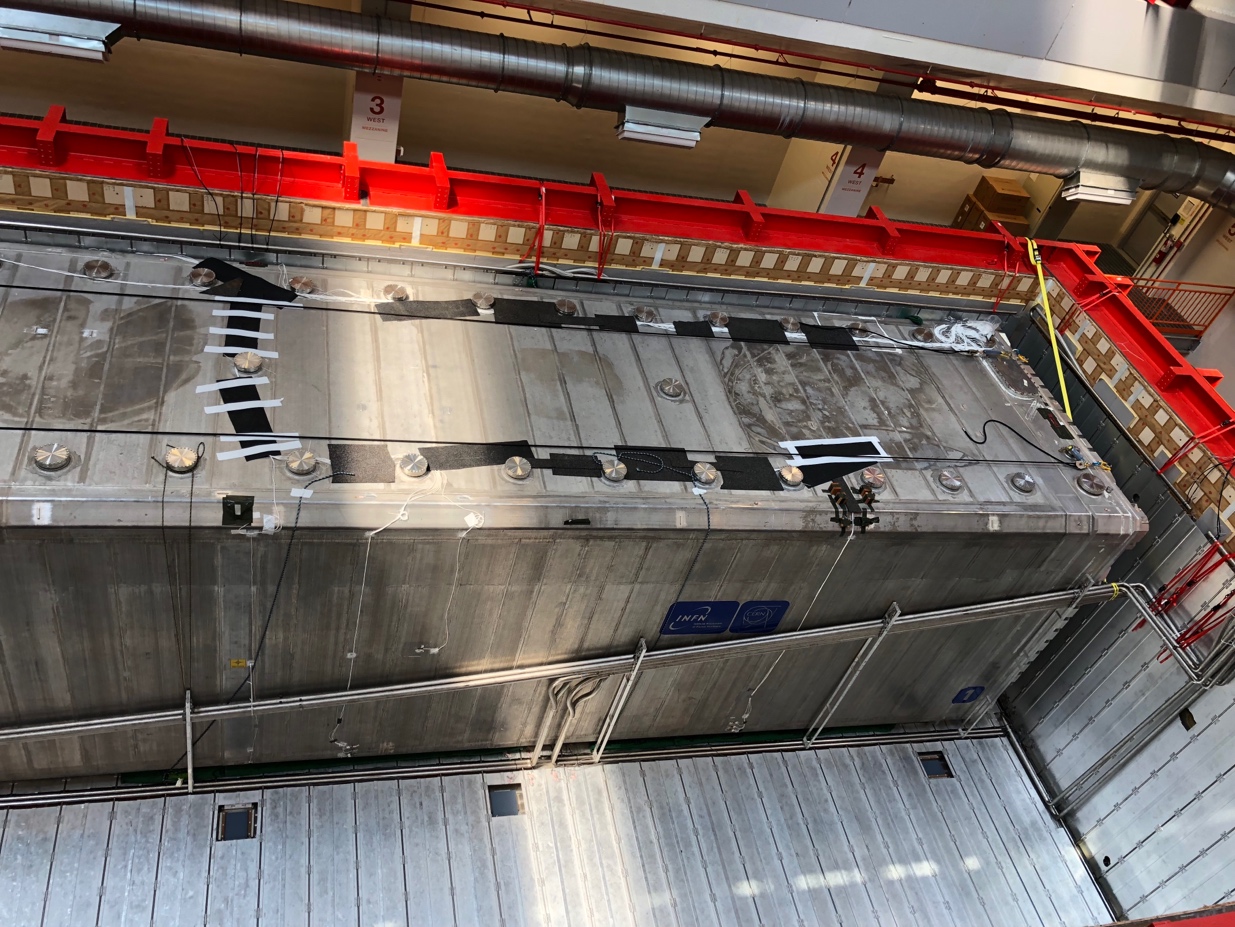}
  \caption{Picture taken after the positioning of the first cold vessel showing also the cold shields inlet pipes.}
  \label{fig:sec5-42}
\end{figure}

\begin{figure}[htbp]
  \centering
  \includegraphics[width=0.75\textwidth]{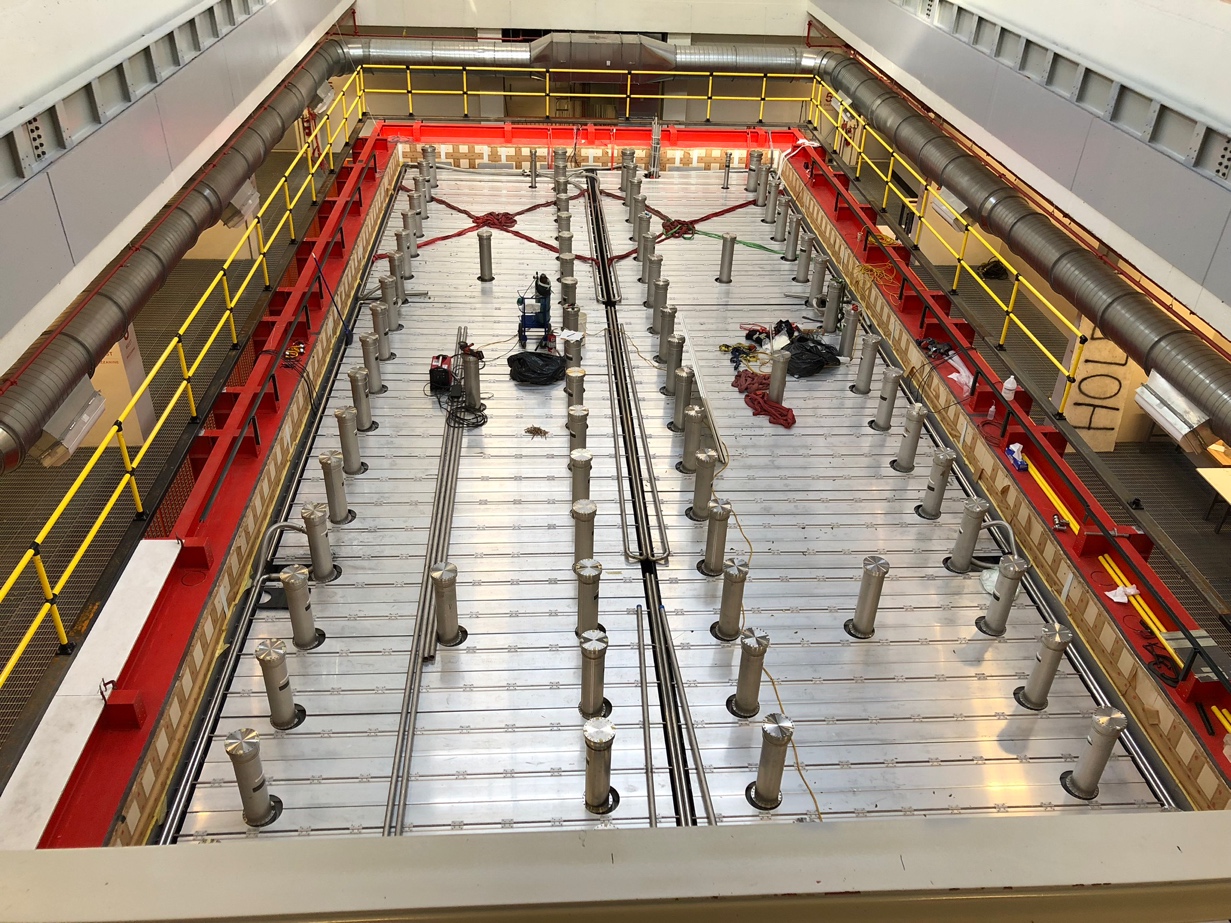}
  \caption{Photograph of the cold shields top part.}
  \label{fig:sec5-43}
\end{figure}

\begin{figure}[htbp]
  \centering
  \includegraphics[width=0.75\textwidth]{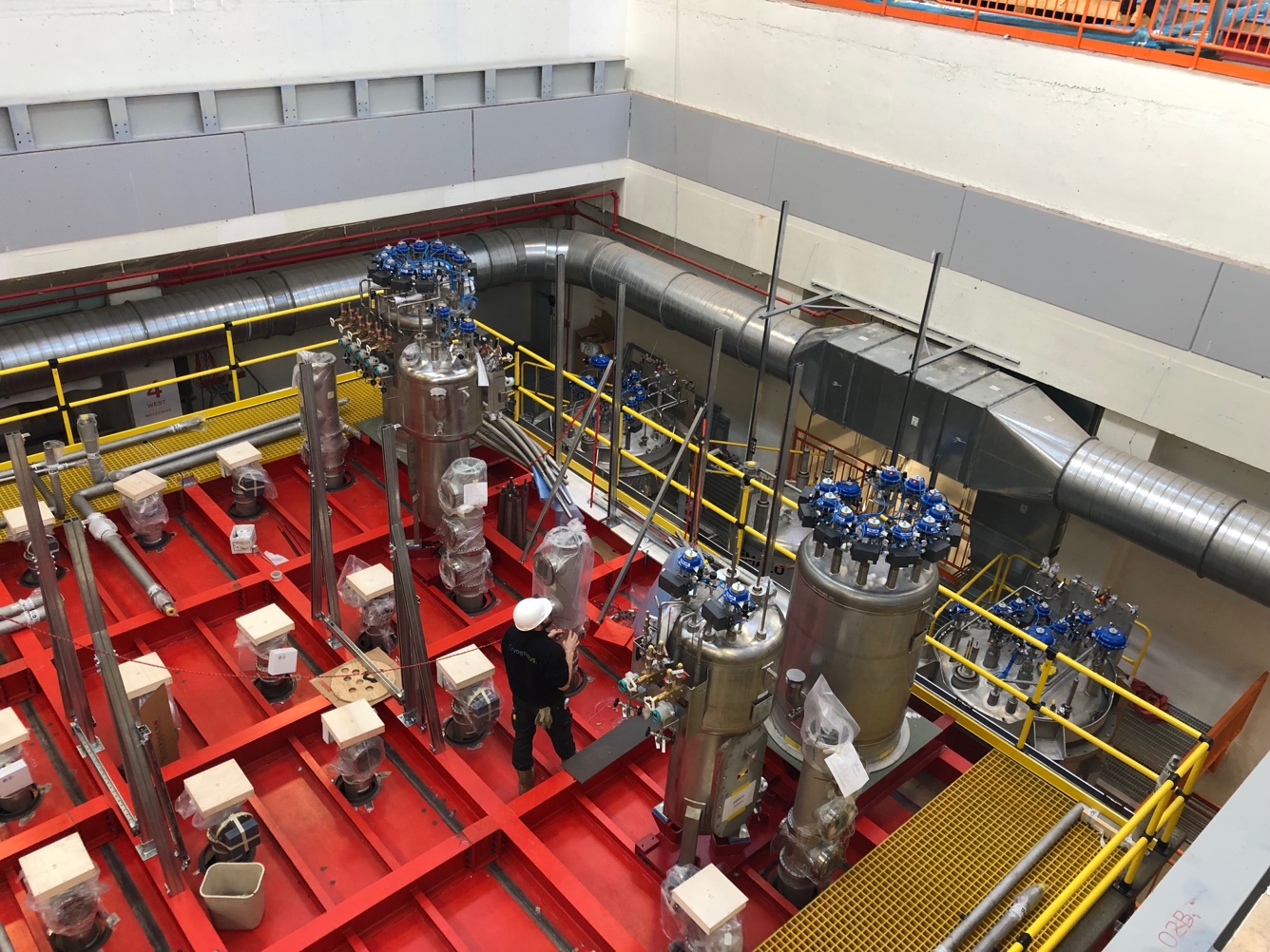}
  \caption{Photograph taken during the installation of the cold shields inlet (right) and outlet (left) valve boxes. In the
picture, the visible outlet pipes are yet to be welded to the ones exiting the corresponding valve box.}
  \label{fig:sec5-44}
\end{figure}

After positioning of the two cold vessels (Figure~\ref{fig:sec5-42}) and installation of the chimney, the top cold shield
panels were put in place and connected with braided hoses to the side panels to complete the cold
shields subsections~3 to~8 of Figure~\ref{fig:sec5-29}. The output lines were installed and welded to the top part of
the panels (Figure~\ref{fig:sec5-43}). The completed circuits were tested for helium leak tightness of the added parts.
They were then pressure tested at 10~bar for 1~day. Welding of the input and output lines to the valve
boxes was performed after installation of the top part of the thermal insulation and of the warm vessel
(Figure~\ref{fig:sec5-44}).

\FloatBarrier

\section{Process controls}
\label{sec:Controls}

The Icarus cryogenics control system is a fully automated system, which allows a continuous cryogenic
operation through the various modes of operation as described in Section~\ref{sec:DescriptionCryo}, while allowing manual
actions by the operator at any moment. It unifies the control of the external, proximity and internal
cryogenics into a single control system. The safety system for ODH is also within scope.

The cryogenics control system was designed, constructed and programmed by Fermilab based on logic
specifications provided by CERN for the overall controls integration and proximity cryogenics, by
Fermilab for the external cryogenics and ODH system and by INFN for the internal cryogenics.

Prior to full operation, the cryogenic control system and associated auxiliary equipment were activated progressively, starting from the control electronics and followed by pumps and actuators, while continuously monitoring the noise levels of the TPC readout electronics. No measurable increase in electronic noise was observed at any stage of this procedure. This operational validation ensures that power supply stability and electromagnetic compatibility of the cryogenic system are adequate for detector operation.

This section describes the selected hardware and software solutions together with the electrical and
control architecture. The software architecture is then presented. Finally, the methodology and the
results for the I/O tests and the functional tests are given.

\subsection{Control system architecture and selected hardware and software solutions}

\subsubsection{System Architecture and Networking}

The cryogenic control system utilizes iFIX by GE Digital, a SCADA software platform. The redundancy is
implemented by installing iFIX on two computers with full redundancy.

\begin{figure}[htbp]
  \centering
  \includegraphics[width=0.90\textwidth]{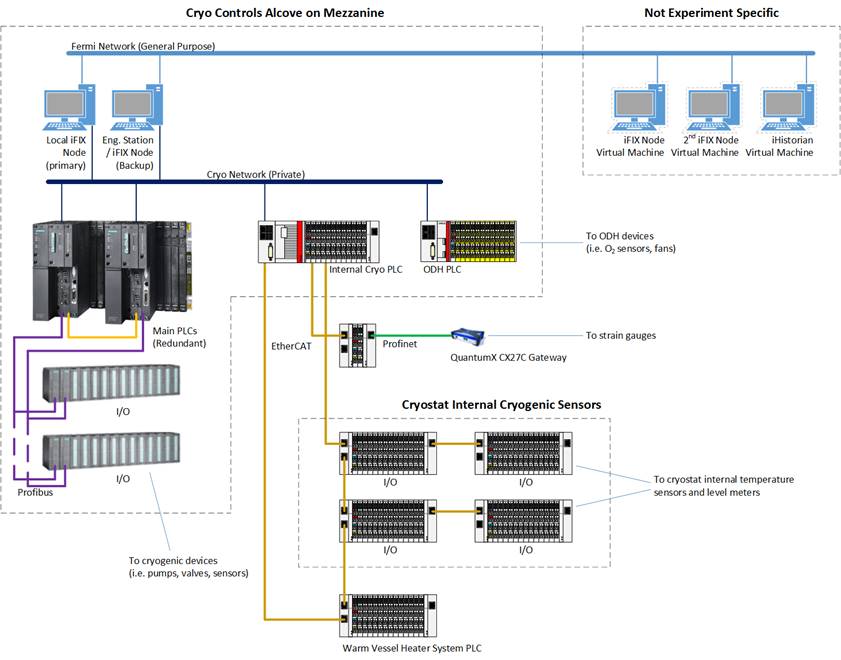}
  \caption{System Architecture}
  \label{fig:sec6-45}
\end{figure}

The primary components of the cryogenic control system include two iFIX computers, one of which is
also an engineering station (for PLC programming), and multiple PLCs (Figure~\ref{fig:sec6-45}).

\subsubsection{Centralized Topology}

The main PLCs provide control and monitoring of most cryogenic equipment. These reside in the first
of a series of five control cabinets in the cryogenic controls alcove on the mezzanine level in the SBN-FD
building. The remaining four cabinets house all the corresponding I/O terminals.

The cryostats’ internal cryogenic sensors have the requirement of being readout by non-switching
signals and thus cannot use Siemens RTD modules. A separate PLC handles readout of these sensors.

\subsubsection{Networks}

The regular Fermi network is available at the SBN-FD building. The iFIX / engineering station computers
make use of this network for connection to the central iFIX and iHistorian virtual machines as well as
remote control system development.

For security, access to the iFIX and iHistorian virtual machines, as well as historical data access through
Seeq, is restricted to machines on the Fermi network (including via VPN connection).

The Cryo Network is private. The iFIX / engineering station computers make use of this network for
communication with the PLCs.

\subsubsection{Supervisory Control and Data Acquisition}

The control system uses Industrial Gateway Server (IGS), also from GE Digital. IGS provides an OPC DA
server that both iFIX and PLCs can connect to and share data. It also provides read-only channels for
the detector control system, which runs on Experimental Physics and Industrial Control System (EPICS),
to collect data.

\subsubsection{Human Machine Interface (HMI)}

The cryogenic control system includes an HMI, using the iFIX Workspace. The HMI provides monitoring
and commanding of all cryogenic equipment. However, the HMI does not contain any control logic.
(PLCs are not dependent on the HMI to operate.)

The design of the HMI incorporates best practices from ANSI/ISA-101.01-2015, Human Machine
Interfaces for Process Automation Systems. This includes implementation of a hierarchy of displays, as
illustrated in Figure~\ref{fig:sec6-46}, Figure~\ref{fig:sec6-47}, Figure~\ref{fig:sec6-48}, Figure~\ref{fig:sec6-49} below.

The above-mentioned hierarchy of displays follows the software architecture described in Section 6.2
including the Process Control Object (PCO) break-down structure, associated modes and status
information, and steppers animation.

\begin{figure}[htbp]
  \centering
  \includegraphics[width=0.90\textwidth]{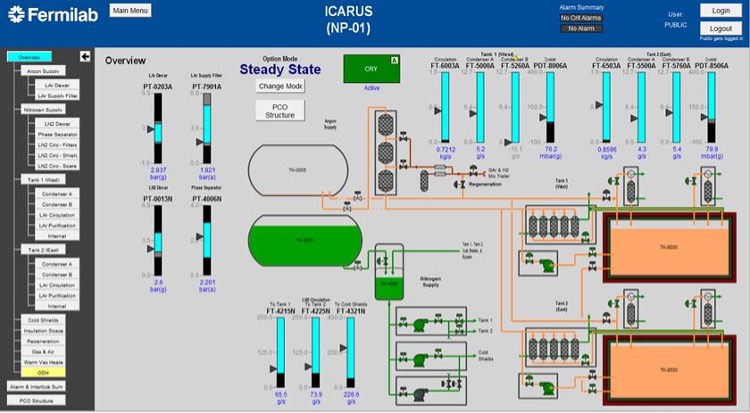}
  \caption{Level 1 Display – Entire System Overview}
  \label{fig:sec6-46}
\end{figure}

The HMI has been developed to enable the display and the simulation of a number of variables in order
to limit the actions needed in the PLC program and the need of controls expert during the
commissioning and operation stages. The simulation of variables is a feature allowing to substitute the
real values read from analogue field devices with a virtual value in order to force the action by the logic
or control loop. These variables include the field inputs and outputs together with associated limits,
ranges, thresholds, the controllers with associated set point and PID parameters, the alarms and
interlocks, a selection of PLC computed variables and the user commands, while the PID parameters of
the controllers remained hard coded into the PLC and the field digital inputs cannot be simulated.

The HMI employs a simple security scheme. It defaults to a “public” user, which allows monitoring only.
To gain access to manipulate the system, users must be granted access by the Admin and log in.

\begin{figure}[htbp]
  \centering
  \includegraphics[width=0.90\textwidth]{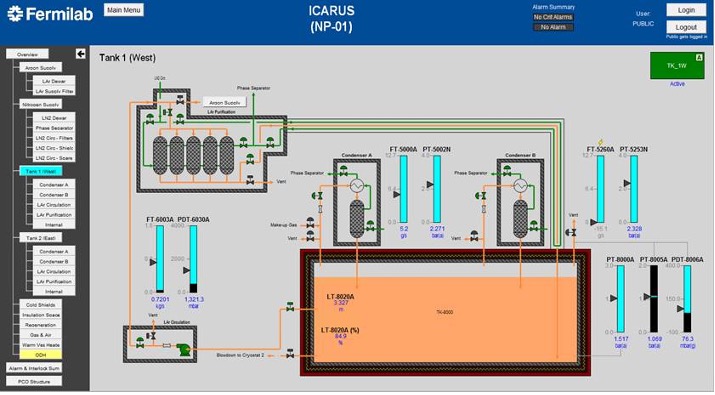}
  \caption{Level 2 Display – Tank 1 (West) Overview}
  \label{fig:sec6-47}
\end{figure}

\begin{figure}[htbp]
  \centering
  \includegraphics[width=0.90\textwidth]{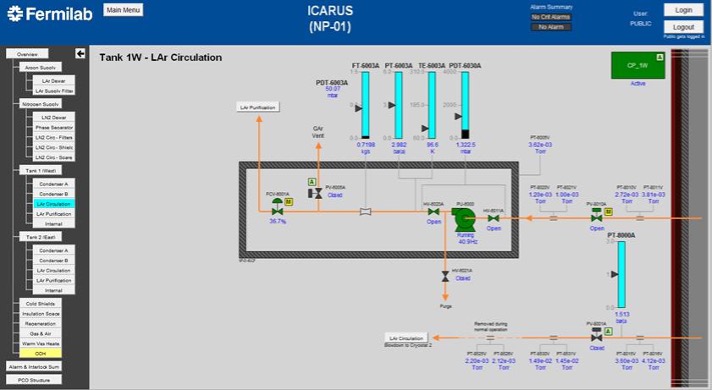}
  \caption{Level 3 Display – Tank 1W LAr Circulation}
  \label{fig:sec6-48}
\end{figure}

\begin{figure}[htbp]
  \centering
  \includegraphics[width=0.90\textwidth]{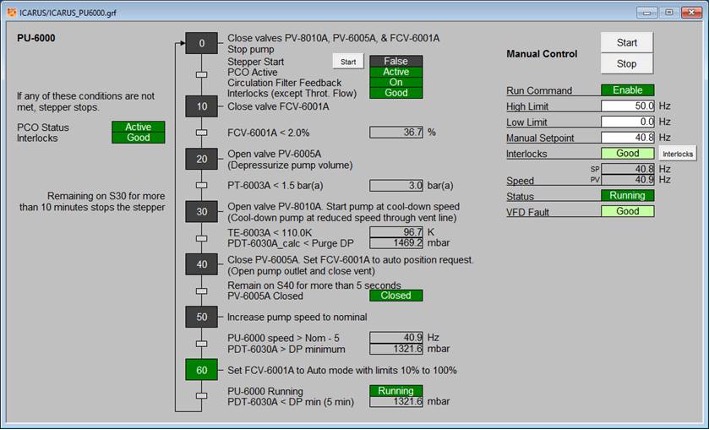}
  \caption{Level 4 Display – Pump PU-6000 Stepper}
  \label{fig:sec6-49}
\end{figure}

\subsubsection{Alarm Management}

\begin{figure}[htbp]
  \centering
  \includegraphics[width=0.75\textwidth]{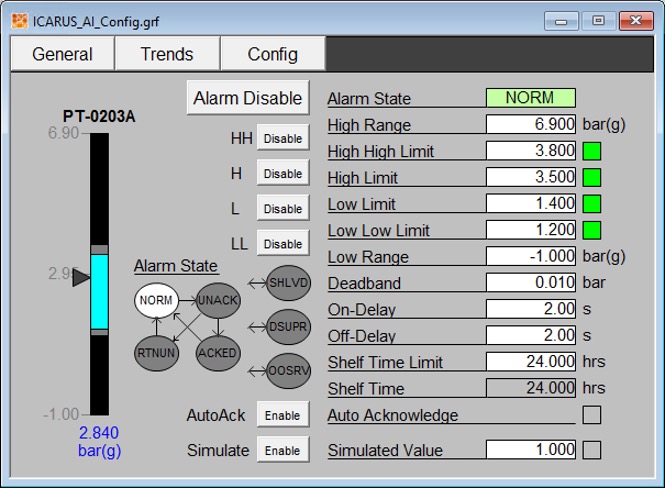}
  \caption{Example Alarm Configuration Display}
  \label{fig:sec6-50}
\end{figure}

Alarm logic is in the PLC, but limits are set, and alarms are acknowledged through the HMI. Alarms
incorporate best practices from ANSI/ISA-18.2-2016, Management of Alarm Systems for Process
Industry. This includes such features as alarms states, dead band, delays, shelving, and auto
acknowledgement (Figure~\ref{fig:sec6-50}).

Alarms are designated as critical or non-critical based on their severity.

Outside of the HMI, remote notification of alarms is implemented by an auto-dialer called WIN-911.
Critical alarms prompt immediate notification at all times, while non-critical alarms only prompt
notification during business hours. (Notifications of non-critical alarms occurring in off hours are sent
when business hours begin.)

One person is designated as the cryo coordinator and holds the cryo coordinator smart phone. Alarm
notifications are sent via text to the cryo coordinator phone as well as by email to a short LISTSERV list
of experts.

Additionally, critical alarms trigger the SBN-FD Cryo Alarm to Fermilab Dispatch via the FACILITY
INCIDENT REPORTING UTILITY SYSTEM (FIRUS) system. They have a prioritized list of system experts to
call.

\subsubsection{Historian (data logging and plotting)}

Historical data collection is accomplished by iHistorian by GE Digital. This system stores data, typically
on a one second interval, to a virtual machine managed for Fermilab Core Computing.

Recent (up to 1 day) historical data pertaining to individual devices is available in the popup window
for the device in iFIX. All historical data, as well as tools for data analytics, are made available via
Industrial Plotting and Analytics Workbench software by Seeq.

\subsubsection{Programmable Logic Controllers}

The cryogenic process control system includes Siemens and Beckhoff PLCs (ODH safety is implemented
separately, see Appendix~\ref{app:ProcessSafety}). Direct control of all cryogenic equipment is done exclusively via PLC. PLCs
contain all control logic necessary to operate all connected equipment. Under normal conditions,
PLCs operate the cryogenic systems autonomously (without input from SCADA).

The engineering station contains the PLC configuration and programming tools. The engineering station
contains the PLC programs (control logic) for all PLCs. Note, the programs do not run on the engineering
station itself.

\begin{figure}[htbp]
  \centering
  \includegraphics[width=0.7\textwidth]{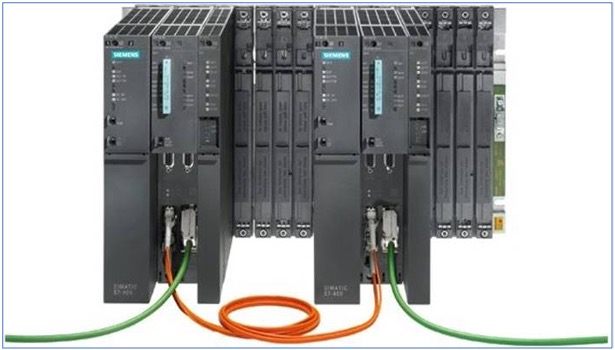}
  \caption{Redundant Siemens S7-400 PLCs}
  \label{fig:sec6-51}
\end{figure}

\begin{figure}[htbp]
  \centering
  \includegraphics[width=0.80\textwidth]{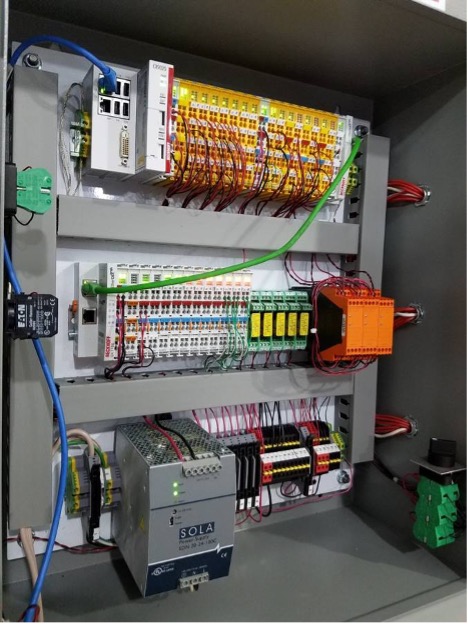}
  \caption{ODH Control System Enclosure}
  \label{fig:sec6-52}
\end{figure}

The main PLCs for the cryogenic control system are a redundant pair Siemens S7-400 PLCs (Figure~\ref{fig:sec6-51}). The internal
cryogenics PLC is a Beckhoff CX590. The ODH system uses a Beckhoff CX990 for standard (non-safe)
operations and an EL6910 safety PLC (Figure~\ref{fig:sec6-52}).

\subsubsection{Field Inputs/Outputs}

The cryogenic system includes numerous devices of many types. This section addresses the I/O
connections to the most common devices, those with several instances throughout the system.

In general, devices are powered by 24V DC from the corresponding I/O terminal. Discrete (on/off)
signals use 24V logic. Most analog signals are 4–20 mA.

\begin{itemize}
\item Pressure control valves (PCVs) and flow control valves (FCVs) are the same type of device, a valve
with a pneumatic positioner. The distinction between PCV and FCV is based on the valve’s role in
the system under normal operation. The valve positioner is provided a single 4–20 mA analog signal
from the PLC. There is no readback to the PLC.

\item Pneumatic valves (PV) are on/off devices actuated by a pneumatic solenoid. The solenoid is
provided a single 24V discrete signal from the PLC. PVs are outfitted with two hall effect sensors for
position readback, one for the open position and the other for closed position. These provide a 24V
discrete signal back to the PLC.

\item Hand valves (HV) are on/off devices actuated manually. Certain HVs, namely those where large
volumes of cryogens flow (e.g. before and after a pump), are outfitted with one or two hall effect
sensors for position readback. They are the same as those for pneumatic valves and provide a 24V
discrete signal back to the PLC.

\item Pressure transmitters (PTs) and pressure differential transmitters (PDTs) provide an absolute
pressure reading or pressure relative to another pressure. These devices provide a single 4–20 mA
analog signal to the PLC.

\item Vacuum pressure transmitters measure the absolute pressure in vacuum spaces. These devices
provide a 0–10V analog signal to the PLC. Additionally, the vacuum pressure transmitters on the side
penetrations of the modules provide a 24V discrete signal to the ODH safety PLC. The additional
signal is for discrete indication of vacuum loss for use in the safety PLC’s discrete only logic.

\item Resistance temperature detectors (RTDs) are essentially resistors that change value based on
temperature. The control system includes RTD terminals made specifically to read them.

\item Variable frequency drives (VFD) provide power to the liquid Argon and Nitrogen pumps. The control
system makes use of one analog input, one analog output, two discrete inputs, and two discrete
outputs per VFD. The PLC provides a speed command via 4–20 mA analog signal to the VFD, which
provides a speed readback signal back to the PLC. The PLC provides a run/stop command 24V
discrete signal to the VFD, which returns a run/stop status back to the PLC. The VFD provides an
internal fault 24V discrete signal to the PLC for which the PLC has a fault reset signal back to the
VFD.
\end{itemize}

\begin{figure}[htbp]
  \centering
  \includegraphics[width=0.85\textwidth]{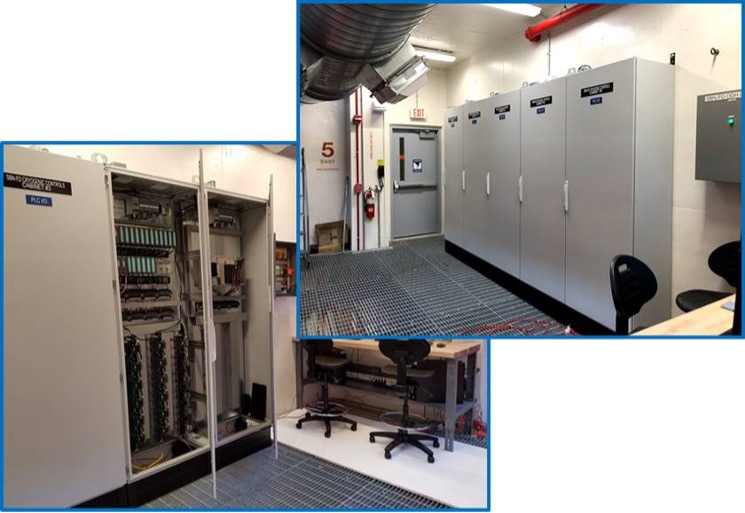}
  \caption{Cryogenic Control Cabinets}
  \label{fig:sec6-53}
\end{figure}

The total number of I/O for main, internal and ODH control subsystems include: 222 analog inputs,
176 RTDs, 40 potentiometers, 94 analog outputs, 184 digital inputs, 124 digital outputs. Provision is
made for system evolution  (Figure~\ref{fig:sec6-53}).

\subsubsection{Backup Power}

The cryogenic control system makes use of the generator and UPS (uninterruptable power supply)
provided by FESS (Facilities Engineering Services Section). The PLCs, I/O, and VFDs for LN$_2$ pumps are
included, as they are necessary for cold shield and condenser operation to maintain modules pressure
and minimize loss of LAr. The VFDs for LAr pumps and heaters are not included.

\subsection{The software architecture}

The software architecture is based on the Functional Analysis (FA) of the cryogenic system, on the
subsequent Product Break-down Structure (PBS), associated Piping and Instrumentation Diagram
(P\&ID) and parts-list (detailing the characteristics of the instrumentation), and on the process logic
specification. The logic specification used as basis for the programming is designed with a modular
structure relying on a master-child control units hierarchy, on associated option modes, sequencers
and interlocks.

\subsubsection{Functional breakdown structure and control units hierarchy}

The functional analysis of the ICARUS cryogenic system was carried out following the ICARUS
requirements described in Section 1.4, in order to define a set of main service functions, technical
functions and constraints. The detailed functional breakdown structure is summarized in Table~\ref{tab:sec6-functions}.

\begin{table}[htbp]
\centering
\caption{Main service and constraint functions of the ICARUS cryogenic system}
\label{tab:sec6-functions}
\begin{tabular}{p{0.06\textwidth} p{0.28\textwidth} p{0.56\textwidth}}
\hline
\textbf{N\textsuperscript{o}} & \textbf{Name} & \textbf{Description} \\
\hline
S2 & To have a stable gas phase density &
To have a stable gas phase density \\
S3 & To control LAr level &
To have a liquid level and level stability according to detector needs \\
S4 & To have a homogeneous temperature &
To have a homogeneous temperature across the detector volume \\
S5 & To prevent gas bubbles &
To prevent argon gas bubble formation in the detector volume \\
S7 & To provide services to support main service functions &
Fluids storage, distribution, purification, regeneration, gas analysis \\
C1 & To guarantee cryostat pressure scale &
To limit argon gas pressure; to avoid opening of magnetic safety devices; to avoid atmospheric
under-pressure (ingress of air) \\
C7 & To cool-down &
To cool down the apparatus before the start of the filling in 5 days while maintaining the prescribed
maximum pressure gradients on the wire chambers \\
\hline
\end{tabular}
\end{table}

The software architecture relies on a ``PCO'' breakdown structure, which itself is derived from the
Functional breakdown structure. The PCO breakdown structure consists in a master-child control units
hierarchy, where the acronym PCO stands for Process Control Object and is borrowed from the CERN
UNICOS CPC terminology. A PCO is a control unit with reference to IEC~61512-1 standard which
represents a collection of field objects (actuators and/or other units) performing a common function or
assembly of functions. For instance, a pump unit is the assembly of a pump, its inlet and outlet isolation
valves, one vent valve and one charge valve, whose function is to circulate the cryogenic liquid. Most
of the time the lower level PCOs correspond to the individual PBS equipment fulfilling the individual
technical functions identified in the functional breakdown structure.

\subsubsection{Option Modes}

Some PCOs have associated option modes defined so that these units can be setup in different
operational states, i.e. in different working modes of the process. For instance, the LN2\_CRY\_SHLD PCO,
which ensures the circulation of liquid nitrogen to the cryostat shield, has 3 associated modes. “With
Gravity” option mode corresponds to the liquid nitrogen being supplied by gravity from the liquid
nitrogen storage tank during cool-down or whenever the nitrogen pump is stopped. “With pump”
option mode corresponds to the liquid nitrogen being circulated by the main pump. “With spare”
option mode corresponds to the liquid nitrogen being circulated by the spare pump in case the main
pump is unavailable. The option modes consist in user commands activated from the HMI. They are
used as a condition either to switch on a set of dependent PCOs fulfilling the required functions at a
given stage of the cryogenic system operation or to set a collection of defined states of the controlled
objects, e.g. open or closed for a pneumatic valve, fixed position or regulation mode for a control
valve, initialization of set-point.

\subsubsection{Sequential Function Charts}

\begin{figure}[htbp]
  \centering
  \includegraphics[width=0.82\textwidth]{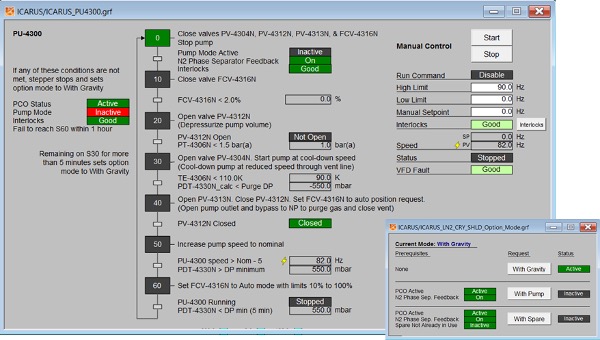}
  \caption{LN2 pump stepper}
  \label{fig:sec6-54}
\end{figure}

The programming is mostly object oriented, while using Sequential Function Charts (SFC) whenever the
process requires a sequence of operations like for the start-up of the cryogenic pumps for instance
(See Figure~\ref{fig:sec6-54}). These SFCs/Steppers are used to automatize the sequence of steps / transitions of a
given process, while the same actions can be alternatively achieved following a manual procedure
limited by interlocks.

\subsubsection{Software interlocks}

Software interlocks allow an actuator or a unit to go to its fail-safe position or they prevent it from
starting in case of abnormal behavior and/or for security reasons. They override any automatic or
manual request. Three types of software interlocks are used in the ICARUS cryogenics control system:
full stops, temporary stops, and start interlocks:

\begin{itemize}[leftmargin=1.5em, labelsep=0.5em, itemsep=0pt, topsep=0pt, parsep=0pt]
    \item Full stops stop the unit/actuator and wait for manual acknowledgement before restarting.
  For instance, in case any ODH / bad vacuum condition is detected the side penetration
  valves and argon pumps are interlocked with a full stop.
     \item Temporary stops stop the unit/actuator and restart automatically when the interlock
  disappears.
    \item Start interlocks prevent the unit from starting (all dependent units/actuators stay in their
  fail-safe position).
\end{itemize}  
  
As described earlier, the interlocks thresholds are set on the HH/LL limits of the corresponding process
variable and can be enabled/disabled from the HMI by expert users.

\subsubsection{Alarming}

Full stops usually (but not systematically) trigger a critical alarm as defined in earlier sections.
Temporary stops applied to the pressure vent valve are also triggering a critical alarm while the other
temporary stops and the start interlocks usually trigger a non-critical alarm or no alarm at all.

Critical alarms are preferably set on low-low or high-high (LL / HH) limits of the corresponding process
variable HMI panel while non-critical alarms are preferably set on low or high (L / H) limits of the
corresponding process variable HMI panel.

\subsubsection{Selected software and programming principles}

If CERN Unified Industrial Control System (UNICOS) partially inspired the HMI development and if its
terminology was sometimes used for the logic specifications, the PLC programming itself does not make
use of, nor reproduces, the UNICOS framework and tooling but uses the Siemens S7 embedded
environment.

The logic specification and programming were achieved and implemented, sequentially, in batches of
option modes and PCOs. This allowed to check-out the above described core functionalities on small
size parts of the program and to spread the eventual adjustments to the other parts of the program.

Substantial programming development and testing efforts have been required to evolve from a control
strategy commonly relying on sets of valve controllers and protection interlocks coupled to manual
actions towards the chosen automated control strategy. Implementing and validating the core
functionalities has been a long iterative process. This development required systematic testing which
led to a smooth cryogenic commissioning phase.

\subsection{Safe and continuous cryogenics operation}

The cryogenic control system includes several features in order to ensure safe and continuous
cryogenics operation, maintainability and flexibility for system evolution.

\subsubsection{Redundancies}

\begin{itemize}[leftmargin=1.5em, labelsep=0.5em, itemsep=0pt, topsep=0pt, parsep=0pt]
    \item All equipment essential to the functioning of the system, like PLCs, I/O modules, LN2 cryogenic
pumps, controllers for vacuum sensors, Ethernet router for the SCADA, OPC Server and EWS are
powered from redundant electrical power supplies including diesel generators and
Uninterruptable Power Supplies (UPS). The cryo system is feed from a combination of building
power/emergency power. Should the building power disappear, the UPS will handle it for a few
seconds, when the generator will kick in. In case of generator failure, the control system may
run on UPS (providing loads are reduced) for several hours.
    \item S7 Siemens PLC have 2 separate CPU’s that are running the same program, which is
synchronized as hot backup through a fiber link. Therefore, a failure of any CPU unit will not
affect the controls, as a standby unit will take control within a few milliseconds. Furthermore,
Interface modules (IM) are connected to different PLC CPU’s, but might cross communicate with
any of 2 CPUs, providing stable control over the connected modules. Signal modules are
connected to the channels through MTA (Marshall Terminal Assemblies), that are powered by
redundant power sources.
    \item All control cabinets have a redundant 24V power. Some have just 2 redundant supplies tied to
one redundancy module; some have 3 power supplies tied to 2 redundancy modules. All power
supply statuses are monitored, and an alarm generated in case of failure. In any case, 24V power
is distributed to a bank of fuses, and a failure of any power supply will not affect 24V distribution.
MTA assemblies are powered from 2 independent fuses, so a single fuse trip will not disturb the
operation.
\end{itemize}

\subsubsection{Safety aspects}

- The vacuum levels of transfer-lines between the side penetrations and the argon pumps are
monitored. Since a degraded vacuum could be a sign of argon potential leak and further ODH risk
created by the accumulation of argon in the lowest locations, such a situation triggers a critical
alarm, the evacuation of the area, as well as the increase of the air extraction speed. At the same
time, interlocks to the cryogenics system are triggered stopping the pump and argon circulation.

- The cryostat modules over-pressure protection is implemented in controls in addition to over-
pressure protection by magnetic valve reliefs. The controls are set for two vent valves per module
to regulate at 150 mbarg and one larger vent valve at 175 mbarg. In the exceptional circumstances
the ultimate high pressure threshold of 200 mbarg is reached, this vent valve is fast opened thanks
to a dedicated electro valve placed on the vent valve actuator supply (hardware interlock based on
redundant pressure switches).

\subsection{Maintainability and flexibility for system evolution}

It is possible to make change to PLC programming online, provided that current set points are backed
up, with no disruptions to the experiment. In certain cases, though, interaction of process, controls and
cryo engineers is needed. It’s also possible to add hardware as needed.

\FloatBarrier

\section{Cryogenic and purification commissioning and operations}
\label{sec:CryoComm}

\subsection{Cryogenic commissioning}

Commissioning and operations of the ICARUS cryogenic system have been divided to 5 different modes
(following the months-long evacuation of the modules to remove atmospheric contamination):
\begin{enumerate}
\item Preparation for cooldown, including activation of the different purifiers, lowering of the
outgassing of the two cold vessels, purging of the different transfer lines and cold boxes, etc.
\item Cool-down of the two cold vessels and the detectors mounted in these vessels.
\item Filling of the two cold vessels by a total of about 400 ton of purified liquid argon per vessel.
\item Stabilization of the filled cold vessels, bringing the cold vessels pressure to stable values, while
lowering the impurity level in the two argon baths.
\item Normal operation, purifying the argon in the two cold vessels in gaseous and liquid phase,
guaranteeing a stable, low, impurity level of the argon in the two cold vessels.
\end{enumerate}

In this sense, the commissioning of the Icarus cryogenic system had a staged approach as shown 
in Appendix~\ref{app:CommStages}. Each stage was preceded by verification of safety and receiving clearance for operations
as per Fermilab safety requirements. The cryogenic commissioning was conducted as a common effort of
the CERN and Fermilab cryogenic teams. The commissioning was started with the CERN team onsite
and then continued with the CERN team supporting remotely the activity of the local Fermilab team.
Meetings have been held almost daily, including weekends, since the beginning of March 2020.

One of the most important requirements for a liquid argon based TPC detector is to allow the
ionization tracks, created by interacting particles within the detector volume, to be transported in
a uniform electric field up to a multi-wire anodic structure placed at the end of the drift path with
reasonably low attenuation of charge $A$. The attenuation of charge $A$ is defined based on the design
drift time $T$ and the achieved free electron lifetime $\tau$ per Eq.~(\ref{eq:attenuation}):

\begin{equation}
A[\%] = 100 \left( 1 - e^{-T/\tau} \right)
\label{eq:attenuation}
\end{equation}

The free electron lifetime in argon is directly related to the electronegative impurity concentration
$\rho$, mostly oxygen, by an inverse linear relationship shown below in Eq.~(\ref{eq:lifetime}):

\begin{equation}
\tau\,[{\rm ms}] \approx
\frac{300~{\rm ms \cdot ppt\ O_2\ equivalent}}{\rho\ [{\rm ppt\ O_2\ equivalent}]}
\label{eq:lifetime}
\end{equation}

Therefore, for the free electron lifetime to exceed design 3~ms with attenuation of 28.3\% for the drift
time of $T = 1$~ms, the impurity ($\rho$) measured in parts of oxygen should be below 100~ppt ($10^{-10}$ parts).

Reaching the same 28.3\% attenuation of charge for the DUNE cryostat to be hosted at the Long-Baseline
Neutrino Facility (B.\ Abi et al., 2020) with 2.5~ms design drift time would already require the lifetime of
7.5~ms, more than twice better than the minimum design lifetime for the Icarus. The original Icarus
installation in Gran Sasso achieved the electron lifetime in excess of 10~ms. With the same design drift
time $T$ between the cathode and the wire chamber being 1~ms, that resulted in attenuation of charge
of 9.5\%.

There are many factors that impact the purity and resulting lifetime for a TPC-based liquid argon
detector, like Icarus. The initial purity depends on the purity of the liquid argon filled into the modules.
As commercial argon has impurities far exceeding 100~ppt, and since the components inside the
modules and piping outgas impurities, the Icarus cryogenic system was designed to remove H$_2$O and O$_2$
from LAr to achieve impurity concentrations below ppb level before TPC could be turned on. In Icarus,
this was achieved by first pumping the empty modules to high vacuum for 9 months to remove a bulk
of contamination from surfaces by desorption. In addition, an elaborate filtration system had to be
installed and prepared~\cite{Ref19} to remove the remaining impurities, mostly from the initial fill and
continuing outgassing of impurities from the materials inside the modules and recapturing them into
liquid argon bath during operations.

Initial activation of the copper in 14 removable cartridges (4 of 5-liter units for gas re-condensers and
10 of 25-liter units for liquid filters) was done in summer 2019. The process was done using the existing
cryogenic system at MicroBooNE. The activation of Cu~0226 was done by initial heating to $\sim$420~K with
pure argon and then introducing mixture of 2.5\%H$_2$--97.5\%Ar to achieve removal of oxygen. The
success of activation process is judged by heating of copper due to exothermic process forming of H$_2$O
from O$_2$ and H$_2$. As the small quantities of copper in the cartridges did not produce significant heat of
exothermic reaction and registered only poorly pronounced temperature spikes during activation, we
measured the absorption of oxygen by activated copper to confirm the success, and then repeated the
activation/regeneration procedure.

\begin{figure}[htbp]
  \centering
  \includegraphics[width=0.90\textwidth]{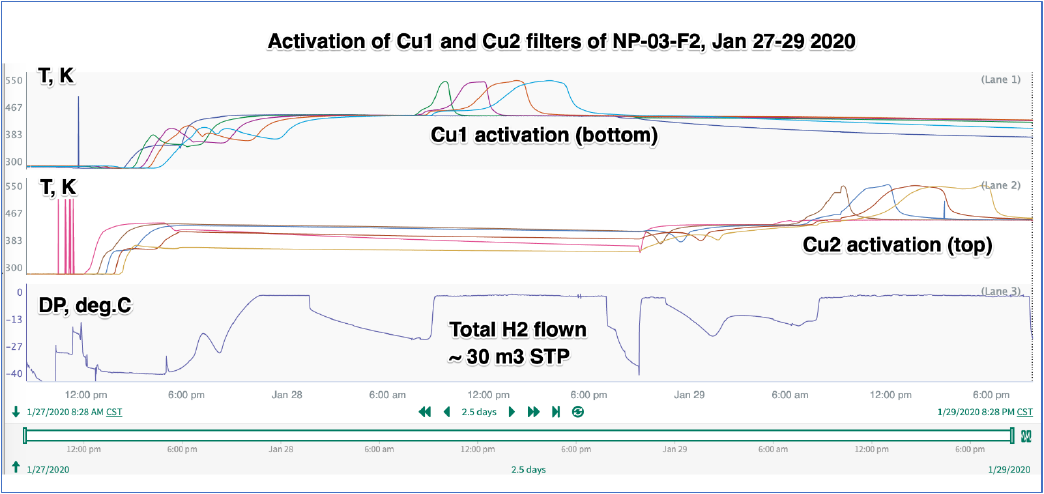}
  \caption{Initial activation of Cu in filter 1 and 2 of the fill purifier}
  \label{fig:sec8-55}
\end{figure}

\begin{figure}[htbp]
  \centering
  \includegraphics[width=0.90\textwidth]{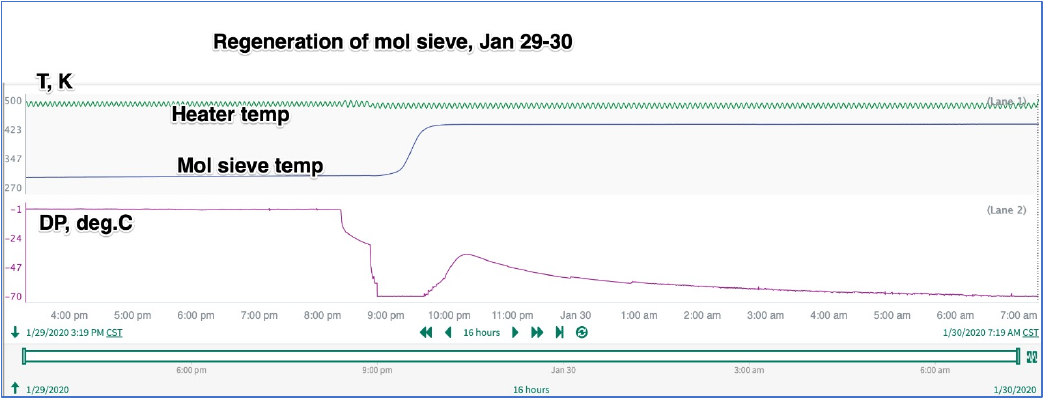}
  \caption{Initial drying of molecular sieve of fill purifier}
  \label{fig:sec8-56}
\end{figure}

Then the activation of the fill purifier filters was done within 3 days starting January 27, 2020. The
purifier contains three vessels, two containing 300~kg of Cu~0226, and one containing 80~kg of molecular
sieve. The activation of copper was done again by using a hydrogen gas mixture to reduce the copper
oxide of the media. Though such activation is typically done at Fermilab using mixture of 2.5\%H$_2$ with
97.5\%Ar, this activation was done with 1.0\%H$_2$ with 99.0\%Ar to replicate the process done at CERN for
NP02/04 cryogenic systems. The procedure involved first heating the media with $\sim$140~m$^3$/hr of pure
argon heated in 10~kW heater to reach $\sim$440~K in the bulk of the copper. Then the standard 2.5\%–
97.5\%Ar mix from the tube trailer was mixed with pure argon to dilute to 1\%H$_2$ concentration and
flown to individual filters Cu1 and Cu2 at the rate of $\sim$140~m$^3$/hr (see Figure~\ref{fig:sec8-55}). Here Fermilab tried,
again, to replicate successful activation of identical filters at CERN where they used $\sim$150~m$^3$/hr rate
for 10~hrs at 1\%H$_2$, thus requiring $\sim$15~m$^3$ H$_2$. The activation process took $\sim$12~hrs per each 300~kg of
copper with temperature peaks up to 547~K and dew point in the exhaust gas spoiled from $-40^\circ$C to
$-2^\circ$C ($\sim$5200~ppm). The drying of molecular sieve was then done next (see Figure~\ref{fig:sec8-56}). At the end of
activation and drying process, the 600~kg of copper had an absorption capacity of $\sim$388~g of O$_2$ while molecular
sieve had a capacity of $\sim$23~kg of H$_2$O.

\begin{figure}[htbp]
  \centering
  \includegraphics[width=0.90\textwidth]{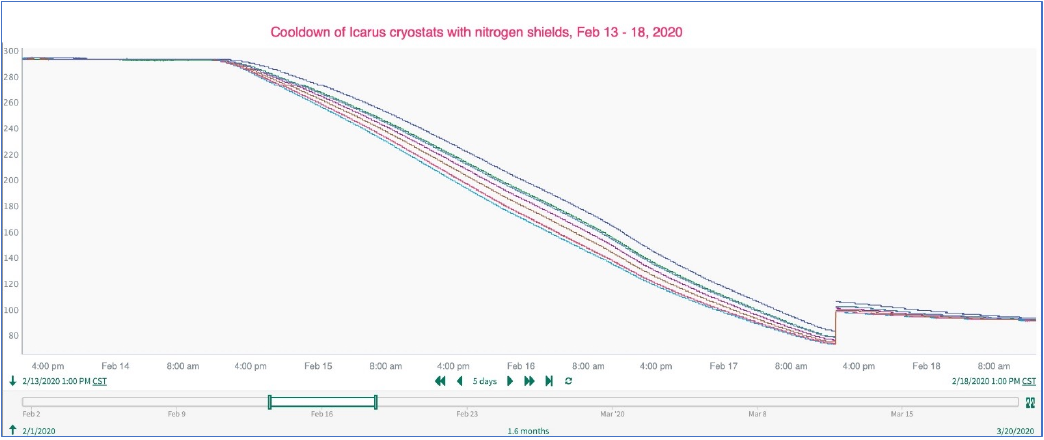}
  \caption{Cooldown of cryostat with LN$_2$ shields. The temperature jump towards the end of the curves is due to a change
in one of the readout algorithm parameters.}
  \label{fig:sec8-57}
\end{figure}

In summary, the cryogenic commissioning of the ICARUS detectors started on February 13, 2020 by
breaking the vacuum in the two main cold vessels with ultra-purified argon gas. Cooldown started on
February 14, 2020 by injecting liquid nitrogen in the cold shields by means of pressure and gravity
transfer from LN$_2$ dewar via phase separator. It took about 4 days to bring the temperature on the wire
chamber below 100~K. The cooling process was continuous and the maximum temperature gradient on
the wire chambers was about 35~K (see Figure~\ref{fig:sec8-57}). The LN$_2$ pumps were cooled down and started
immediately after gravity transfer; it took several days to stabilize their operations due to failure of the
bypass regulation valve for the LN$_2$ pump dedicated to the shields. The repairs proved to be
unsuccessful, and the pump was left to operate with fixed 10\% opening of the bypass valve.

On February 19, 2020, the gas recirculation condenser units were put into operation to purify the argon
gas before the start of the liquid filling, but their operations were not stable without sufficient liquid
level in the modules and therefore their operations were delayed until the end of the fill.

\begin{figure}[htbp]
  \centering
  \includegraphics[width=1.00\textwidth]{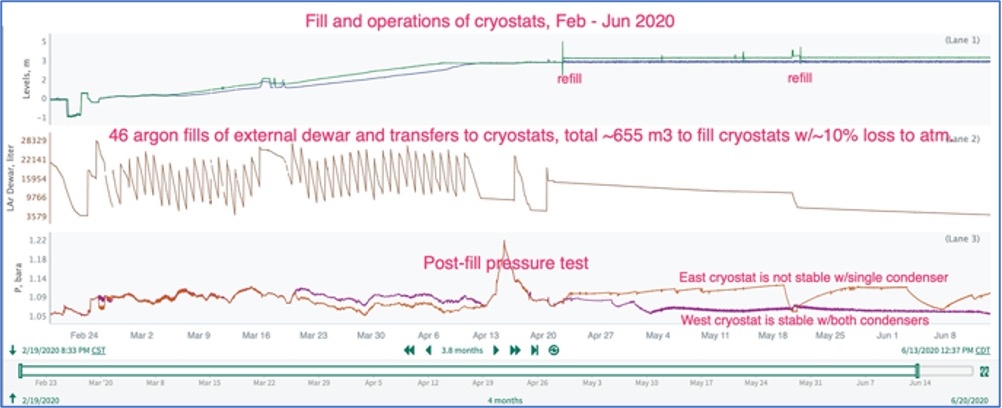}
  \caption{Regeneration of fill purifier filters during fill campaign}
  \label{fig:sec8-58}
\end{figure}

\begin{figure}[htbp]
  \centering
  \includegraphics[width=1.00\textwidth]{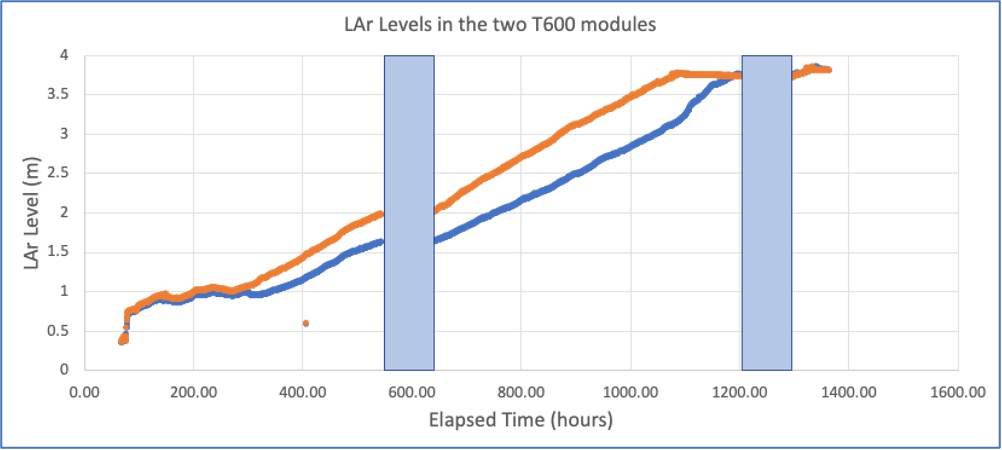}
  \caption{LAr fill campaign -- daily deliveries and purification periods}
  \label{fig:sec8-59}
\end{figure}

The filling with LAr started on February 19, 2020, by initiating a total of 46 daily deliveries by Air
Products. Once the transfer lines and liquid purifier filters were cooled down, the transfer of ultra-
purified liquid argon from LAr dewar via fill purifier to the modules ran 24/7 from February 24 until April
20 with interruption at midpoint on March 16--20 for the regeneration of fill purifier filters
(see Figure~\ref{fig:sec8-58} and Figure~\ref{fig:sec8-59}). The filling was stopped again when the liquid reached the
$-6$~cm below the nominal level to perform the final pressurization qualification test of the two cold
vessels. During the fill, the boiloff gas of $\sim 10\%$ was vented to atmosphere to control the pressure in the
modules. The impurities downstream of the fill filter were measured during the fill campaign to be in
ppb range.

On April 21, 2020, the liquid recirculation was started. The recirculation path for the West module took
LAr from the top side penetration, but the recirculation for the East module had to take LAr from the
bottom side penetration in order to provide stable supply to the suction of the argon pump. That was
due to the combination of length and gooseneck shape of the supply piping penetrating the insulation
space from the East module. While both modules have identical top and bottom side penetrations, the
piping for the East module is longer than the piping for the West module. The recirculation rates are
1.85~m$^3$/hr in the West module and 2.25~m$^3$/hr in the East module.

A general cryogenic stabilization phase was completed around the end of May 2020. That was mostly
related to achieving desired level of ullage pressure in the modules at $\sim 1.070$~bara by
(a) adjusting the temperature of the nitrogen in the shields and  
(b) setting the gas purifier condenser units at their maximum capacity (adjusted for module pressure).  
The nitrogen shields pressures were varied between 2.85 and 2.95~bara with downstream control valves
to adjust pressure of nitrogen from 4~bara supply from the pump to 2.2~bara in the phase separator.

After the initial fill and during the operations, most impurities are coming to the argon gas phase in the
ullage from outgassing of materials (mostly cables) and from micro-leaks possibly developed in some
of the 48 feedthrough flanges per module, each having 1152 wires, located at room temperature on
top of the detector. To prevent these impurities from diffusing into the main LAr volumes of the
modules, the Icarus filtration system is provided with gas recirculation condenser-filter units each
containing an Ar/N$_2$ heat exchanger and a single 5-liter cartridge with activated Cu. To keep the modules
above atmospheric pressure and prevent air back-streaming from leaks, these condenser-filter units,
two per module, work by condensing gaseous argon with liquid nitrogen. The Cu filter media was
activated prior to the start of Icarus commissioning.

Finally, during the operations, the Icarus cryogenic purification system provides circulation of liquid
argon at $\sim 0.8$–1.0~kg/s per module with dedicated cryogenic pumps through the liquid argon filter.
Each filter unit is dedicated to the module and contains five 25-liter cartridges filled with molecular
sieve and Cu in 25\%/75\% fraction. The circulation rate is defined by the capacity of the largest liquid
argon pumps (models BNCP-32C and BNCP-32E) available from Barber-Nichols at the time of the Icarus
design. This flow rate defines the average time $T_r \approx 104$~hours to recirculate the full volume of
each module.

\begin{figure}[htbp]
  \centering
  \includegraphics[width=0.70\textwidth]{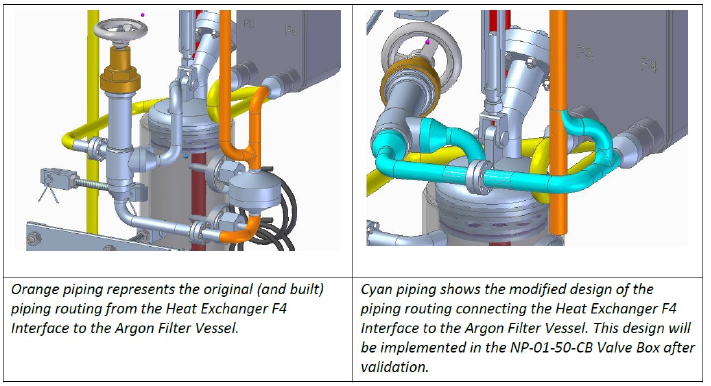}
  \caption{Modification of re-condensers}
  \label{fig:sec8-60}
\end{figure}

One of the issues that resulted in lower-than-expected initial purity in both Icarus modules was due to
the difficulties in commissioning the gas condenser-filter units responsible for filtering the impurities
from the gaseous ullage space of the module. At least two of the units kept developing a gas vapor lock
in their internal argon piping, thus preventing operation at full design 6.5~g/s per condenser-filter unit.
The vapor locking was due to a so-called “gooseneck’’ in the cryogenic piping where the fluid was at
conditions close to saturation. The internal piping for all four condenser-filter units was modified in
Spring 2021 to remove the gooseneck, which improved the capacity up to 7~g/s (see Figure~\ref{fig:sec8-60}).

At the start of the cryogenic commissioning, all activities in the detector building not related to
cryogenics were suspended and the building was put in a higher safety condition, with strong
limitations on the presence of people onsite. At the end of the liquid argon filling, after the final pressure
test, the standard ODH safety conditions were restored and regular activities on top of the detector
could be re-started.

The gas recirculation rate through the gas condenser-filter units must match the boiloff rate of argon
in the modules in order to efficiently remove impurities and to properly control the pressure in the
ullage and have enough flexibility to operate with wide margin. Unlike a passively insulated cryogenic
vessel, the boiloff rate for the Icarus modules is defined by the temperature of the ten actively operated
LN$_2$ cooling circuits inserted between the warm and cold vessels. The pressure in, and thus temperature
of, each of the individual circuits is regulated by a valve placed in each of the ten circuits, with the
liquid/gaseous mixture returning to the nitrogen phase separator. The nitrogen system includes two
liquid nitrogen pumps (“filters’’ and “shields’’ pumps, BNCP-68 models) installed in vacuum-shielded
valve boxes.

Unfortunately, one of the control valves maintaining the regime of the “shields’’ pump failed stuck in
mid-position after the fill of the modules, preventing reliable shifting of the pump's operating point on
its curve and reaching the desirable shields temperature and therefore maximum boiloff and removal
of impurities from the ullage. While both nitrogen pump units could be replaced with a separate spare
pump unit, the control valves of each pump were designed to be shared and in service at all times,
preventing their maintenance or replacement without losing nitrogen circulation. With insufficient
boiloff and gas recirculation, and in order to increase bulk purity of liquid argon, periodic venting of
gaseous argon from ullage was started in late 2020. The automatic venting has been exercised up to
four times a day, resulting in loss of $\sim 12$~L/day or 1~cm of liquid level per module every two months.
This required topping the modules in March 2021, August 2022 and December 2023. Though the refills
were less than 2\% of the total module volume and were transferred via regenerated fill purifier, they
still introduced additional contamination and therefore additional loading of the gas and liquid filters
with impurities.

The purity of the argon in East and West modules was the goal of investigations since completion of
commissioning in 2020. As the purity depends on many contributing factors—initial purity after fill,
ongoing outgassing from materials inside the modules, recirculation rate of liquid argon, recirculation
and condensing rate of gaseous argon from the ullage, etc.—these investigations focused on optimizing
conditions to achieve the design lifetime of 3~ms.

\begin{figure}[htbp]
  \centering
  \includegraphics[width=0.80\textwidth]{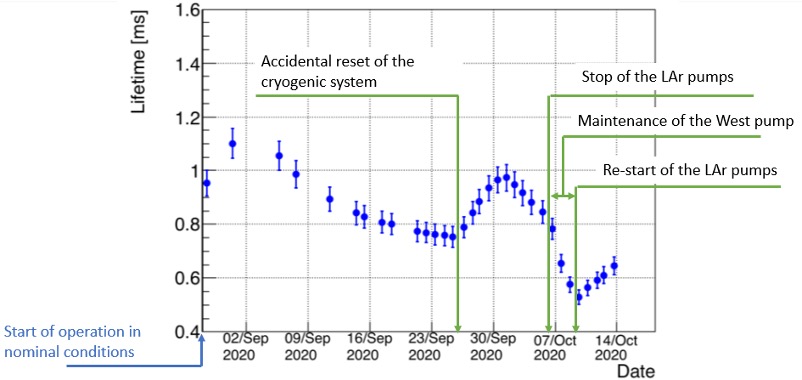}
  \caption{Investigation of purity dependence on operations}
  \label{fig:sec8-61}
\end{figure}

\begin{figure}[htbp]
  \centering
  \includegraphics[width=0.80\textwidth]{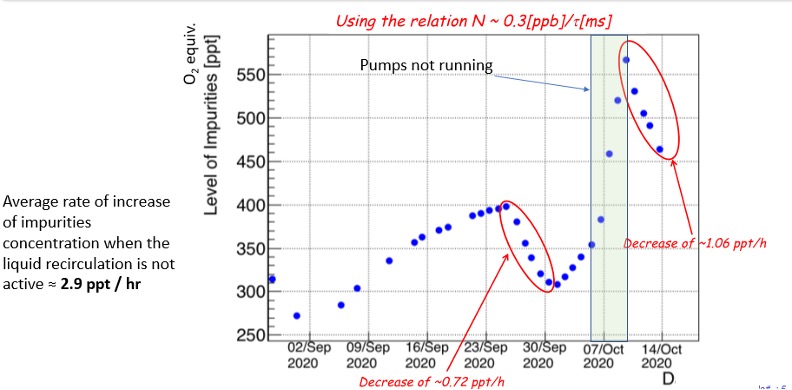}
  \caption{Investigation of purity dependence on operations}
  \label{fig:sec8-62}
\end{figure}

\begin{figure}[htbp]
  \centering
  \includegraphics[width=0.80\textwidth]{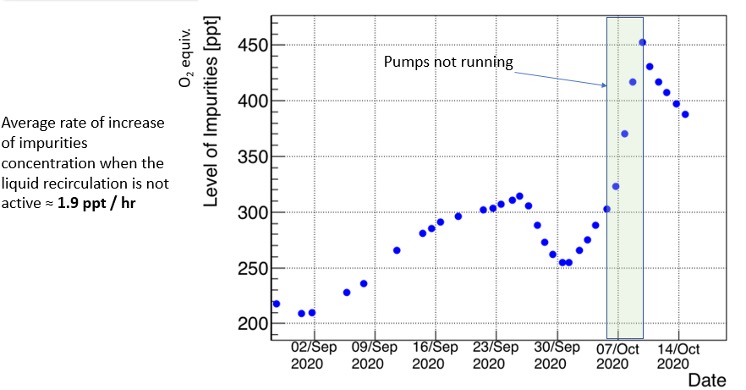}
  \caption{Investigation of purity dependence on operations}
  \label{fig:sec8-63}
\end{figure}

Starting in Sep–Oct 2020, the studies established how much the bulk purity depends on recirculation
rates in the West module (see Figures~\ref{fig:sec8-61}, \ref{fig:sec8-62}, \ref{fig:sec8-63}).

Liquid recirculation studies show that the expected asymptotic impurities concentration $C$ is given by
the equilibrium condition:
\[
0 = \frac{dN}{dt} = L - C S \;\Rightarrow\; C = \frac{L}{S}
\]
where $L$ is the rate of impurities entering the liquid (leak rate) and $S$ is the nominal recirculation speed
($\approx 2.35$~m$^3$/hr for the East module and $\approx 2.09$~m$^3$/hr for the West one). A 100\% filtering capacity
of the recirculated liquid is assumed. The leak rate $L$ can be measured from the observed decrease of
free-electron lifetime when the liquid recirculation pump is off. From this measurement, the calculated
asymptotic values are $\approx 320$~ppt for the East module and 236~ppt for the West one (corresponding
lifetimes: 0.94~ms and 1.27~ms). The fact that the calculated values are very similar to those measured
immediately before stopping the pumps (356~ppt and 300~ppt) tells us that purity is not limited by a
lower‐than‐expected recirculation speed or purification capacity.

The nominal adsorption capacity of each filter cartridge of the gas recirculation units is about 1.5~g of
O$_2$. The influx of impurities from gas to liquid can be calculated by measuring the drop of lifetime in
the argon bath when the circulation pump is off for a specific time period while impurities are still being
removed from the ullage via gas recirculation. This influx was determined to be $\sigma \approx 0.40$~ppt/hr
for the original Icarus at Gran Sasso. The value was measured to be $\sim 2$–3~ppt/hr for Icarus at Fermilab
at the start of commissioning and stabilized at $\sim 0.55$~ppt/hr after one year of operations when the
initial outgassing influx subsided and gas recirculation filtration improved.

The theoretical lifetime $\tau$ achieved in liquid argon at time $t$ can be calculated using the following equation, based on
$\sigma$ (ppt/hr), the full-volume recirculation time $T_r$ (hr), and initial concentration $C_0$ (ppb):

\begin{equation}
\tau(t)\,[{\rm ms}] =
\frac{0.3}{
\sigma\,T_r \;+\; \left(10^9 C_0 - \sigma\,T_r\right)\,e^{-\,t/T_r}
}
\label{eq:sec8-lifetime-evolution}
\end{equation}

From this perspective, a slight increase of the West pump flow from 0.8~kg/s to 1.0~kg/s in Fall 2022,
which decreased the recirculation time from 128~hr to 104~hr, resulted in an increase of the theoretical
lifetime from 4~ms to 5~ms for the same influx $\sigma = 0.55$~ppt/hr and initial impurities concentration
$C_0 = 2$~ppb.

\begin{figure}[htbp]
  \centering
  \includegraphics[width=0.70\textwidth]{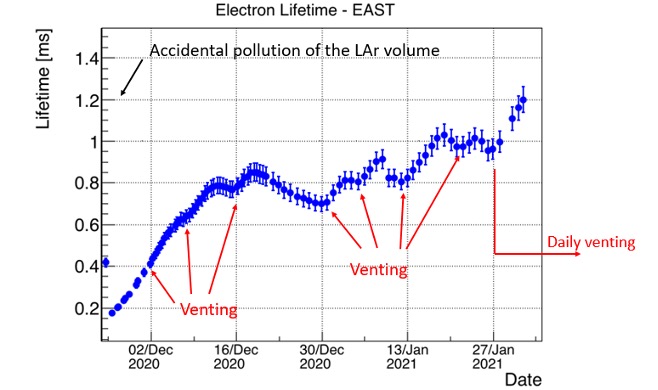}
  \caption{Investigation of purity dependence on venting (East module)}
  \label{fig:sec8-64}
\end{figure}

\begin{figure}[htbp]
  \centering
  \includegraphics[width=0.70\textwidth]{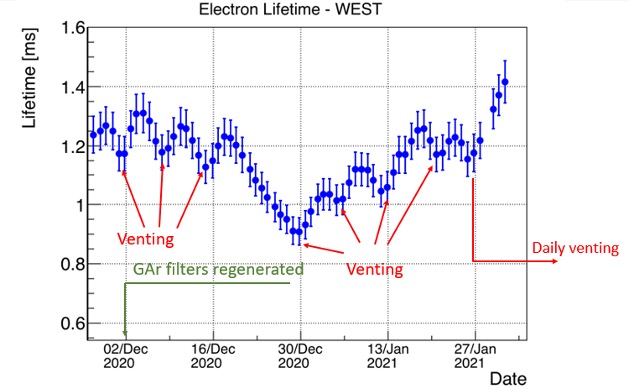}
  \caption{Investigation of purity dependence on venting (West module)}
  \label{fig:sec8-65}
\end{figure}

Starting from November 2020, the argon gas from ullage in both modules was vented at variable
frequency and duration (up to four times a day for 15~min at 50\% opening of the vent valve).
The conclusion is that the present gas recirculation rate is insufficient to efficiently prevent 
impurities to enter the liquid phase. The periodic venting is flushing the impurities that accumulate 
in the chimneys (Figures~\ref{fig:sec8-64}, \ref{fig:sec8-65}).

One of the issues that may have impacted purification was the design decision to replace Oxysorb media
(Messers-Griesheim GmbH), previously used in Icarus at Gran Sasso and at CERN, with copper media
Cu-0226 from BASF. Oxysorb is a chromium-based media that binds oxygen by chemical reaction, while
Cu-0226 is a highly dispersed copper oxide impregnated on high-surface-area alumina which serves to
remove O$_2$ and, to a lesser extent, H$_2$O. While Oxysorb and Cu have similar absorption capacities at
room temperature, leading to re-use of the original volumetric sizing, the Cu media appeared to exhaust
sooner than expected at liquid argon temperature.

Therefore, we first regenerated all four gas condenser-filter cartridges and then added a 10-liter filter
with molecular sieve (30\%) and Cu (70\%) upstream of each condenser-filter unit. This increased the O$_2$
adsorption capacity by a factor $\approx 20$ and added H$_2$O adsorption capacity but did not produce the
desired improvement in lifetimes.

Specifically, the lower lifetime in the West module at the same liquid argon circulation rate pointed to a
potential deficiency in adsorption capacity of the West liquid argon filter. Causes could include:
ineffective activation of the media or unequal flow distribution among the five cartridges, causing one
to load close to breakthrough.

Typically, molecular sieve requires heating above $250^\circ$C for full bake-out, and Cu media requires
heating to at least $165^\circ$C before admitting Ar/H$_2$ mixture to remove O$_2$. Unfortunately, we could not
reach those preheating levels with the original regeneration setup in 2020, so the media may have had
less than 100\% adsorption capacity.

Therefore, the West argon pump was turned off in Fall 2022 and the cartridges in the West argon filter
were regenerated in the improved setup, allowing preheating above $200^\circ$C. After reinstallation of
the filter cartridges and restarting LAr recirculation, the lifetime in the West module stabilized at a level
above 6.5~ms, almost twice higher than before the regeneration and 1.3 times higher than the 
theoretical value for the measured influx of 0.55~ppt/hr

As the recirculation rate remained the same, it must be attributed to the drop in influx $\sigma$ from
0.55~ppt/hr to 0.43~ppt/hr due to venting of the modules four times a day.

Final stages of stabilizations started in May 2021 after the 40CB condenser was modified and brought
online with increased capacity of up to 7.9~g/s. By mid–2023, the following state of the system was
achieved.

\textbf{LN$_2$ circulation system:}

\begin{itemize}
\item \textbf{LN$_2$ “filters’’ pump 60-CF PU-4100}, which supplies LN$_2$ to all shields of the liquid filter boxes,
transfer lines and condensers, is set to run at 82~Hz, 4~bara discharge while maintaining a differential
of 1.2~bara. The “filters’’ pump supplies LN$_2$ to 40FC and 50FC purifier boxes, intercepting heat
from the filter cartridges (five per filter box) and from the long argon transfer lines from the filter
boxes located on the north side to the module entries on the south side. Presently, inlets are opened
36–38\%, while returns are regulated based on temperature between 2.75–2.95~bara.  
The “filters’’ pump also supplies LN$_2$ to 40CA, 40CB, 50CA and 50CB condenser boxes. Each
condenser is regulated individually. LN$_2$ supply from the pump is common to each condenser pair
(40 or 50), while the 2-phase nitrogen returns flow back to the phase separator (PS) through
individual transfer lines.  
There are no isolation valves in the delivery transfer line upstream of each condenser; the only isolation is via the
control valve located 750~mm deep in the condenser box, which results in accumulation of ice inside
40CA and 50CA during maintenance.

\item \textbf{LN$_2$ “shields’’ pump 60-CT PU-4300}, supplying LN$_2$ to all shields of the liquid filter boxes,
transfer lines and condensers, is set to run at 84~Hz, 4.2~bara discharge while maintaining a
1.6~bara differential. Ten individual LN$_2$ shield circuits are fed via inlet valves (60-TI) and returned
via outlet valves (60-TO).  
While inlet valves are kept in manual position (adjusted to maintain pump flow in the desired
range), outlet valves regulate pressure (and therefore temperature) of the shields.  
To maximize boil-off and cleaning of the argon ullage, the shields are operated at $\sim 3.6$~bara.  
Control-valves limits are set to allow for appropriate regulation range.
\end{itemize}

\textbf{LAr circulation system:}

\begin{itemize}
\item \textbf{West 40-CP PU-6000 pump (BNCP-32C)}  
Set to 42~Hz, 1.33~bara discharge, maintaining $\sim 0.96$~kg/s flow to the West module via 40FC filter box.

\item \textbf{East 50-CP PU-6500 pump (BNCP-32E)}  
Set to 49.1~Hz, 2.0~bara discharge, maintaining $\sim 0.96$~kg/s flow to the East module via 50FC filter box.
\end{itemize}

\textbf{Condensers:}

Gaseous argon is collected from the modules’ ullages via individual connections to each chimney. There are 10+10 chimneys
on the north side feeding condensers 40CA/50CA, and 10+10 on the south side feeding 40CB/50CB.
The goal is to collect and re-circulate gas to control pressure in the module while re-circulating (and therefore filtering) 
max amount of gas. Therefore, the objective is achieved by setting the LN2 shields at max pressure (temperature) 
to maintain max boiloff, while setting the condensers at max flow rates.
In mid–2023, the following operating point produced a total condensing rate of 14.2–14.5~g/s per module:

\begin{itemize}
\item LN$_2$ shields at 3.53~bara (PS at 2.2~bara, pump at $\sim 250$~g/s), all inlet valves 100\% open.
\item Purifier boxes and Ar transfer lines tuned to 2.9~bara, with inlets at 42\% and 40\%.
\item The condensers are tuned to their almost max capacity:
\begin{itemize}
\item 40CA: $\geq 6.80$~g/s at 100\% Ar, 57\% N$_2$ in, 2.285~bara N$_2$ out
\item 40CB: $\geq 7.35$~g/s at 100\% Ar, 70\% N$_2$ in, 2.297~bara N$_2$ out
\item 50CA: $\geq 6.8$~g/s at 100\% Ar, 57\% N$_2$ in, 2.260~bara N$_2$ out
\item 50CB: $\geq 7.5$~g/s at 100\% Ar, 55\% N$_2$ in, 2.280~bara N$_2$ out
\end{itemize}
\end{itemize}

To improve purity, vent valves are opened 2–3 times per day for a predefined duration at 50\% opening, venting
$\sim 10$~L/day to atmosphere.

The most time consuming and disruptive operations for the Icarus cryogenic system are those
related to the startup of the cryogenic pumps and maintaining stability of the liquid argon and
nitrogen flows to the loads, e.g.\ filters and shields, via reasonably long transfer lines. The liquid
argon and nitrogen pumps by Barber--Nichols require replacement of the bearings every
6000 to 9000 hours. While the service is typically done by Fermilab qualified technicians by using
OEM parts, such service results in interruption of circulation for the argon pumps, which do not
have an inline spare, for at least 72~hours with the deterioration of bulk LAr purity by
$\sim 40$~ppt. As for the service for the nitrogen pumps, it involves cooldown and switchover to the
spare inline pump, which typically requires restart of the entire nitrogen system. That is due to
the design of the common suction manifold from the phase separator to all three pumps. It
manifests in disruption of the nitrogen flow in one pump’s suction and gas vapor locking when
the other pump suddenly starts pumping liquid after being primed. It is also exacerbated by the
loss of the control valve in the shields valve box. Loss of a nitrogen pump then leads to vapor
locking in the long vacuum-jacketed transfer lines leading to the filters or shields. Use of
intermediate phase separators and cooldown valves would improve startup operations.

\begin{figure}[htbp]
  \centering
  \includegraphics[width=0.75\textwidth]{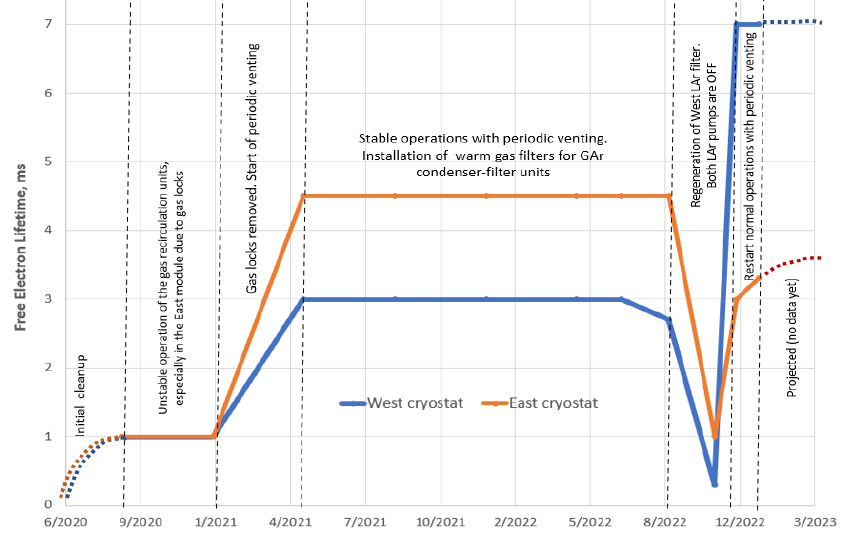}
  \caption{Electron lifetime achieved for the Icarus LAr-TPC detectors (2020–2022)}
  \label{fig:sec8-66}
\end{figure}

As we learned valuable lessons in design and operations of the world’s largest LAr--TPC detector,
the Icarus--T600 demonstrated a remarkable success in taking scientific data in the Fermilab
Booster Neutrino Beam. Already in August~2020, just a few months after the fill, the purity in the
Icarus detectors reached 0.3~ppb, equivalent to 1~ms lifetime, and the detectors started taking
scientific data with cosmic neutrino radiation. The plot in Figure~\ref{fig:sec8-66} shows the electron
lifetime in the Icarus detectors as related to the technical improvements and scientific data taking.

\subsection{Detector operation}

\begin{figure}[htbp]
  \centering
  \includegraphics[width=0.60\textwidth]{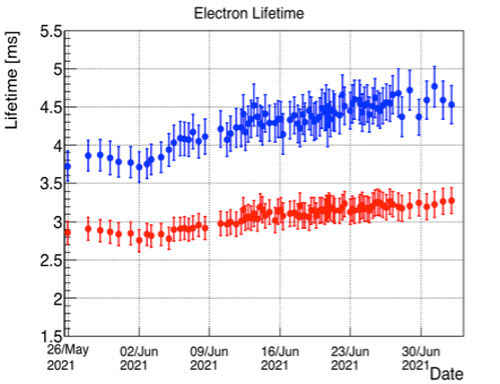}
  \caption{Free-electron lifetime during the first neutrino run (RUN0).  
  Blue: East module; Red: West module.}
  \label{fig:sec8-67}
\end{figure}

In May 2021, the free-electron lifetime was about 3~ms in both modules, and it later stabilized at
about 3.2~ms in the West module and 4.5~ms in the East module (Figure~\ref{fig:sec8-67}). The corresponding
concentrations of impurities (O$_2$ equivalent) are 0.09~ppb and 0.067~ppb, respectively. This allowed
the start of a first continuous neutrino run (RUN0) toward the end of 2021, coinciding with
Fermilab’s beam operation. Some neutrino events collected during this first run are shown in
Figure~\ref{fig:sec8-68} to illustrate the quality of the recorded data.

\FloatBarrier

\begin{center}

  \includegraphics[width=0.90\textwidth]{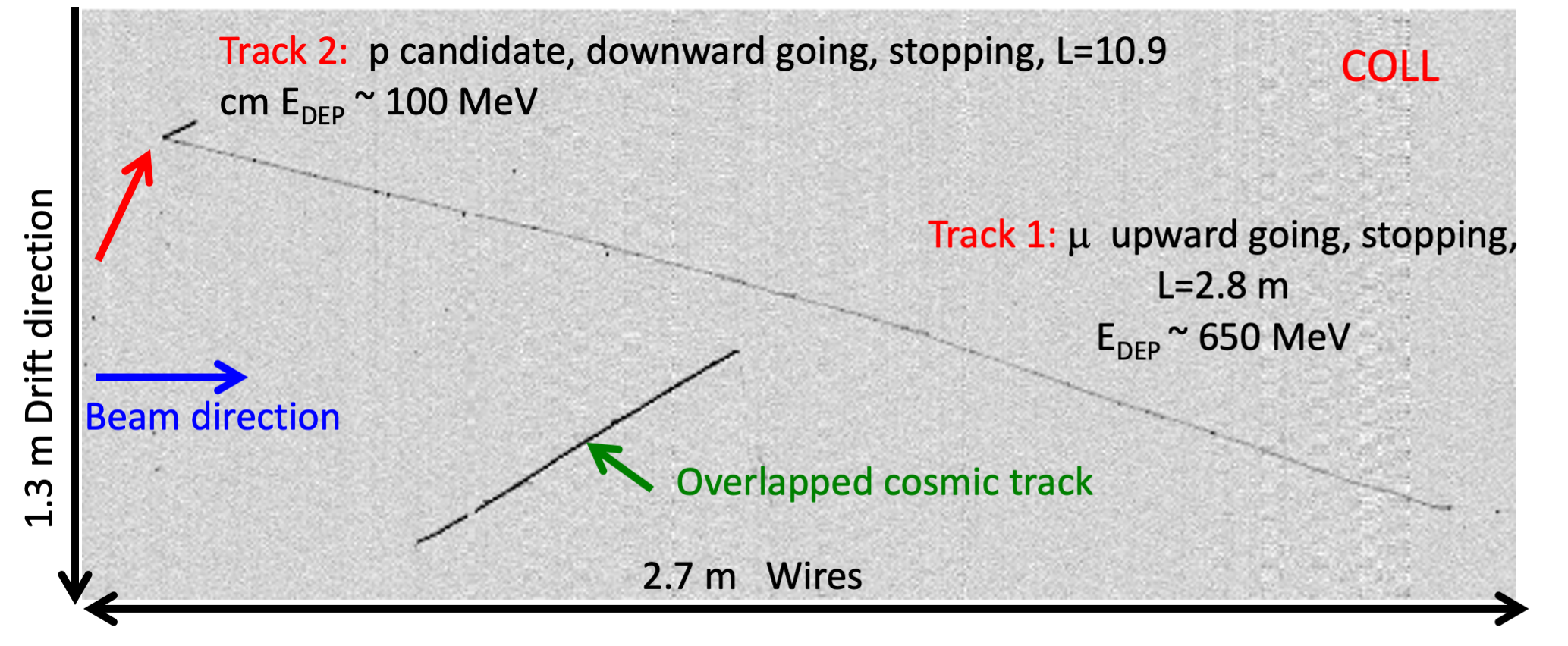}
  
  \includegraphics[width=0.70\textwidth]{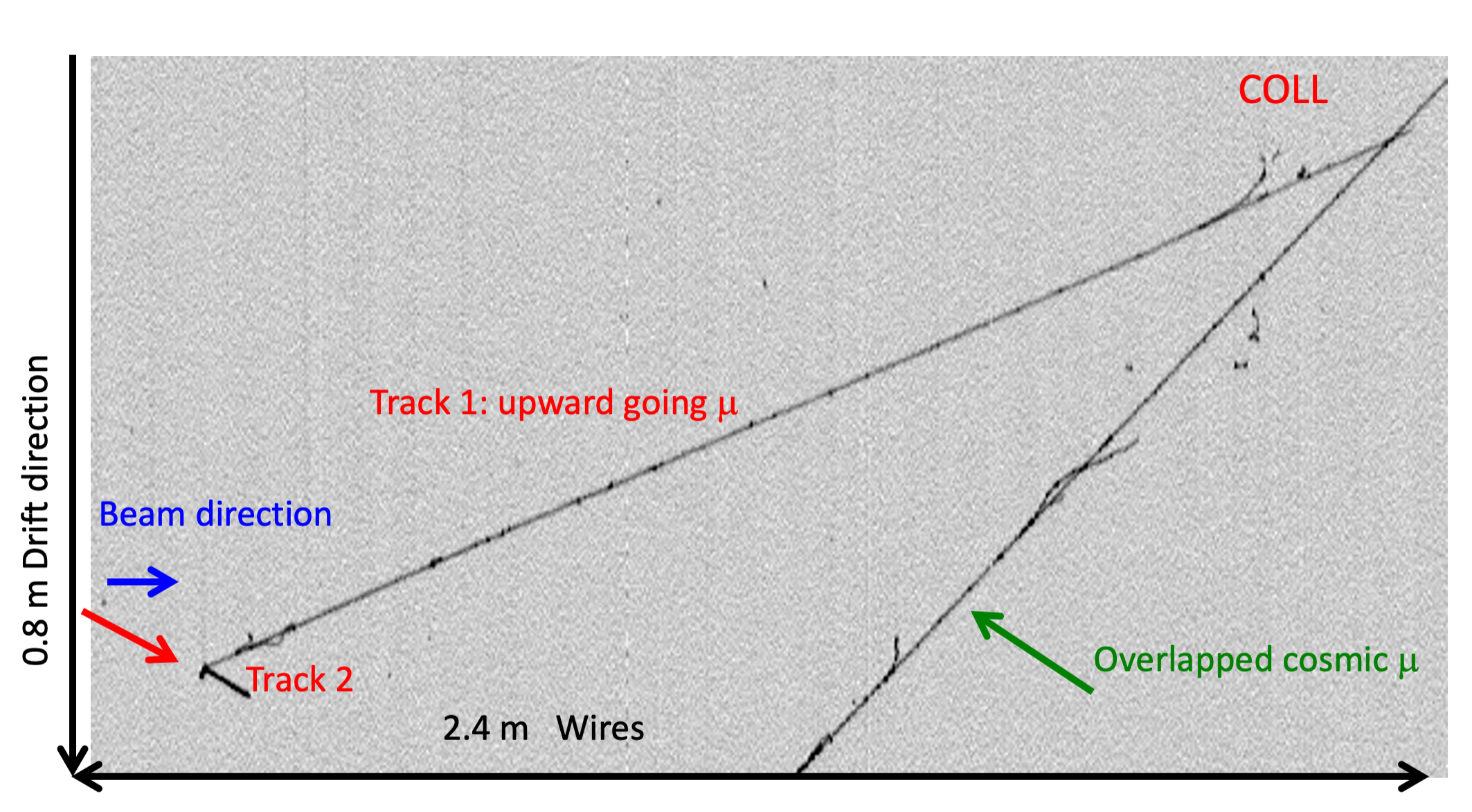}
  
  \includegraphics[width=0.70\textwidth]{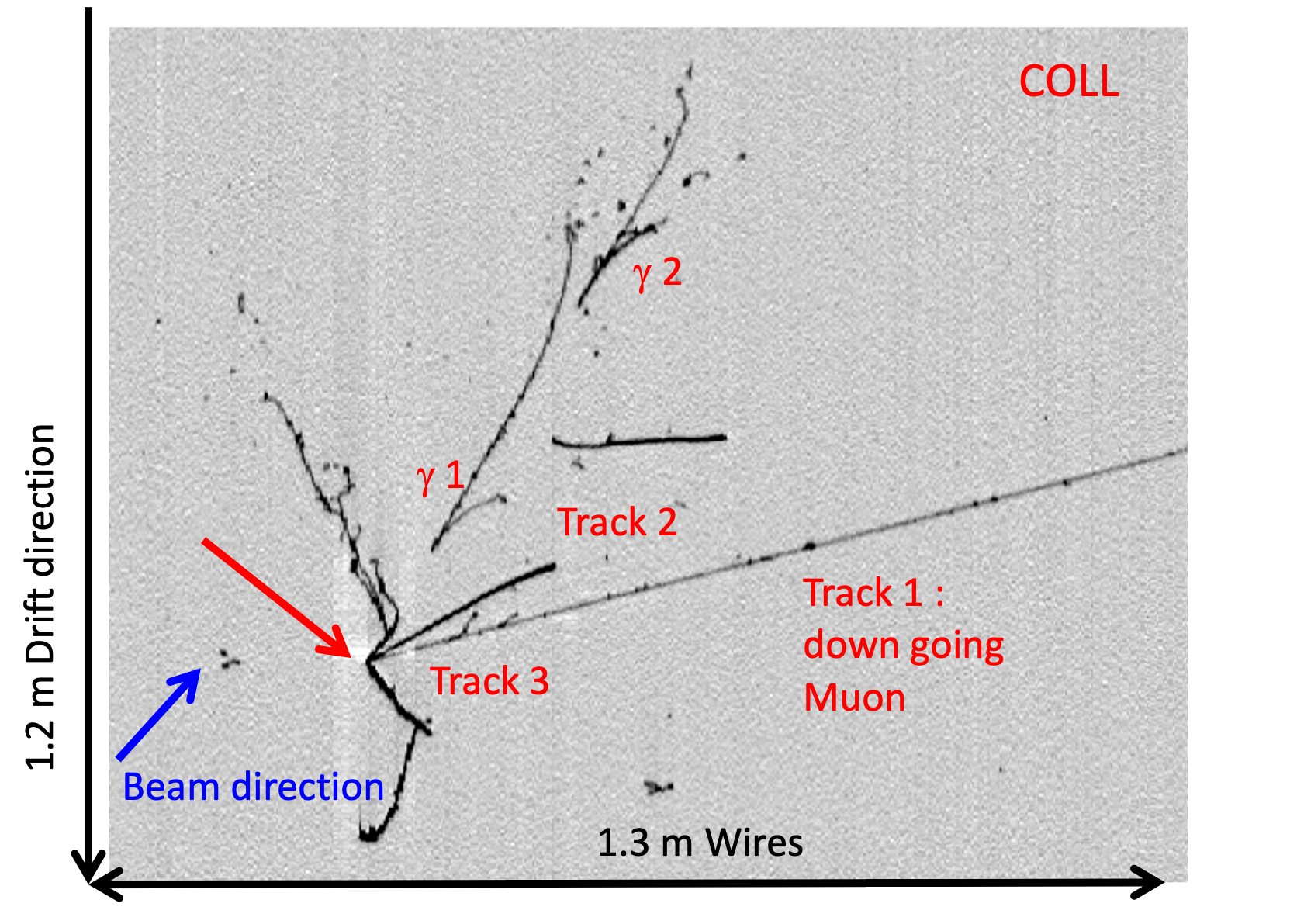}
  
  \includegraphics[width=0.70\textwidth]{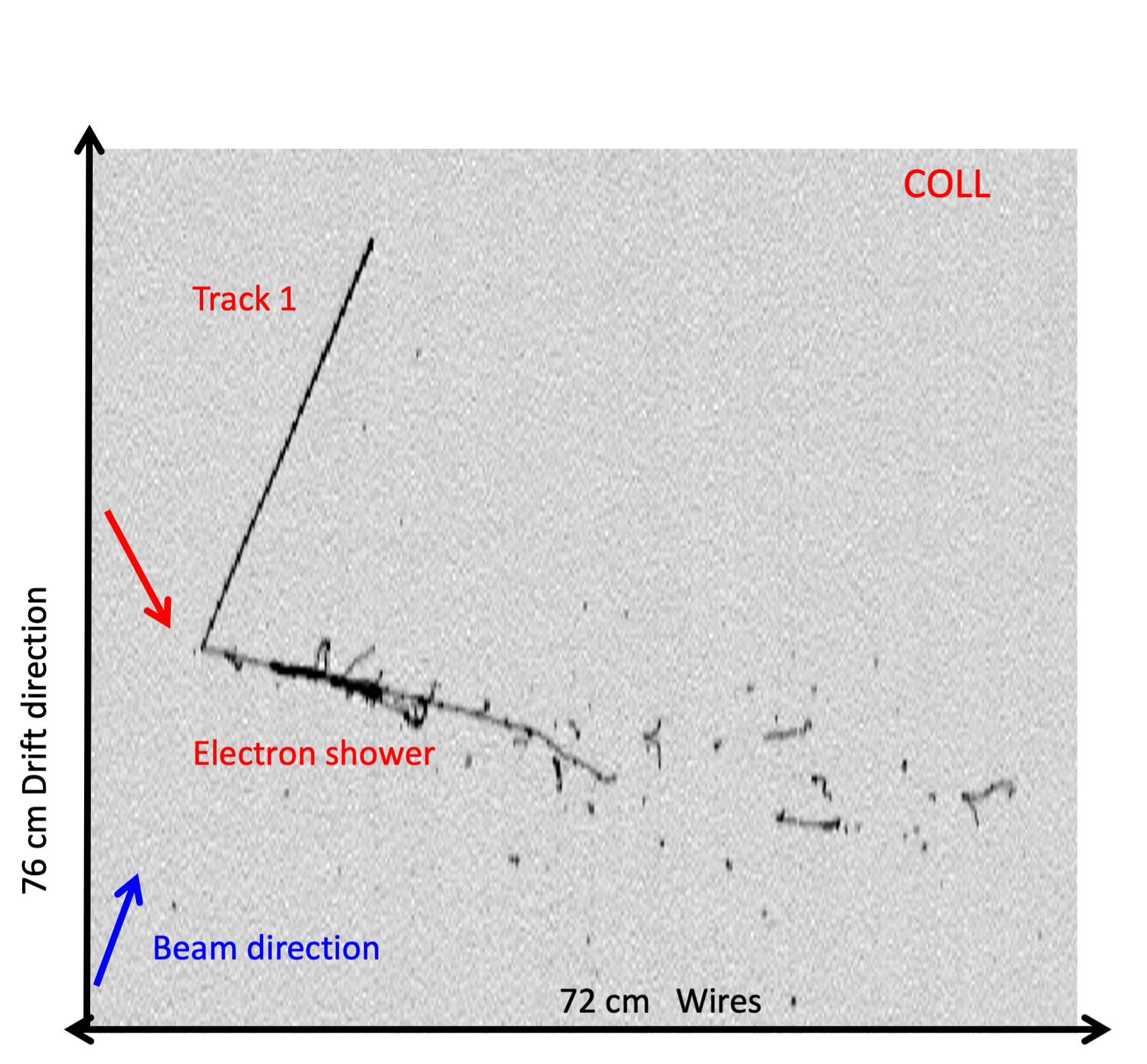}
  
  \captionof{figure}{\small Small sample of neutrino events collected during the first neutrino run (RUN0).  
  The red arrow indicates the neutrino interaction vertex. Only the collection view is shown.  }
  \label{fig:sec8-68}
\end{center}

\FloatBarrier

\begin{figure}[htbp]
  \centering
  \begin{subfigure}[t]{0.95\textwidth}
    \centering
  \includegraphics[width=0.50\textwidth]{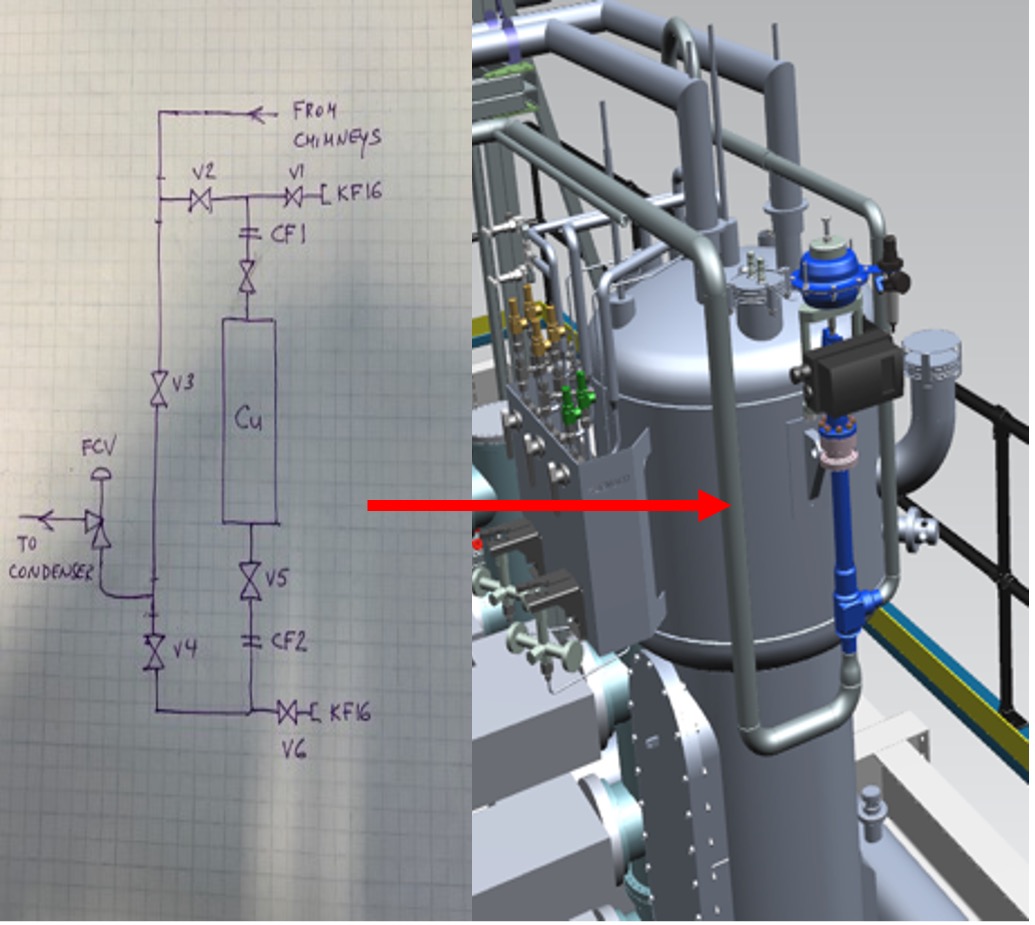}
  \end{subfigure}
  \begin{subfigure}[t]{0.95\textwidth}
    \centering
  \includegraphics[width=0.75\textwidth]{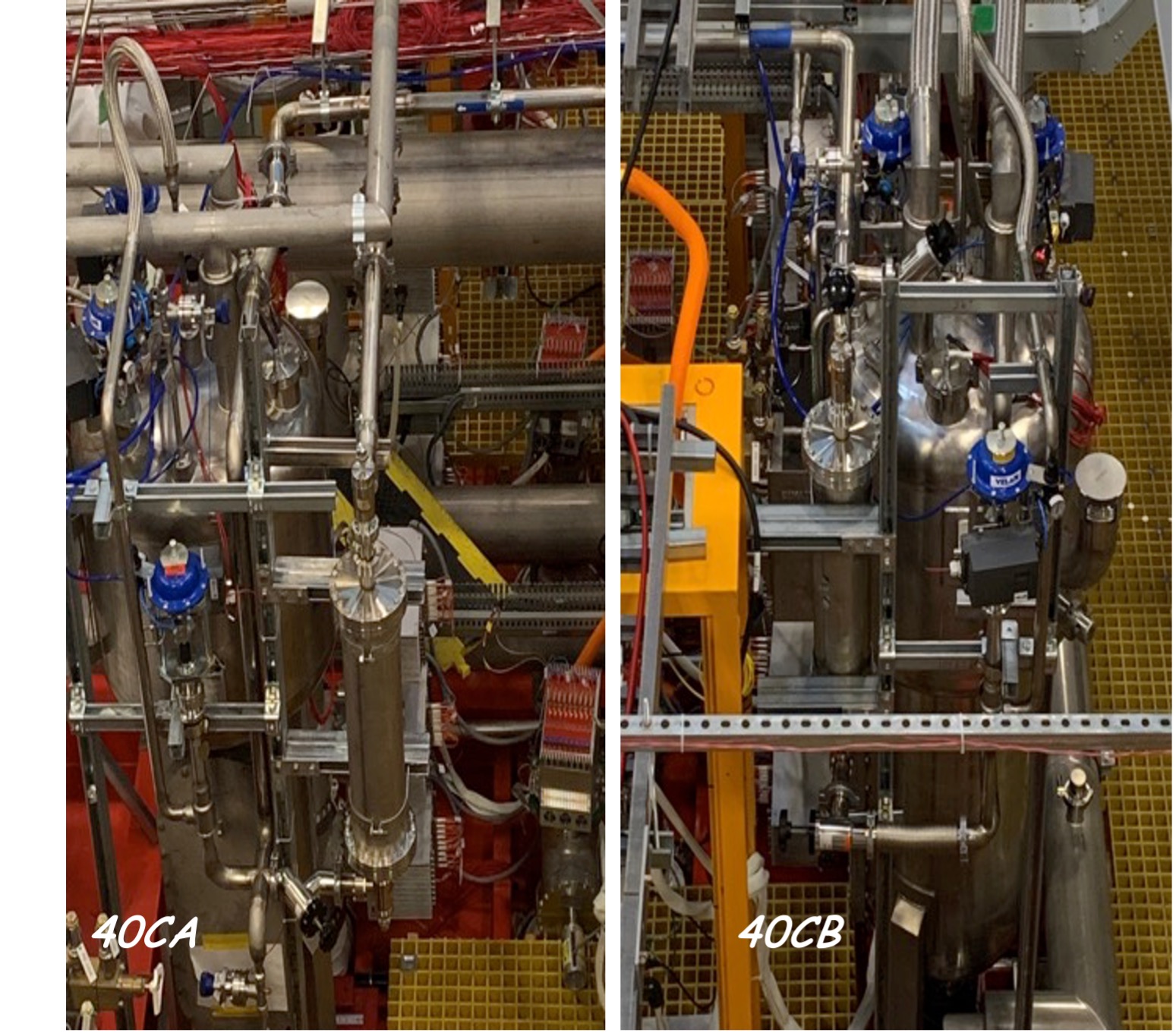}
  \end{subfigure}
  \caption{Top: schematics of warm filters added to gas recirculation circuits.  
  Bottom: photos of the West-module GAr units after installation of warm filters.}
  \label{fig:sec8-69}
\end{figure}

During the summer of 2021, the capacity of the argon gas recirculation system was further increased by
inserting filter cartridges before the inlet of the condensers (Figure~\ref{fig:sec8-69}). The cartridges are filled
one-quarter with molecular sieves 4A and three-quarters with copper medium, operate at room
temperature, and have a total volume of about 11~L of filtering media. The additional filters enhanced
H$_2$O removal from the gas recirculation system and increased filtering capacity by about a factor of 20,
eliminating the need for periodic regeneration of the cold cartridges, with considerable advantages in
terms of reduced maintenance effort, reduced operational risk, and improved continuity of operations.
Insertion of the warm filters into the GAr recirculation circuit resulted in a pressure drop of about
10~mbar, with minimal effect on recirculation speed.

During the summer of 2022, after the first long-duration neutrino run (RUN1) and following a slow but
steady decrease of the free-electron lifetime in the West module, it was decided to regenerate the
corresponding filters of the liquid-argon recirculation system. The regeneration—also affected by a
small accident during this nontrivial operation—resulted in a stop of liquid recirculation in the West
module for about two months and about three weeks in the East module, where the filters were not
regenerated. Normal operation was restored at the beginning of October 2022 in the East module and
about one month later in the West module, where the liquid-recirculation pump was also upgraded.

\begin{figure}[htbp]
  \centering
  \begin{subfigure}[t]{0.95\textwidth}
    \centering
  \includegraphics[width=0.9\textwidth]{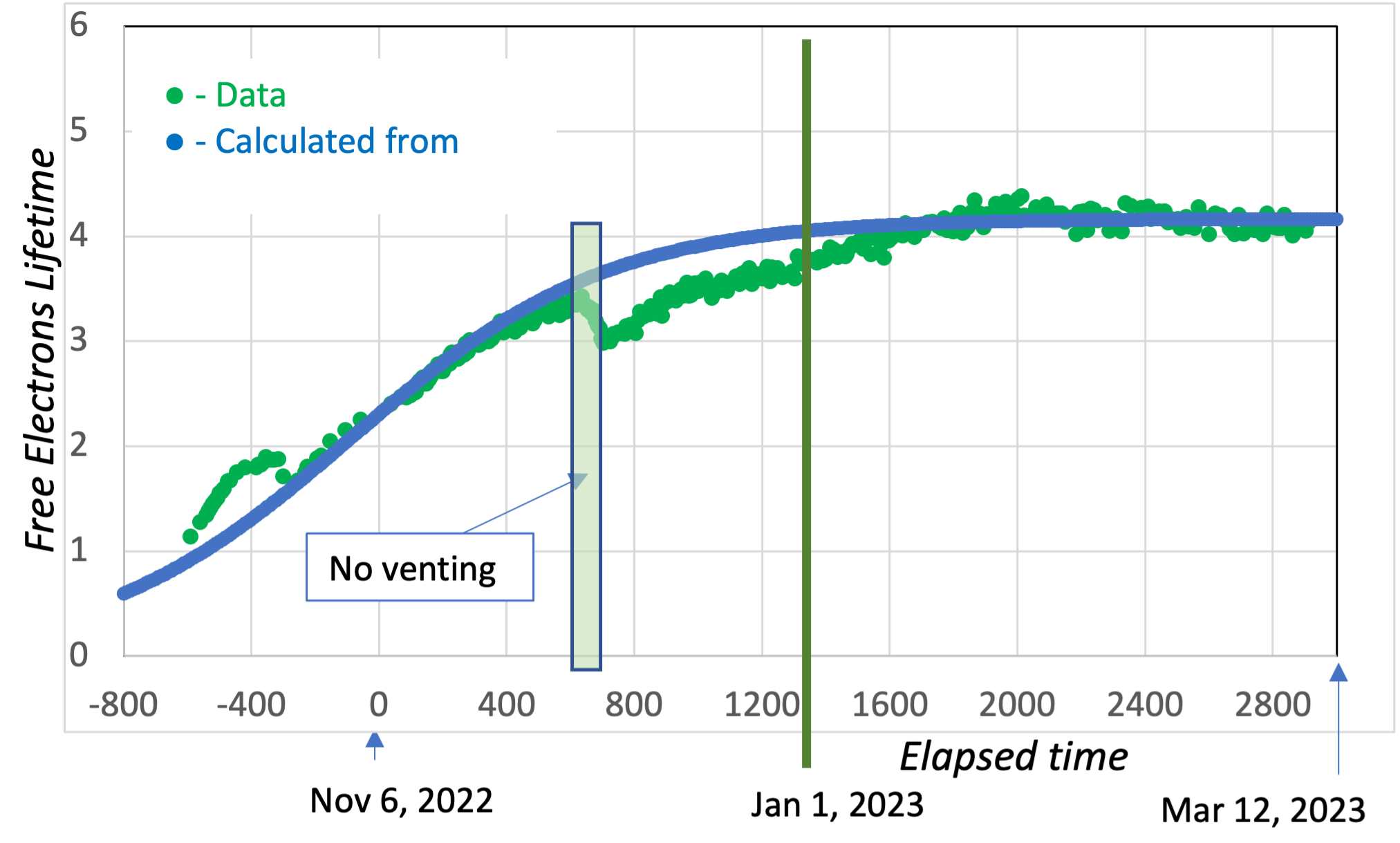}
  \end{subfigure}
   \begin{subfigure}[t]{0.95\textwidth}
    \centering
  \includegraphics[width=0.9\textwidth]{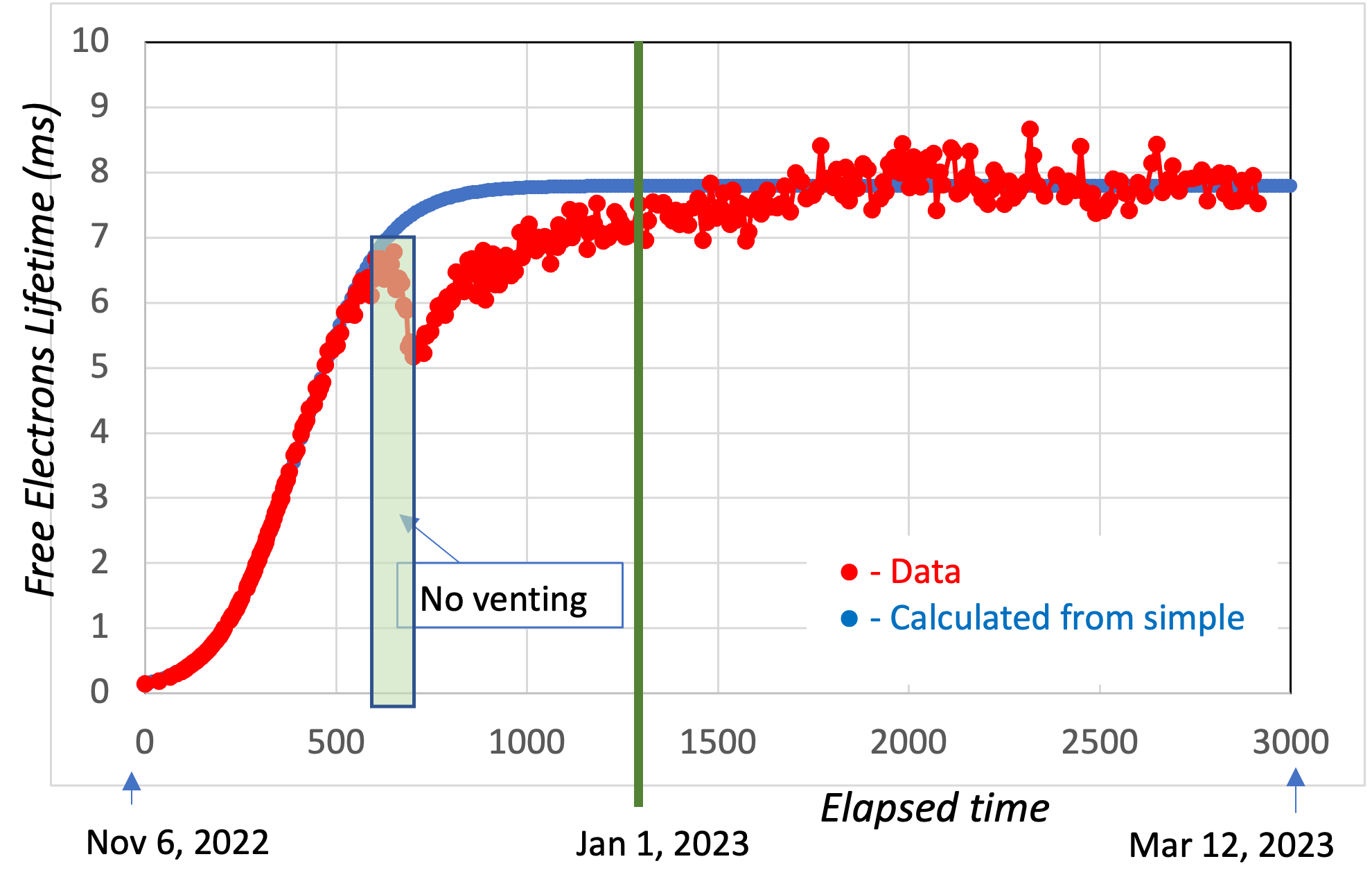}
  \end{subfigure}
 \caption{Free-electron lifetime evolution after restart of LAr recirculation at end of 2022.  
  Top: East module; Bottom: West module.  
  A simple recirculation model prediction is superimposed.}
  \label{fig:sec8-70}
\end{figure}

\begin{figure}[htbp]
  \centering
  \includegraphics[width=0.9\textwidth]{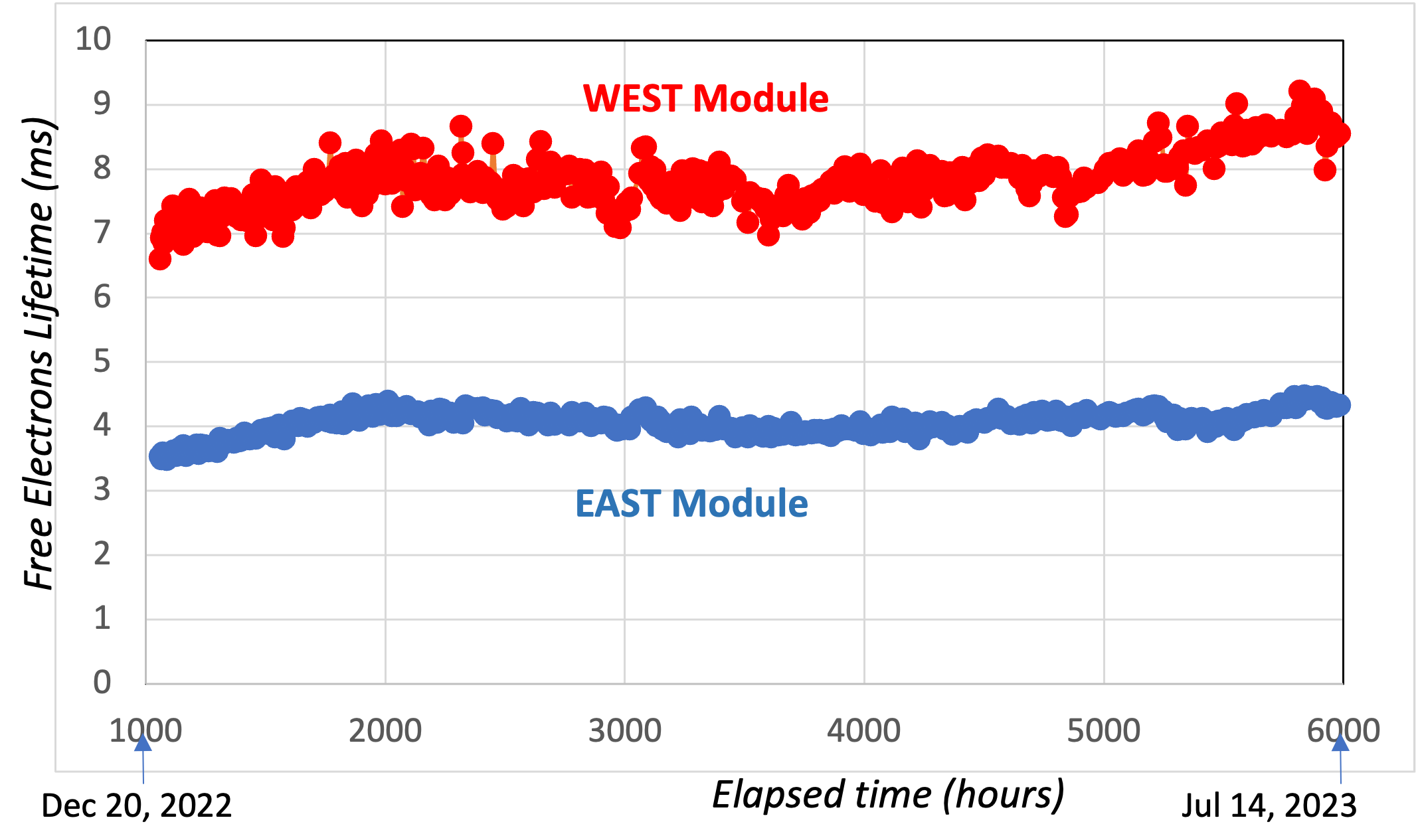}
  \caption{Free-electron lifetime evolution during neutrino RUN2}
  \label{fig:sec8-71}
\end{figure}

After restart of liquid recirculation, the free-electron lifetime in the West module increased rapidly,
whereas the improvement in the East module was significantly slower (Figure~\ref{fig:sec8-70}). The
free-electron lifetime stabilized at about 8~ms in the West module and 4~ms in the East module during
January 2023, allowing a very successful second long-duration neutrino run (RUN2) (Figure~\ref{fig:sec8-71}).

The liquid filters of the East module were also regenerated in late summer 2023. After restart of liquid
recirculation, the purity in the East module increased rapidly, following a trend very similar to that of
the West module. The free-electron lifetime exceeded 6~ms in less than one month and, at the time of
this writing, is still increasing.

\FloatBarrier

\section{Conclusions}

The cryogenic and purification systems are now fully operational to support scientific mission of
the ICARUS-T600 detector in the Booster and NuMI neutrino beams at Fermilab. Many useful
lessons have been learned and improvements have been made to achieve this success. The plans
include regeneration of the East liquid argon filter in 2023 and adding an additional external filter
to support refills of argon needed due to vented argon.

It must be noted that, after the initial hiccups following the cold commissioning, concentrated on
the stabilization of the gas recirculation systems, whose solution, mainly due to the rather severe
limitations of operating during the hot phases of the COVID-19 pandemic, required several months,
the ICARUS-T600 cryogenic and purification systems have shown a remarkable stability and
performance.

Also considering that the initial cryogenic stabilization occurred during the detector
commissioning, no significant data loss has occurred due to cryogenic interventions and the quality
of the argon, both in terms of purity and thermal stability, has been well within the requirements
for physics quality data since May 2021.

\section{Acknowledgements}

This manuscript has been authored by Fermi Forward Discovery Group under Contract No.\
89243024CSC000002 with the U.S. Department of Energy, Office of Science, Office of High Energy
Physics. L. Pasqualini supported by the research contract per Law 240/2010, Art.~24 (3)(a), and 
D.G.R.~693/2023 (REF. PA:2023-20090/RER -- CUP: J19J23000730002) by FSE+ 2021--2027.

\clearpage
\appendix
\titleformat{\section}
  {\normalfont\large\bfseries}
  {Appendix~\Alph{section}}
  {1em}
  {}

\section{Process Safety}
\label{app:ProcessSafety}

Fermilab is managed by the Fermi Forward Discovery Group for the U.S. Department of Energy Office 
of Science under Contract No. 89243024CSC000002. As such, Fermilab must abide by the Appendix I of
DOE Orders and Notices, List B, which is so called Fermilab Work Smart Standards. The Work Smart
Standards lists a large number of governing documents applicable to the management and operation
of Fermilab, for example, specific Codes of Federal Regulations, ANSI/ASME standards and Fermilab
ES\&H Manual (FESHM) chapters.

As the components of the cryogenic systems were delivered by three different institutions and their
contractors but installed at Fermilab, which is operated by U.S.\ Department of Energy, one of the main
challenges was to demonstrate compliance to all governing codes and standards that are part of
Fermilab Work Smart Set of Standards. As a result, Fermilab conducted extensive studies for
equivalency between applicable U.S.\ National Consensus Codes and E.U.\ International Codes. That
allowed Fermilab to approve use of cryogenic equipment designed and constructed per E.U.\
International Codes, such as pressure vessels, piping, reliefs, etc. See Table~\ref{tab:sec7-equivalency}.

In addition, it was demonstrated, but not officially approved, that minimum required pressure relieving
capacities required for vacuum insulated cryogenic vessels per EN 21013-3 (previously 13648) equal or
greater than the minimum required relieving capacities required for same (process or storage) vessels
per U.S.\ standards API~521 and CGA~S-1.3. It was also demonstrated, but not officially approved, that
the requirements per EN~13458 are commensurate with requirements per CGA~A-1.3 as applied to the
cryogenic static vacuum insulated vessels.

The path to receive Operations Readiness Clearance for the Icarus cryogenic system and its pressurized
equipment, such as cryostat modules, vessels and piping, was based on verification of conformity of
the installed equipment to the governing standards (see Table~\ref{tab:sec7-equivalency}). The verification process,
which is established at Fermilab through FESHM, involves describing each pressure or vacuum vessel,
piping system, or structure and their safety in a so-called Engineering Note. The Engineering Notes are
then reviewed by a group of experts and safety professionals before issuing permission for final
validation tests and recommendation for operations. Such reviews were straight-forward for the
equipment and systems that were either marked by the manufacturer with U stamp (ASME) or CE stamp
(PED), e.g.\ cryogenic dewars, filter vessels, nitrogen phase separator, etc.

\begin{table}[htbp]
\centering
\caption{Established equivalency between U.S.\ National Consensus Codes and E.U.\ International Standards}
\label{tab:sec7-equivalency}
\begin{tabular}{p{0.08\textwidth} p{0.28\textwidth} p{0.27\textwidth} p{0.27\textwidth}}
\hline
\textbf{Study} & \textbf{Component Type} & \textbf{National Consensus Code} & \textbf{International Code} \\
\hline
1 &
Pressure Vessels &
ASME BPVC VIII &
EN13445 \\
2 &
Process Piping &
ASME B31.3 &
EN13480 \\
3 &
Pressure Relief Devices &
ASME BPVC VIII and API521 &
EN4126 \\
4 &
Structures &
IBC/ASCE7, AISC360, ADM1 &
Eurocode EN1990, EN1991, EN1993, EN1998 \\
5 &
Electrical Equipment for Measurement, Control and Laboratory use &
UL 61010 &
IEC 61010 \\
\hline
\end{tabular}
\end{table}

Validation of safety for the Icarus cryostats was the most challenging aspect of the safety validation.
The cryostats are fully described in Section~\ref{sec:LArContainers}. They are designed to be filled with $\sim$380~tons of liquid
argon at maximum 320~mbarg gas pressure in the ullage (based on relief set pressure).

Due to ullage pressure being below 0.5~barg (though additional $\sim 0.5$~bar of hydrostatic pressure at the
bottom of the vessel), the vessels were reviewed per Fermilab FESHM~5031.5 ``Low-Pressure Vessels
and Fluid Containment''. The vessels are operated under internal pressure during ``normal'' operations,
but they also were pumped down to full vacuum for leak checking and cleanup before commissioning.
The vessels were successfully tested at full mechanical vacuum at CERN and then at Fermilab. Since the
cryostat vessels were not ASME- or CE-marked and used material not listed in the EN~13445 design
code/standard, Fermilab and CERN followed an elaborate path to ensure and validate the pressure
safety of the cryostats. That included redundant protection from overpressure at 320~mbarg with three
(per module) Velan magnetic safety devices and additional VELAN control valve (one per module) at
200~mbarg or below. Only two out of three magnetic devices are sufficient to handle all pressurization
scenarios and are expected to be never opened. The control valve has a flow capacity verified to be
more than enough to relieve all pressurization scenarios except fire and is also interlocked from a
pressure switch set at 200~mbarg.

As CERN demonstrated that design and construction of the cryostat vessels followed Sound Engineering
Practice (SEP) per appropriate design standard EN~13445, which is accepted as equivalent to ASME
BPVC Section~VIII per FESHM, and Fermilab demonstrated independent evaluation by FEA, the vessels
were allowed to be validated by pressurization per approved Risk Assessment. The first phase of
validation by pressurization was done in April~2019 by pneumatic pressurization of each module to
350~mbar (1.1~$\times$ design pressure) while measuring strains with strain gauges mounted in various
locations on the cold vessels. The strains were shown to be about $-20\%$ to $+5\%$ than predicted (relative
to predicted values) with the maximum strain $\sim 0.1\%$ and demonstrated symmetric and linear structural
behavior of the cold vessel. The second phase of validation was done in April~2020 by filling the
modules with liquid argon and additional pressurization of ullage to 225~mbarg while reading the strain
gauges and comparing them with the theoretical values from the FEA analyses. Again, the strains
demonstrated very good correlation with the predicted values under the same loading conditions and
allowed to finalize validation of structural integrity of the modules under combined hydrostatic and
pneumatic loads at cold conditions.

Validation of safety for the Proximity cryogenic equipment presented lesser, but still a challenge. The
Proximity cryogenic equipment, which was manufactured by Demaco, was only categorized where PED
allowed such; most of the vessels and piping were not categorized due to small diameters and relatively
low design pressure and therefore were executed per Sound Engineering Practice (SEP) per PED,
article~4, par.~3. As CERN/Demaco delivered all equipment packaged in vacuum-jacketed valve boxes and
transfer lines, Demaco guaranteed that QA/QC for the design and manufacturing processes are certified
by a Notified Body, Lloyds Register, to conform to Modules H/H1. Demaco issued a statement that it
had one universal quality system that has been determined to comply with Module H/H1 by the
Notified Body (Lloyds Register) for the categorized products. All Demaco PED CE-marked products and
those products that were not marked (SEP) were designed and manufactured to the same assessment
procedures. As the two guiding standards, EN~13445 and EN~13480, per which Demaco designed,
manufactured and tested the equipment, were already accepted in FESHM as being equivalent to ASME
BPVC Section~VIII and B31.3, the Proximity cryogenic equipment met the requirements of FESHM. The
Proximity cryogenic equipment is protected with dual switchable Herose pressure reliefs sized for all
pressurization scenarios, including the fire conditions and loss of vacuum.

Since the Icarus cryogenic system is mostly housed in the SBNFD detector building, the release of
cryogens in building can cause depletion of oxygen and an Oxygen Deficiency Hazard (ODH) for
personnel. The total anticipated fatality rate from spilling cryogens into building was assessed per
FESHM~4240. This ODH assessment is a quantitative assessment of the increased risk of fatality from
exposure to reduced atmospheric oxygen due to failure of components and release of inert gases into
building. It is a methodology of calculating a total fatality rate and assigning an appropriate ODH class
for each of the areas of the building. The analysis typically uses most conservative assumptions, for
example release to $t \rightarrow \infty$, conservative assumption for failure rates, leak rates and escape paths.

The ODH assessment for SBNFD building resulted in calculating the total fatality rate (combined for all
areas and operations) for mezzanine, pit and top of cryostat as $5.76\times 10^{-8}$~1/hr, therefore requiring
assigning ODH0 class. Still, while the area above the grade level is assigned classification of Engineered
ODH Class~0, all areas below grade level (mezzanine, pit, top of cryostat) are administratively assigned
a higher ODH Class~1. Therefore, the following ODH control measures were implemented for the SBNFD
building before introducing any inert gases into building:

\begin{itemize}[leftmargin=1.5em, labelsep=0.5em, itemsep=0pt, topsep=0pt, parsep=0pt]
    \item Placing ODH equipment on emergency generator power, including ODH monitors, ventilation
  equipment and controls equipment, including PLC.
    \item Detection with six (6) ODH sensors installed at strategic locations where the spilled cryogens may
  propagate or settle.
    \item Installation of vacuum sensors in the insulating space of the cryogenic transfer lines connecting to
  the side penetrations. The side penetrations are closed on loss of vacuum or ODH alarming at 18\%.
    \item Ventilation with two (2) ODH fans of total capacity of 15,000~scfm (7~m$^3$/s), plus additional fans for
  the access stairways. Minimum ventilation rate of $\sim 20\%$ is always sustained, while ramping up to
  full capacity activated on ODH alarming at 19.5\% oxygen. The fans are automatically tested for full
  capacity once every 24~hours and set to alarm if failing to demonstrate full capacity.
    \item Alarming with 4 strobes and 4 horns placed at strategic locations throughout SBNFD building.
  Personnel is required to evacuate the building immediately when ODH alarms.
    \item Configuring all argon cryogenic supplies and pumps to be shut off at 18\% oxygen.
    \item Providing crash buttons installed for each module on the pit level close to the argon pumps.
    \item Managing entry to SBNFD building via keycard readers, therefore allowing administrative control
  for access. Allowing only ODH-qualified and trained personnel to enter levels below ground while
  maintaining two-person rule.
\end{itemize}

A final Operations Readiness Clearance for the Icarus cryogenic system was obtained in February 2020.
The timeline and details of cryogenic and purification commissioning are described in Section~\ref{sec:CryoComm}.

\section{Icarus Cryogenic system commissioning stages}
\label{app:CommStages}

\begin{table}[H]
\centering
\label{tab:sec8-table6}
{\small
\begin{tabular}{p{0.25\textwidth} p{0.45\textwidth} p{0.25\textwidth}}
\hline
\textbf{Stage of commissioning} & \textbf{Objective} & \textbf{When} \\
\hline
Fill of LN$_2$ and LAr dewars &
Demonstrate reliable leak-tight operations, measure heat losses &
Aug -- Oct 2019 \\[4pt]

Initial pumping mode for modules &
Pumping modules to high vacuum to remove contamination &
May 2019 -- Feb 2020 \\[4pt]

Proximity cryo system purification &
Purification of the valve boxes and transfer lines &
Oct -- Nov 2019 \\[4pt]

Insulation space purge for warm structure &
Purge warm structure and insulation space with GN$_2$ to remove atmospheric moisture and
prepare for cooldown of shields &
Jan -- Feb 2020 \\[4pt]

Starting gas analyzers &
Calibrate and demonstrate reliable operations of gas analyzers &
Jan 2020 \\[4pt]

Activation of cartridge purifiers &
Removal of O$_2$ and H$_2$O from 14 removable cartridges (four 5-liter units for gas re-condensers
and ten 25-liter units for liquid filters) &
Jun -- Jul 2019 \\[4pt]

Initial activation of fill purifier &
Removal of O$_2$ and H$_2$O from 3 filters of the fill purifier (two vessels containing 300 kg of
Cu~0226, and one vessel containing 80 kg of molecular sieves) &
Jan 2020 \\[4pt]

Start of cryogenics control system &
Verification of I/Os, loops, alarms, interlocks, algorithms &
Sep 2019 -- Feb 2020 \\[4pt]

Fill of modules with GAr &
Letting the vacuum up with GAr and stabilizing pressure control &
Feb 2020 \\[4pt]

Cooldown of modules, transfer lines and valve boxes with LN$_2$ &
Cooldown of modules to $\sim$90 K by flowing LN$_2$ by pressure feed and gravity through shields &
Feb 2020 \\[4pt]

Start of LN$_2$ circulation system &
Starting of LN$_2$ pumps and stabilizing LN$_2$ flows &
Feb 2020 \\[4pt]

Fill of modules with LAr &
Receiving 46 loads, verifying for conformance to specification and 24/7 transfer to both modules
while controlling pressure and temperature gradients &
Feb -- Apr 2020 \\[4pt]

Start of gas re-condensers and stabilization of modules’ pressures &
Starting to circulate, condense and purify GAr boiloff from ullage with gas re-condensers &
May 2020 \\[4pt]

Start of LAr circulating pumps &
Starting circulation of LAr pumps via LAr purifiers &
Apr 2020 \\[4pt]

Verification of initial purity &
Use HV drift chambers and TPC to verify initial purity of argon &
Jul -- Aug 2020 \\[4pt]

Improvements for argon purity and pressure control &
Investigations for improvements for the non-functional gas re-condensers (SE and SW units)
finally resulting in internal piping modification &
Aug 2020 -- May 2021 \\[4pt]

Final tuning for pressure stability and design purity &
Tuning shields for maximum temperature and maximum gas formation in ullage to match
maximum capacity of condensers. Periodic venting of gas from ullage to increase bulk purity
of argon in modules. &
Jan 2021 -- May 2021 \\
\hline
\end{tabular}
}
\end{table}

\end{document}